\documentclass[useAMS,usenatbib]{mn2e}
\usepackage{epsfig}
\usepackage{float}
\usepackage{placeins}
\usepackage{graphicx}
\usepackage{epsfig}
\usepackage{bm}
\usepackage{amsfonts}
\usepackage{amssymb}
\usepackage{times}
\usepackage{natbib}
\usepackage{color}
\title[Multi-band variability of blazars]{Multi-band optical-NIR variability of blazars on diverse timescales}

\author[Agarwal et al.]
{Aditi Agarwal$^{1,2}$\thanks{E-mail: aditi@aries.res.in}, 
Alok C.\ Gupta$^{1,2,3}$\thanks{CAS Visiting Fellow}, 
R. Bachev$^{4}$, 
A. Strigachev$^{4}$, 
E. Semkov$^{4}$,
\newauthor  Paul J. Wiita$^{5}$, 
M.\ B\"ottcher$^{6}$,
S. Boeva$^{4}$,
H. Gaur$^{3,7}$,
M. F. Gu$^{3}$,
S. Peneva$^{4}$, 
S. Ibryamov$^{4}$,
\newauthor U. S. Pandey$^{2}$
\\ 
$^{1}$Aryabhatta Research Institute of Observational Sciences (ARIES),
Manora Peak, Nainital -- 263002, India\\
$^{2}$Department of Physics, DDU Gorakhpur University, Gorakhpur - 273009, India \\
$^{3}$Key Laboratory for Research in Galaxies and Cosmology, Shanghai Astronomical Observatory,
Chinese Academy of Sciences, 80 Nandan Road, Shanghai 200030, China \\
$^{4}$Institute of Astronomy and National Astronomical Observatory,
Bulgarian Academy of Sciences, 72 Tsarigradsko Shosse Blvd., 1784 Sofia, Bulgaria \\
$^{5}$Department of Physics, The College of New Jersey, P.O.\ Box 7718, Ewing, NJ 08628, USA\\
$^{6}$Centre for Space Research, North-West University, Potchefstroom, 2520, South Africa \\
$^{7}$Inter-University Centre for Astronomy and Astrophysics (IUCAA), Ganeshkhind, Pune 411 007, India \\
}

\begin{document}

\date{Accepted ....... Received  ......; in original form ......}

\pagerange{\pageref{firstpage}--\pageref{lastpage}} \pubyear{2014}

\maketitle

\label{firstpage}

\begin{abstract}
To search for optical variability on a wide range of
timescales, we have carried out photometric monitoring of two flat spectrum radio quasars, 3C 454.3 and 3C 279, plus one BL Lac, S5 0716+714, 
all of which have been exhibiting remarkably
high activity and pronounced variability at all wavelengths. CCD magnitudes in B, V, R and I pass-bands were determined 
for $\sim$ 7000 new optical observations from
114 nights made during 2011 -- 2014, with an average length of $\sim$~4~h each,
at seven optical telescopes: four in Bulgaria, one in Greece, and two
in India. We measured multiband optical flux and colour variations on diverse timescales.
Discrete correlation functions were computed among B, V, R, and I observations,
to search for any time delays. We found weak correlations in some cases with no significant time lags.
The structure function method was used to estimate any characteristic
time-scales of variability. We also investigated the spectral energy distribution of the three blazars 
using B, V, R, I, J and
K pass-band data.  We found that the sources almost always follows a bluer-when-brighter trend.
We discuss possible physical causes of the observed spectral variability. 
\end{abstract}


\begin{table*}
\caption{ Details of telescopes and instruments}
\textwidth=6.0in
\textheight=9.0in
\vspace*{0.2in}
\noindent
\begin{tabular}{p{2.2cm}p{2.2cm}p{2.3cm}p{2.2cm}p{2.3cm}p{2.3cm}p{2.5cm}} \hline
Site:             &A                               & B                             & C                                    & D/G                                        &E                               &F                                 \\\hline
Scale:            &0.2825\arcsec/pixel             &0.258\arcsec/pixel             & 1.079\arcsec/pixel                   & 0.330\arcsec/pixel$^{\rm a}$             &       0.535\arcsec/pixel       &0.37\arcsec/pixel                          \\
Field:            &$9.6\arcmin\times9.6\arcmin$    &$5.76\arcmin\times5.59\arcmin$ & $73.66\arcmin \times 73.66 \arcmin$  & $16.8\arcmin\times16.8\arcmin$           & $18\arcmin\times18\arcmin$     & $13\arcmin\times13\arcmin$                   \\
Gain:             &2.687 $e^-$/ADU                 &1.0 $e^-$/ADU                  & 1.0 $e^-$/ADU                        & 1.0 $e^-$/ADU                            & 1.4 $e^-$/ADU                  &10 $e^-$/ADU                                 \\
Read Out Noise:   &8.14 $e^-$ rms                  &2.0 $e^-$ rms                  & 9.0 $e^-$ rms                        & 8.5 $e^-$ rms                            & 4.1 $e^-$ rms                  &5.3 $e^-$ rms                                 \\
Typical seeing :  & 1\arcsec to 2\arcsec           & 1.5\arcsec to 3.5\arcsec      & 2\arcsec to 4\arcsec                 & 1.5\arcsec to 3.5\arcsec                 &   1.2\arcsec to 2.0\arcsec                               & 1\arcsec to 2.8\arcsec                   \\\hline
\end{tabular} \\
A  : 1.3-m Ritchey-Chretien telescope at Skinakas Observatory, University of Crete, Greece \\
B  : 2-m Ritchey-Chretien telescope at National Astronomical observatory Rozhen, Bulgaria \\
C  : 50/70-cm Schmidt telescope at National Astronomical Observatory, Rozhen, Bulgaria   \\
D  : 60-cm Cassegrain telescope at Astronomical Observatory Belogradchik, Bulgaria \\
E  : 1.30 meter Ritchey-Chretien Cassegrain optical telescope, ARIES, Nainital, India \\
F  : 1.04 meter Sampuranand Telescope, ARIES, Nainital, India  \\
G  : 60-cm Cassegrain telescope at National Astronomical Observatory Rozhen, Bulgaria \\
$^{\rm a}$ With a binning factor of $1\times1$
\end{table*}

\section{Introduction} \label{sec:intro}
Some of the brightest radio-loud Active Galactic Nuclei (AGN), called blazars, are understood to
have relativistic jets viewed at an angle of $\leq$ 10$^{\circ}$ from the line of sight (LOS) 
(e.g., Urry \& Padovani 1995), which amplifies chaotic flux variability spanning the entire electromagnetic (EM) 
spectrum. The blazar class includes BL Lacertae objects (BL Lacs) and flat spectrum radio quasars 
(FSRQs). BL Lacs exhibit non-thermal continuum emission with essentially featureless 
optical spectra, whereas FSRQs have the prominent emission lines characteristic of quasars. 
These sub-classes differ in wavelength dependent optical polarization properties, with BL Lacs 
displaying increased polarization
toward blue which could be due to some intrinsic phenomenon related to the jet emitting region, while the FSRQs show
an opposite trend, likely arising from the contribution of unpolarized thermal radiation from 
the accretion disc and the surrounding regions
at shorter wavelengths (Smith 1996; Raiteri et al.\ 2012). The observational properties of the blazar 
class of AGN include  broad-band continuum 
spectral energy distributions (SEDs) dominated by non-thermal emission from radio through $\gamma$-rays, 
a relativistic jet, extreme luminosity due to Doppler boosting of the relativistic jet emissions, a 
high degree of polarization, superluminal motion of some of their radio components 
(e.g., Aller et al.\ 1992; Aller et al.\ 2003; Andruchow et al.\ 2005), 
and flux variability on timescales of hours to years at all wavelengths. 

The pronounced variability throughout the electromagnetic spectrum (from radio 
to GeV or even TeV energies) at all accessible time scales ranging from few minutes through months to 
decades, has been generally
divided into three classes: flux changes observed from few tens of minutes to less than a day are commonly 
called intra-day variability (IDV; Wagner \& Witzel 1995, Kinman 1975; Clements, Jenks, \& Torres 2003; 
Rector \& Perlman 2003; Xie et al.\ 2004; Gupta et al.\ 2008b) or intra-night 
variability or micro variability; those from several days to few weeks are known as short time variability 
(STV; Lainela et al.\ 1999; Kranich et al.\ 1999), while long term variability (LTV) covers flux changes 
from several months to many years (e.g., Gupta et al.\ 2004; Jurkevich 1971; Sillanp{\"a}{\"a} 
et al.\ 1988; Liu, Xie, \& Bai 1995; Fan \& Lin 2000; Terrell \& Olsen 1972).

Blazar SEDs exhibit double-peaked structures (Fossati et al.\ 1998).
The low-frequency component is typically observed to peak in the near-infrared (NIR) or optical bands 
for low-synchrotron-peaked blazars (LSPs, consisting of FSRQs and low-frequency-peaked BL Lac objects [LBLs])
and in the UV or X-rays for high-synchrotron-peaked blazars (HSPs, generally belonging 
to the class of high-frequency-peaked BL Lac objects, HBLs; Giommi, Ansari \& Micol 1995). Blazars whose SED peaks 
are located at intermediate frequencies sometimes are termed intermediate-synchrotron-peaked blazars 
(ISPs; Sambruna, Maraschi, \& Urry 1996). Specifically, Abdo et al.\ 2010 defined non-thermal-dominated AGN 
based on the peak frequency of their synchrotron hump, $\nu_{\rm peak}$, such that LSP sources are
those with $\nu_{\rm peak} \leq 10^{14}$~Hz; ISPs have $10^{14} \, {\rm Hz} < \nu_{\rm peak} < 10^{15}$~Hz, 
and HSPs have $\nu_{\rm peak} \geq 10^{15}$~Hz. The second SED component extends up to the \(\gamma\)-ray 
bands, peaking at GeV energies for LBLs and FSRQs, and at TeV energies for HBLs.
The low energy component is commonly ascribed to synchrotron radiation from ultra relativistic electrons in the
relativistic jet. The high-energy component is often interpreted as due to Compton
up-scattering of low energy photons by the same electrons in the jet (leptonic models; Padovani \& Giommi 1995)
although hadronic models, in which protons are assumed to be accelerated to ultra relativistic energies and 
$\gamma$-ray emission results from proton synchrotron radiation and photo-pion production induced processes, 
remain viable (e.g., B{\"o}ttcher et al.\ 2013). \par

The key motivation of this work was to search for optical flux and colour variability of 3C 454.3, S5 0716+714 and
3C 279 on diverse
time scales, including analyses of colour-magnitude variations, inter-band cross correlations,
and optical/NIR 
SEDs. Here we report optical observations of the targets in the B, V, R, and I optical bands monitored  
 from 2011 to 2014.

\section{\bf Observations and Data Reductions}
\label{observations}
Our optical photometric observations of the blazars were performed in the B, V, R, and I pass-bands,
using seven telescopes, two in India, one in Greece, and four in Bulgaria, all equipped with CCD detectors. 
The details of the telescopes, instruments and other parameters used are given in Table 1. Complete details 
of the observations are listed in Tables 1, 2 and 3 of supplementary material available for this article.
IDV light curves (LCs), covering nights when the 
observation runs were at least $\sim$~4 hrs in the B, V, R, and I passbands, are displayed in Figures 1, 2 and 3.
Photometric observations obtained with telescopes listed in Table 1, 
were bias subtracted and twilight flat fielded followed by cosmic ray removal
using standard packages (MIDAS\footnote{ESO-MIDAS is the acronym for the European Southern Observatory 
Munich Image Data Analysis System which is developed and maintained by European Southern Observatory},
IRAF\footnote{IRAF is distributed by the National Optical Astronomy Observatories, which are operated
by the Association of Universities for Research in Astronomy, Inc., under cooperative agreement with the
National Science Foundation.}). Aperture photometry was performed using standard 
IRAF and Dominion Astronomical
Observatory Photometry (DAOPHOT II) software (Stetson 1987; Stetson 1992) routines.
The calibrated LCs are displayed in Figs.\ 1, 2 and 3.
Data from the Steward Observatory (SO) spectropolarimetric monitoring 
project\footnote{\tt http://james.as.arizona.edu/$\sim$psmith/Fermi/} led by P.\ Smith were also used in the optical V 
band.
The SMARTS\footnote{\tt http://www.astro.yale.edu/smarts/glast/3C454.3lc.html} photometric data and 
LCs for all the blazars are publicly available on the web (Bailyn et al.\ 1999).We have taken data from the SMARTS 
web archive for each of the blazars in the B, V, R, and I pass-bands for a few nights for the purpose of investigating 
STV.  We also used archived data from the J and K bands for studying variations in the optical/NIR SED.

\begin{figure*} 
 \epsfig{figure= 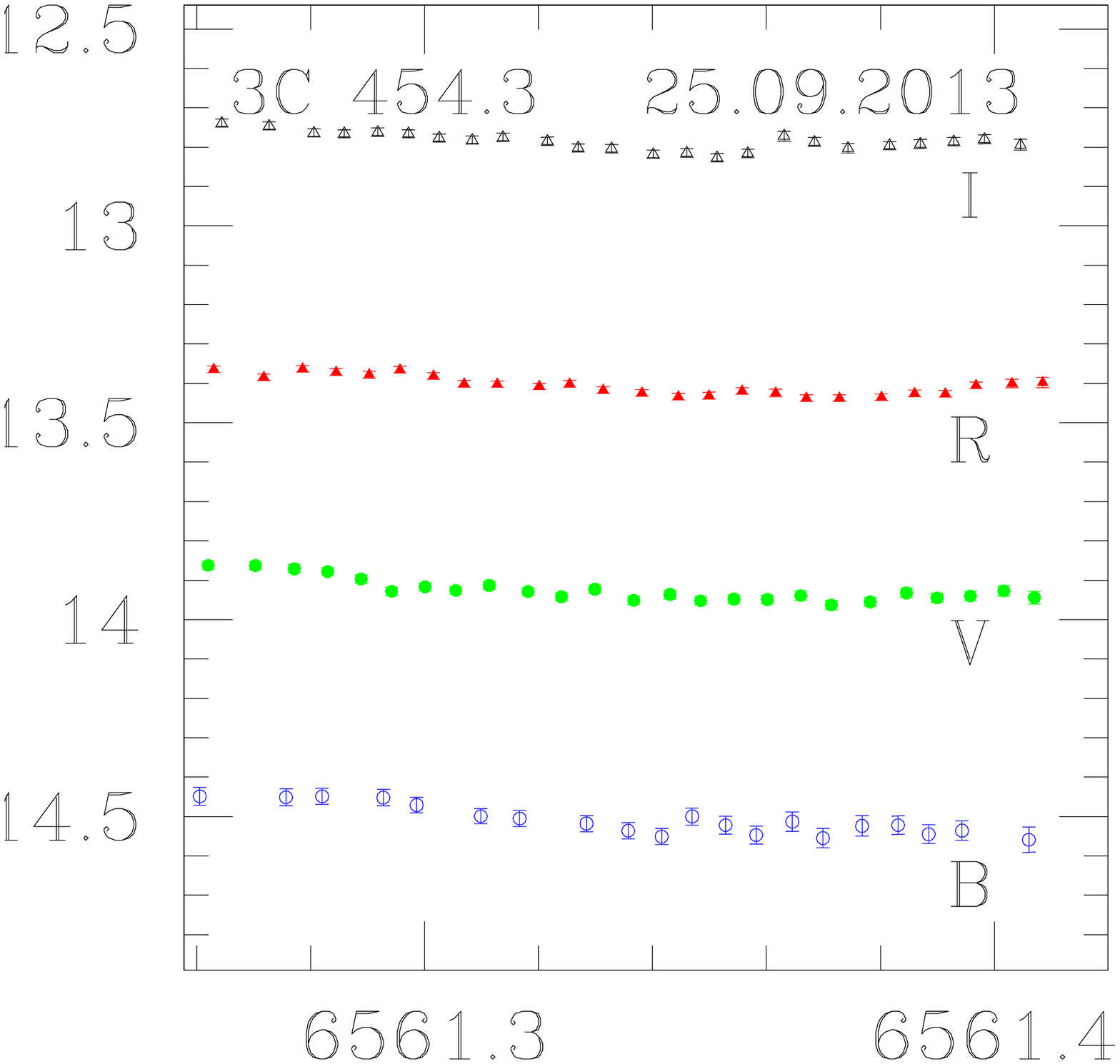 ,height=1.567in,width=1.59in,angle=0}
\epsfig{figure= 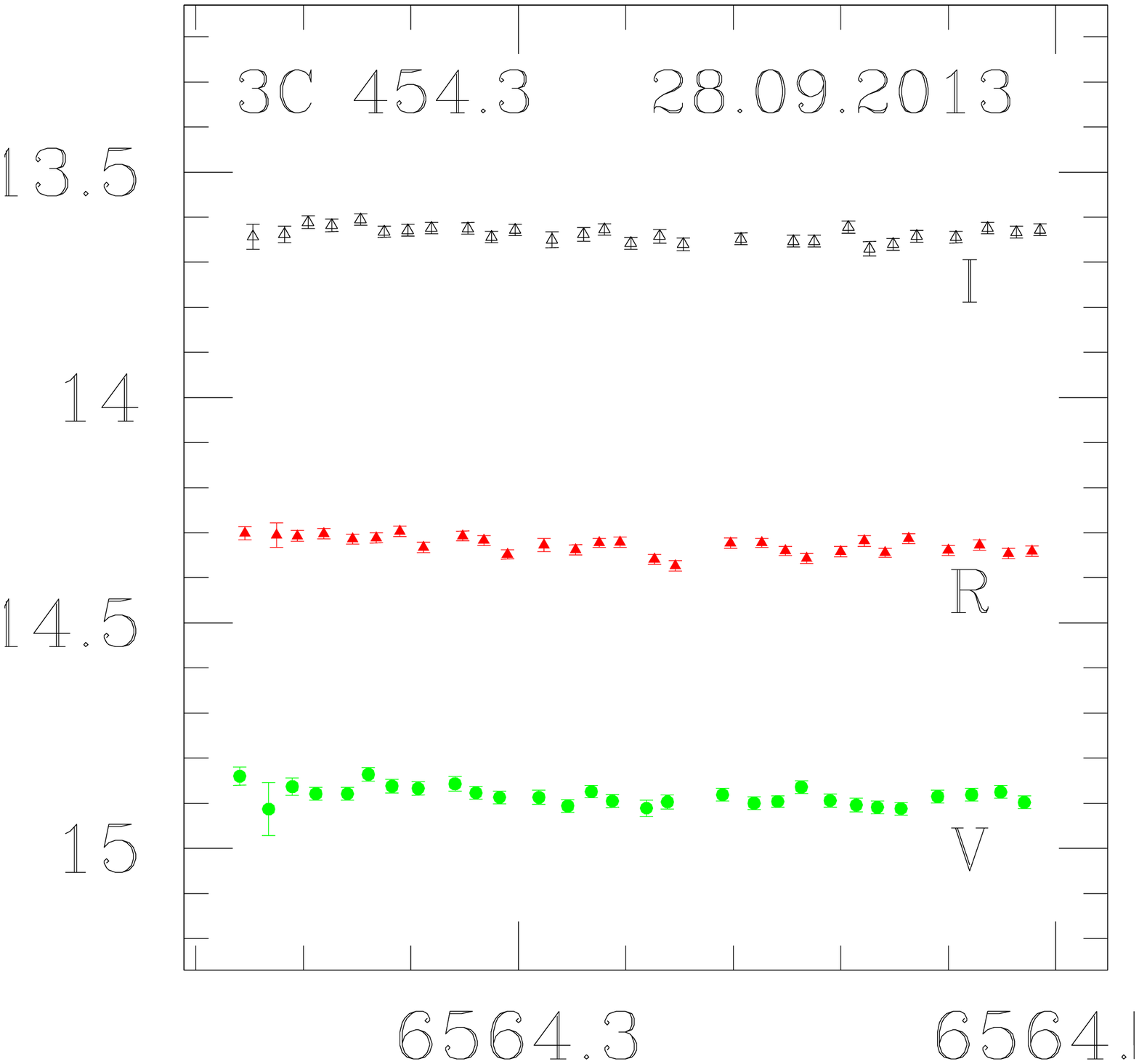 ,height=1.567in,width=1.59in,angle=0}
\epsfig{figure= 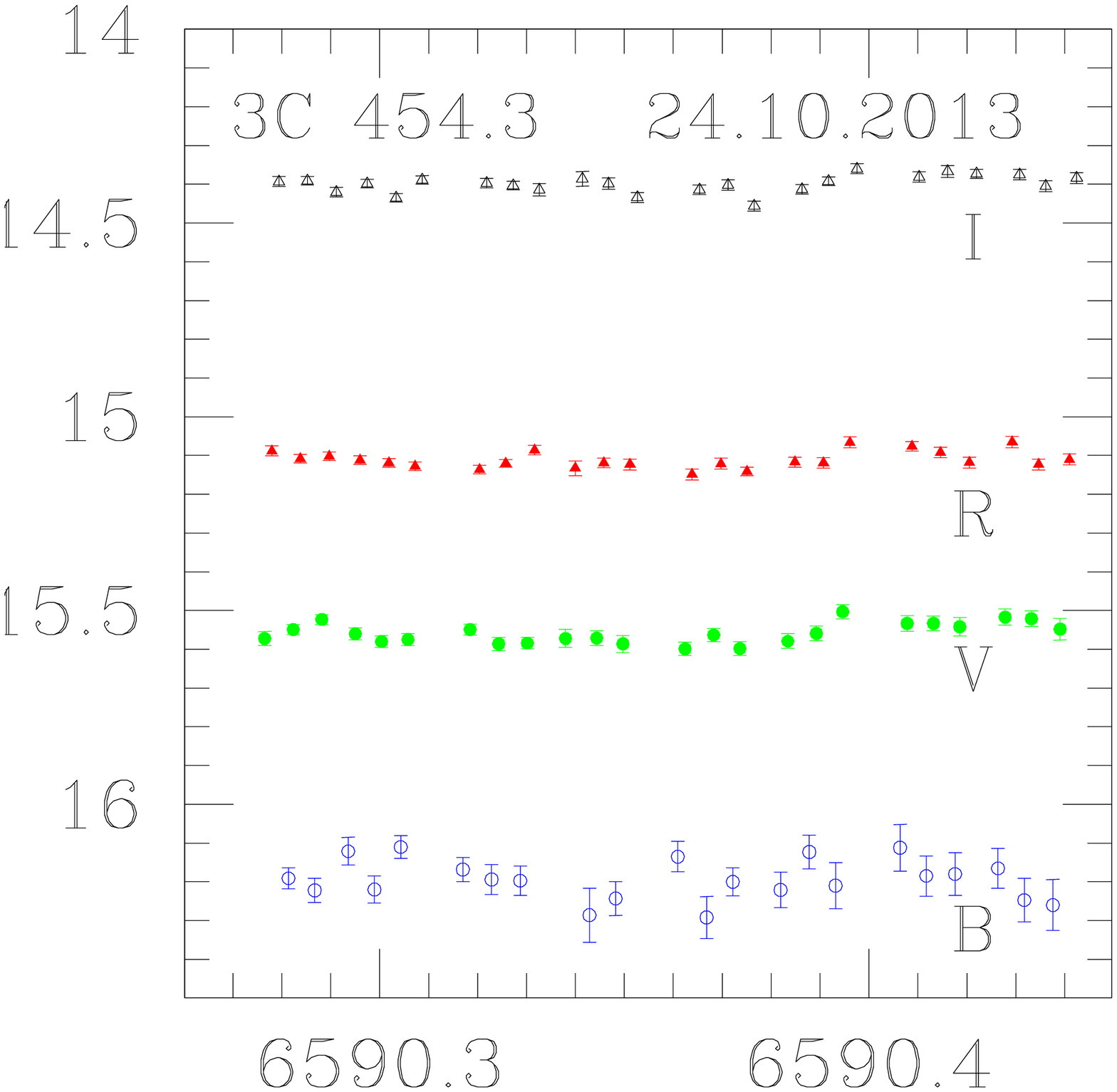 ,height=1.567in,width=1.59in,angle=0}
\epsfig{figure= 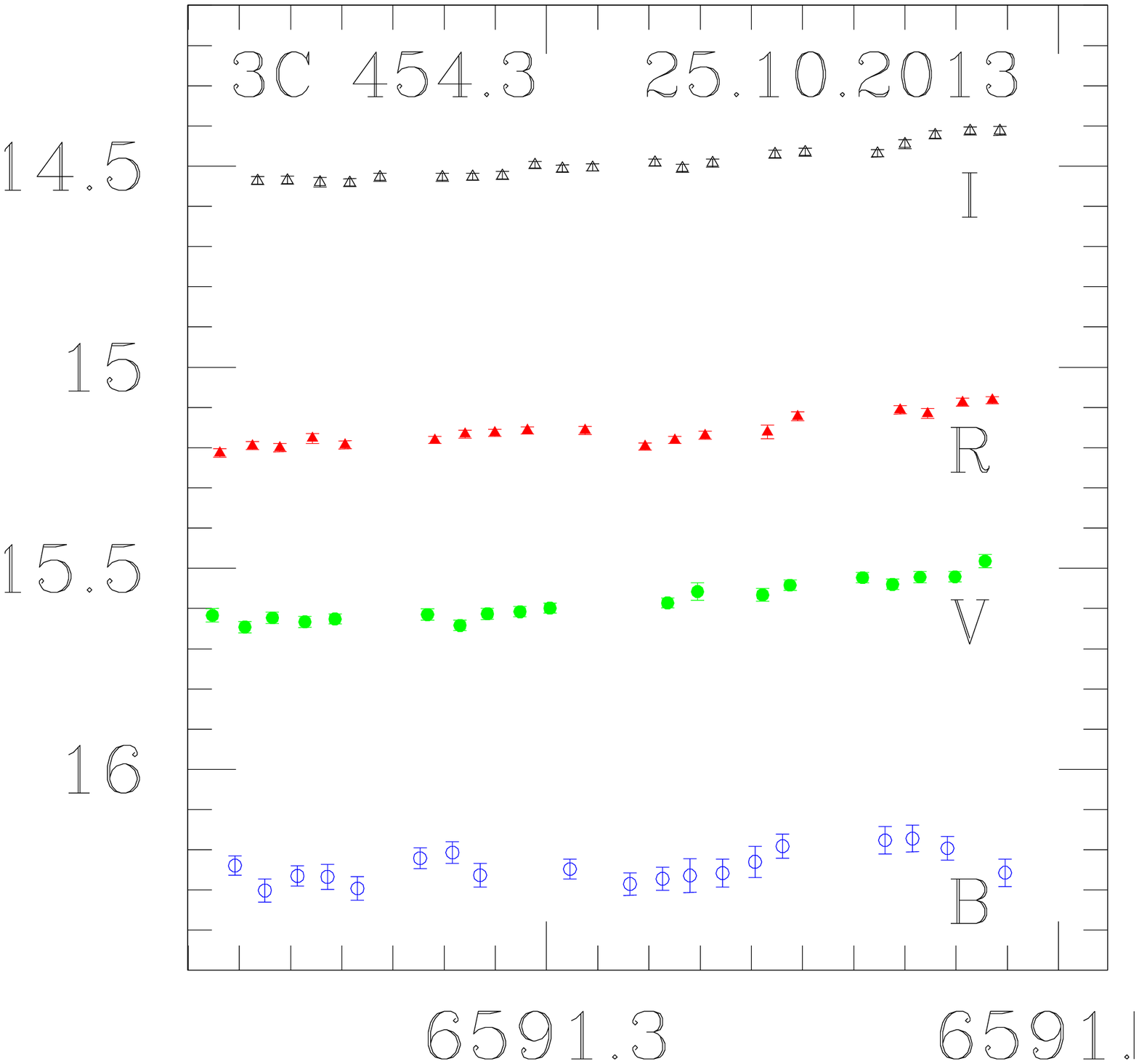 ,height=1.567in,width=1.59in,angle=0}
\epsfig{figure= 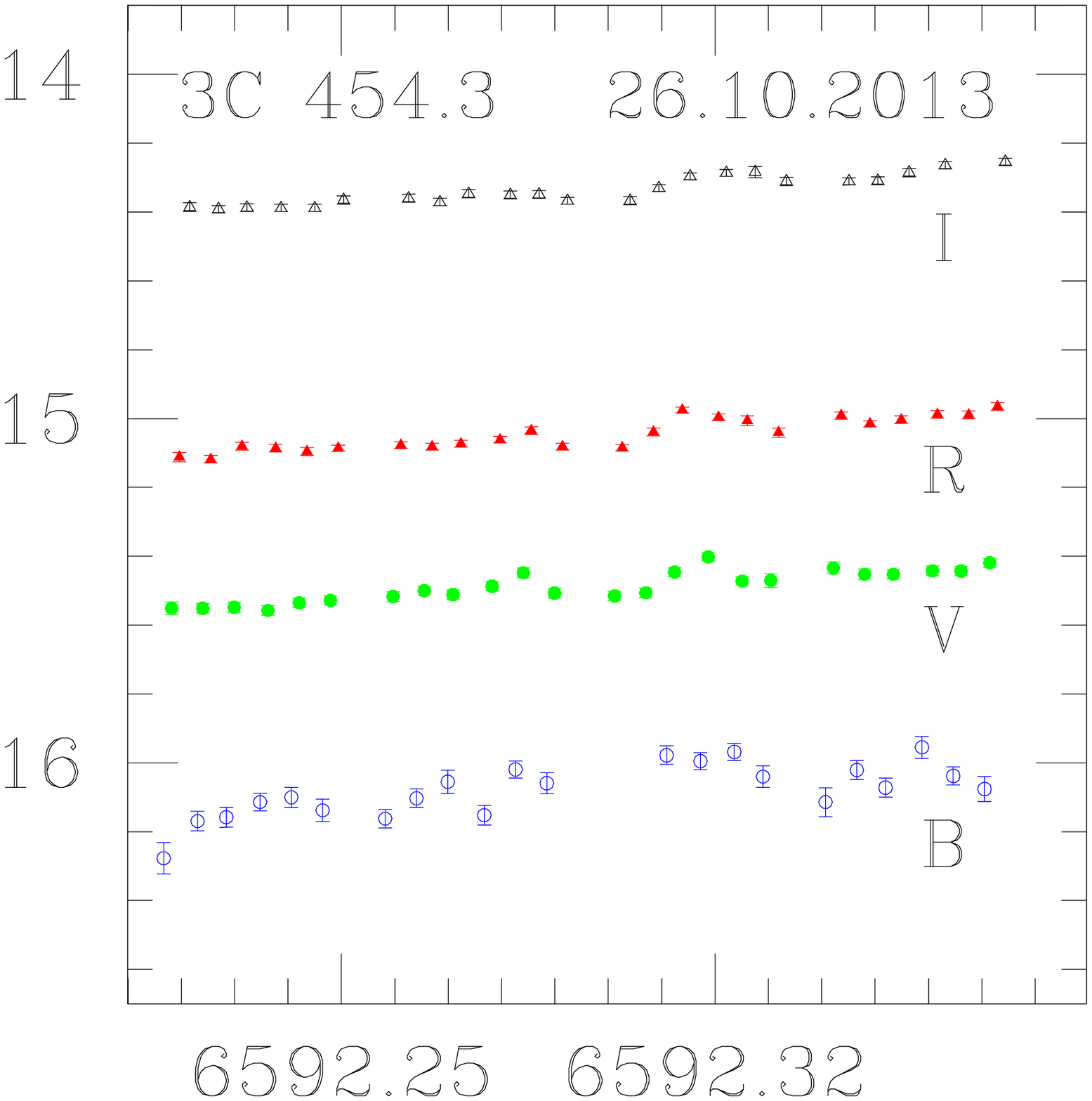 ,height=1.567in,width=1.59in,angle=0}
\epsfig{figure= 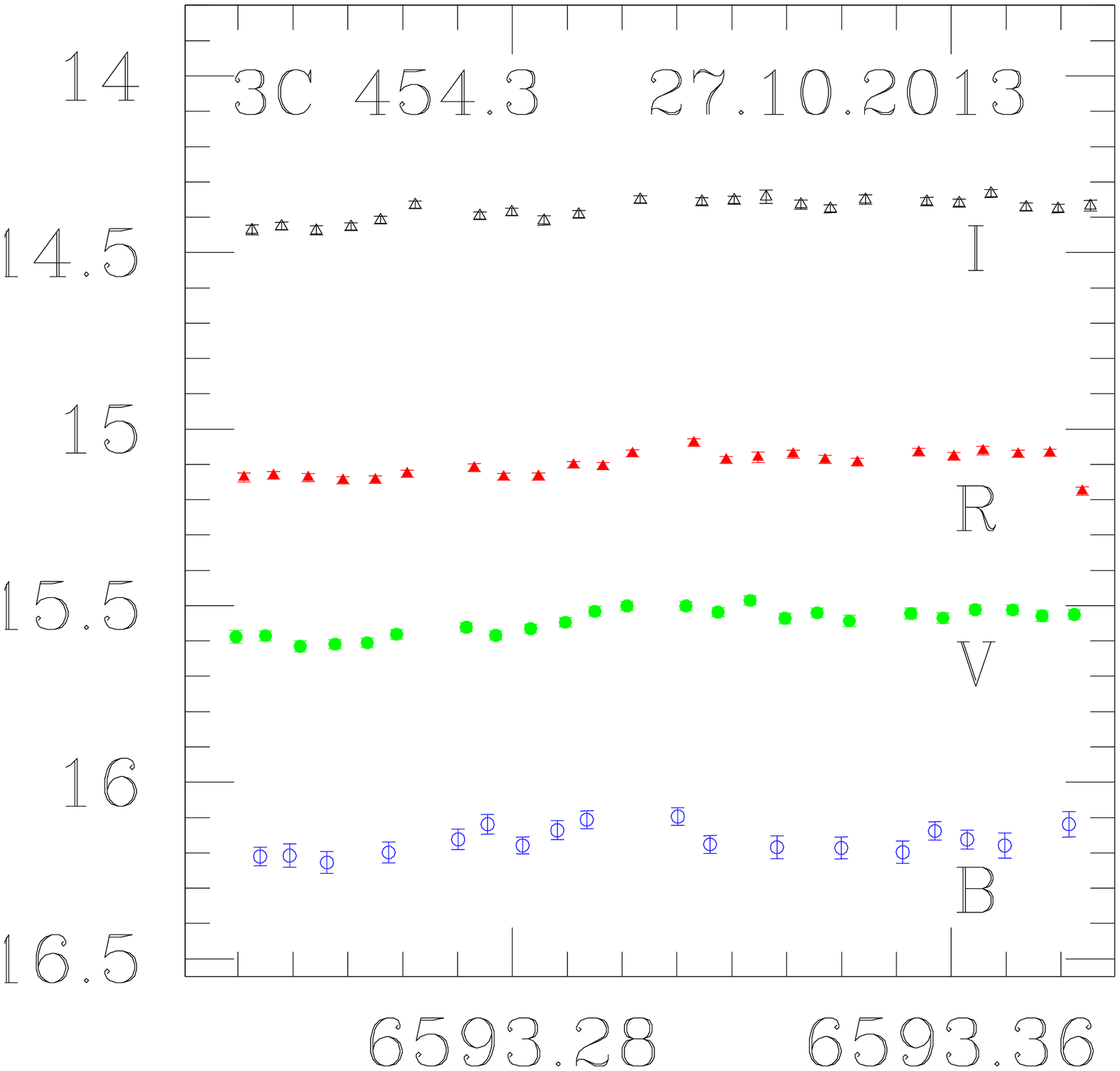 ,height=1.567in,width=1.59in,angle=0}
\epsfig{figure= 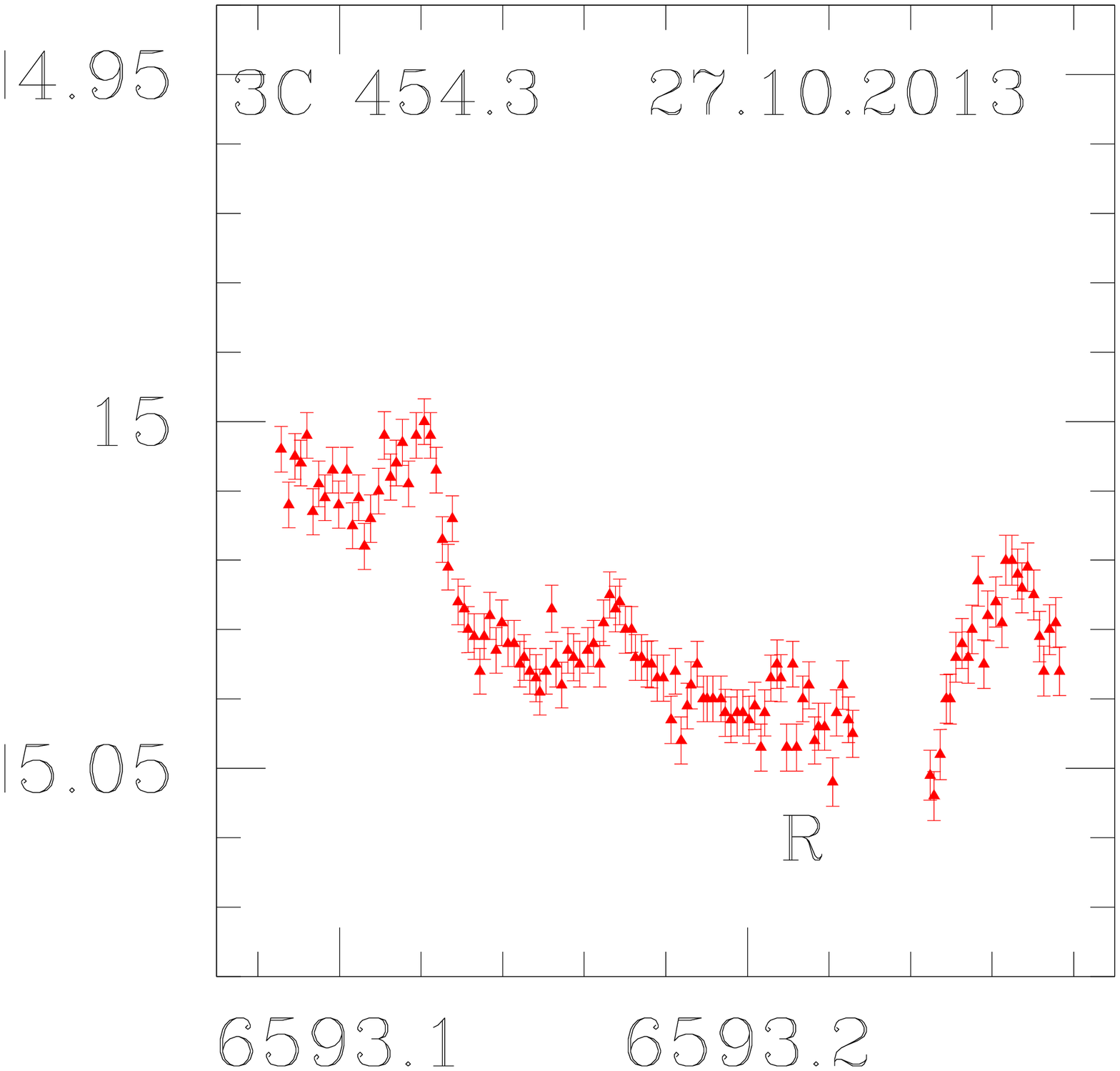 ,height=1.567in,width=1.59in,angle=0}
\epsfig{figure= 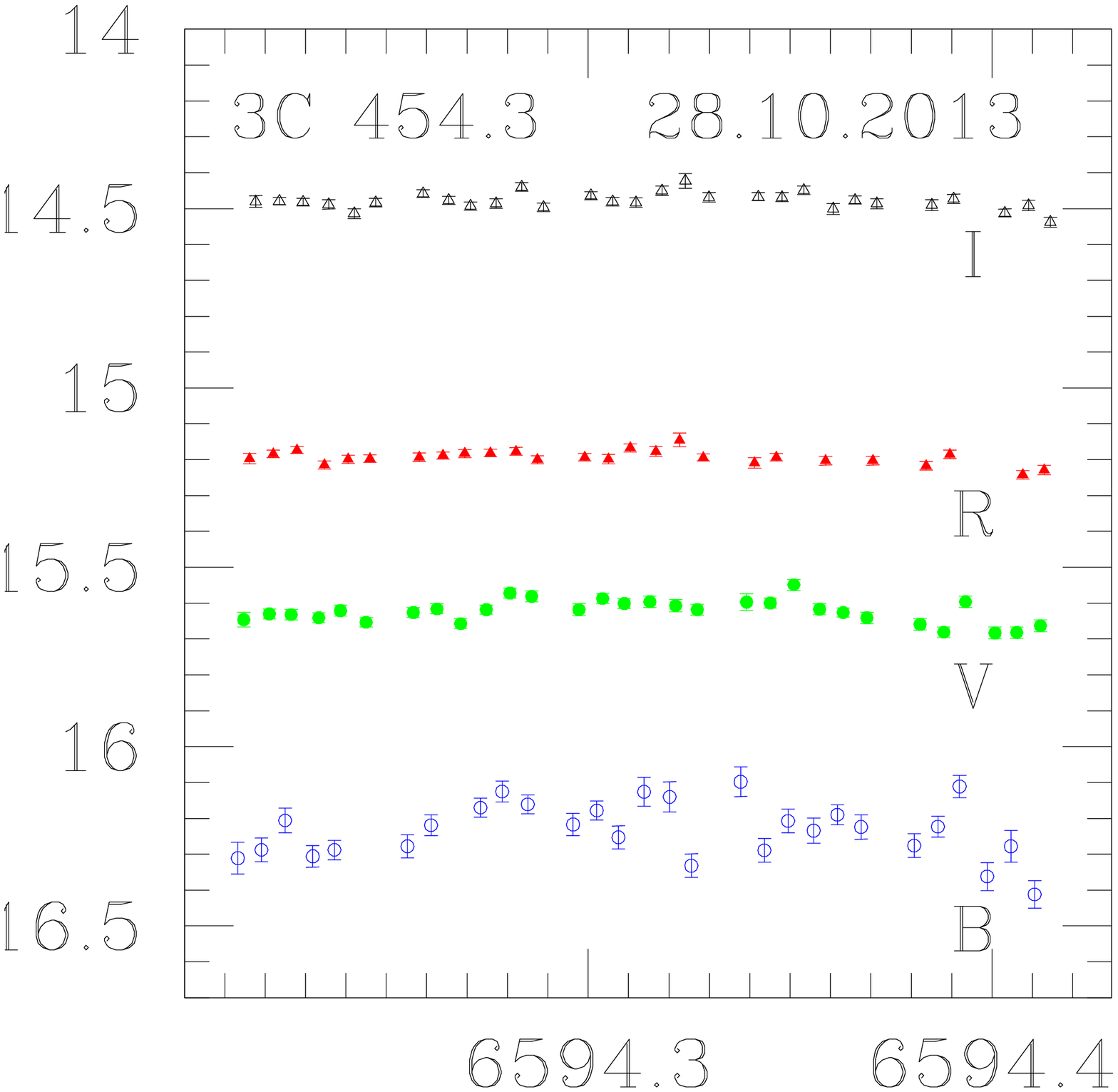,width=1.59in,height=1.567in,angle=0}
\epsfig{figure= 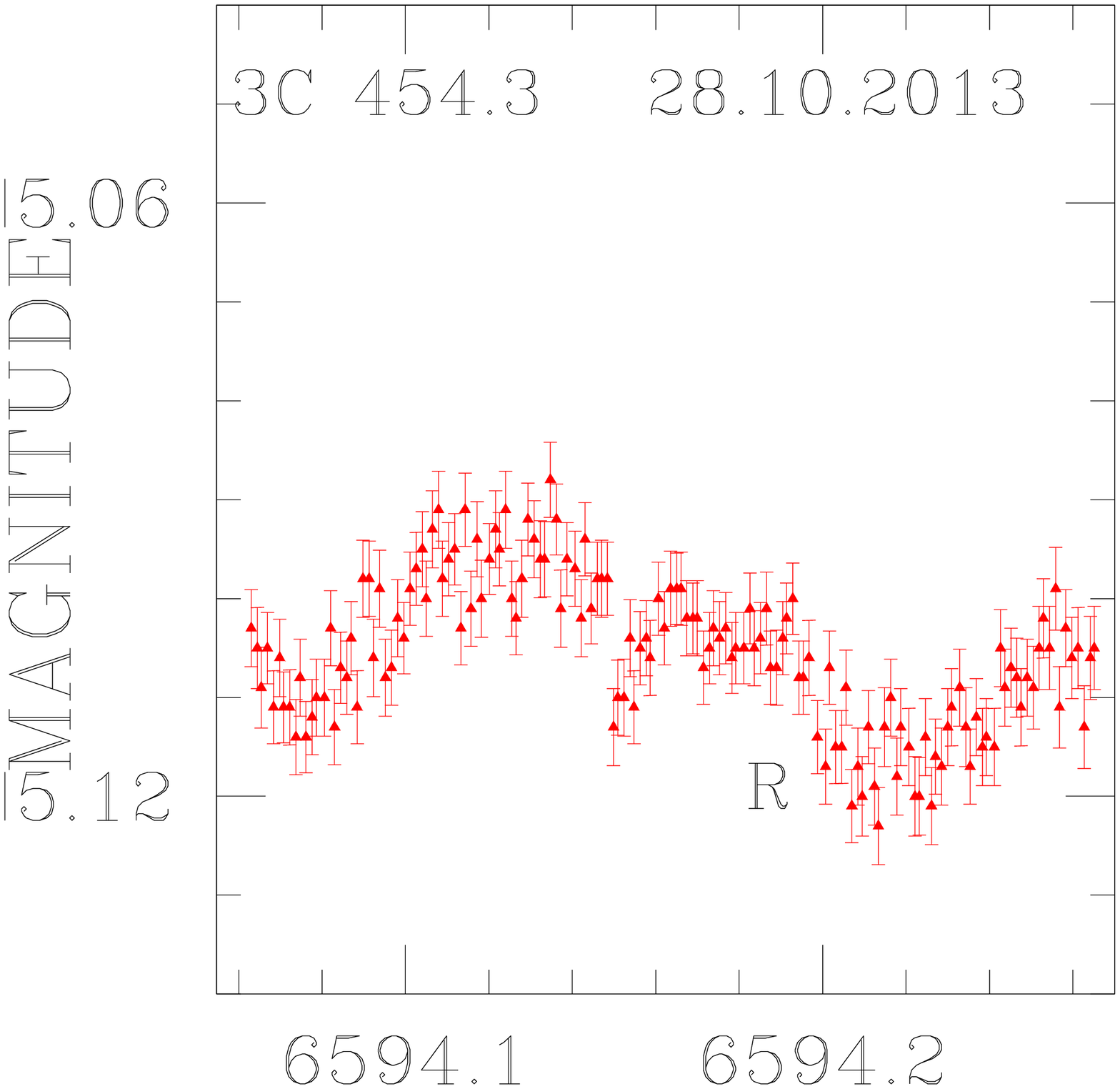 ,height=1.567in,width=1.59in,angle=0}
 \epsfig{figure= 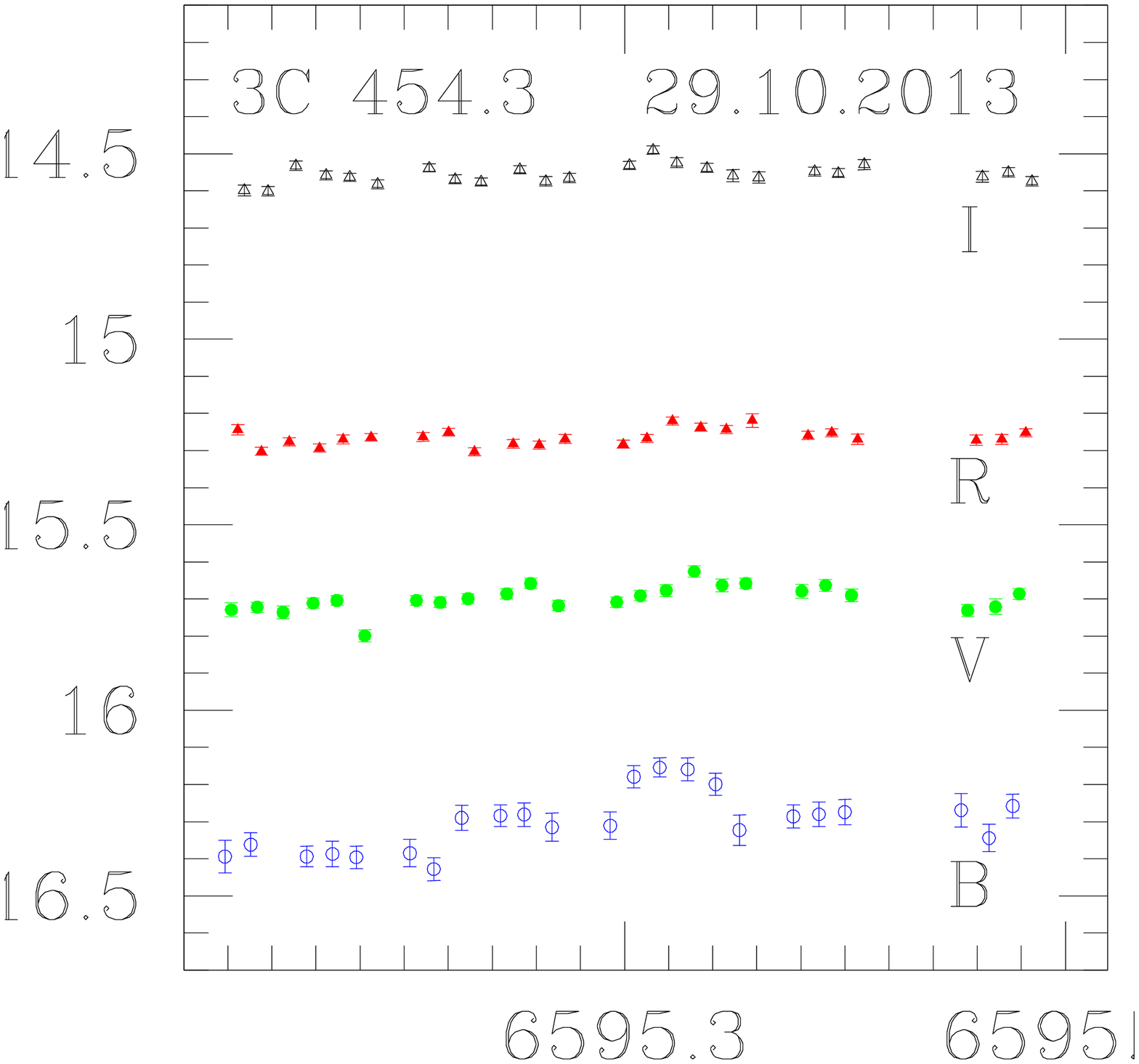,width=1.59in,height=1.567in,angle=0}
  \epsfig{figure= 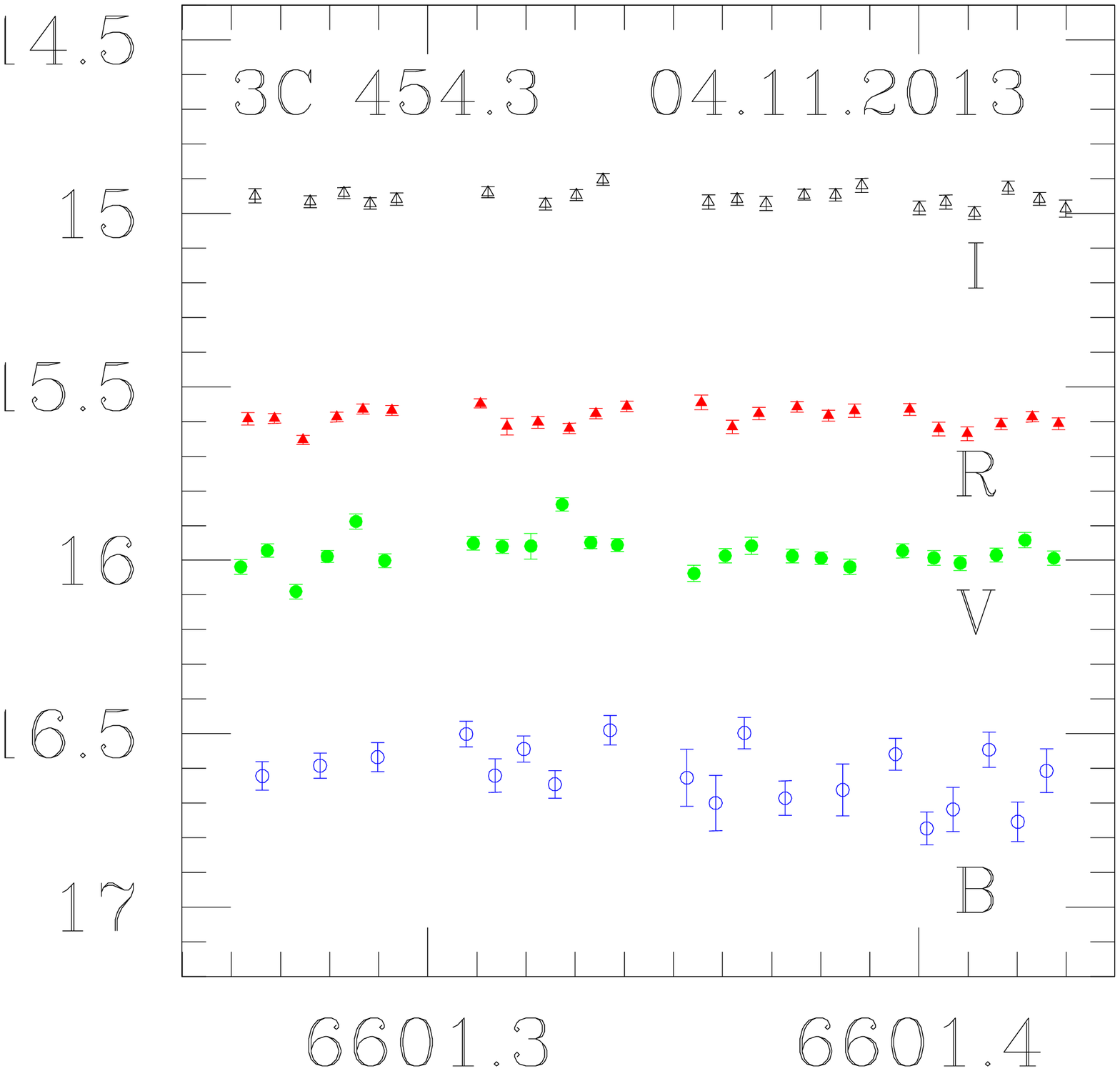,width=1.59in,height=1.567in,angle=0}
\epsfig{figure= 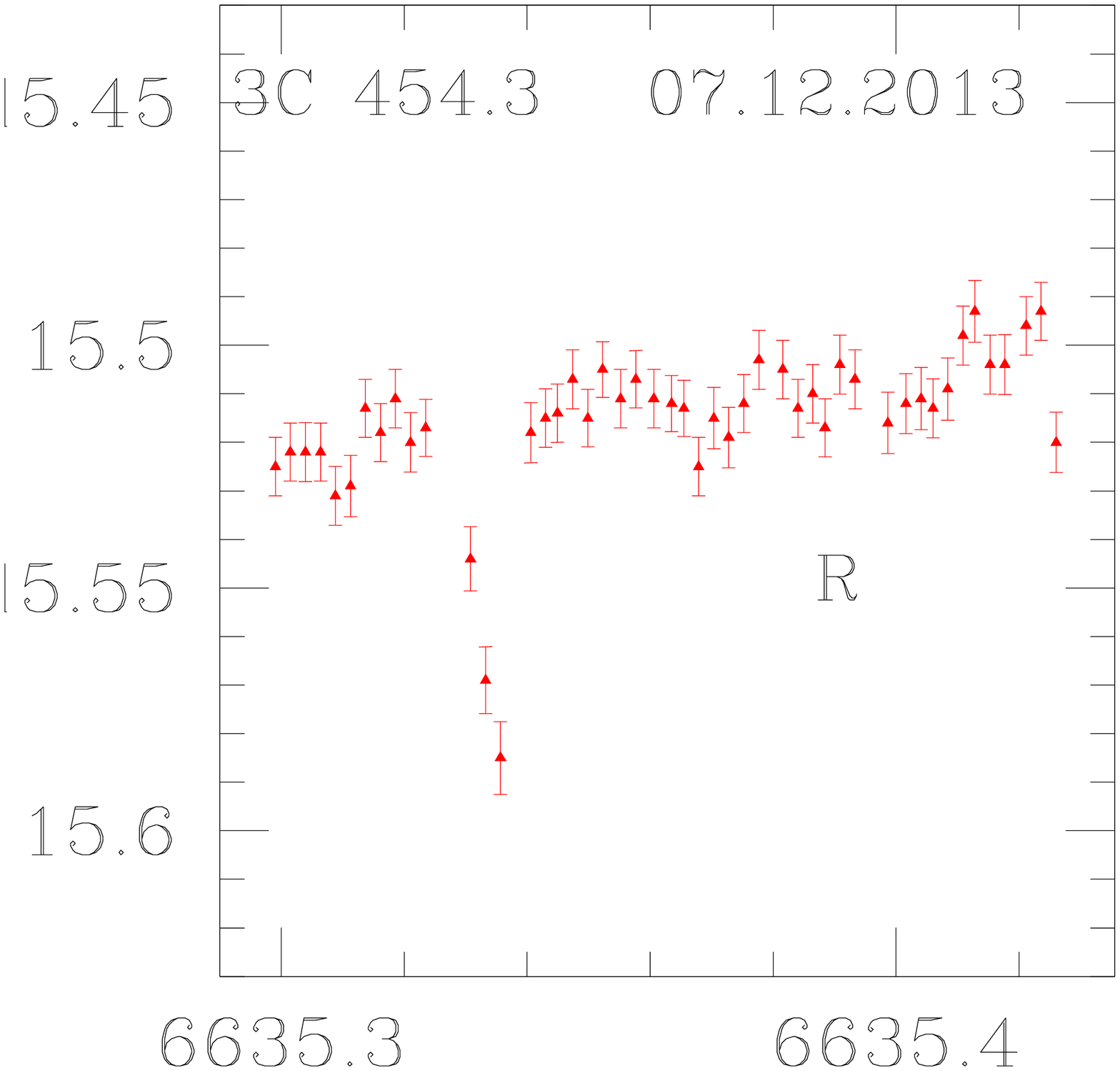 ,height=1.567in,width=1.59in,angle=0}
\epsfig{figure= 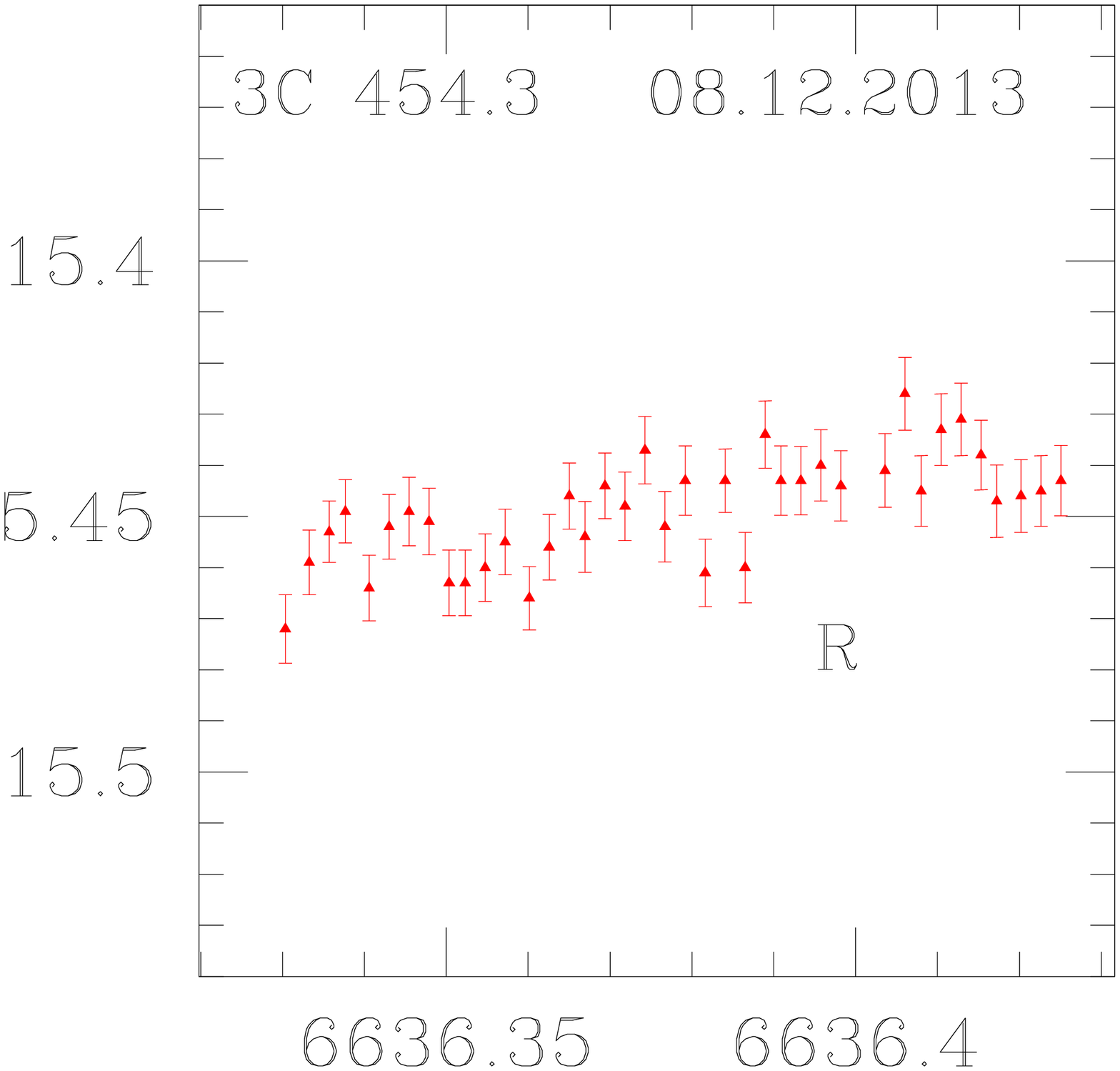 ,height=1.567in,width=1.59in,angle=0}
\epsfig{figure= 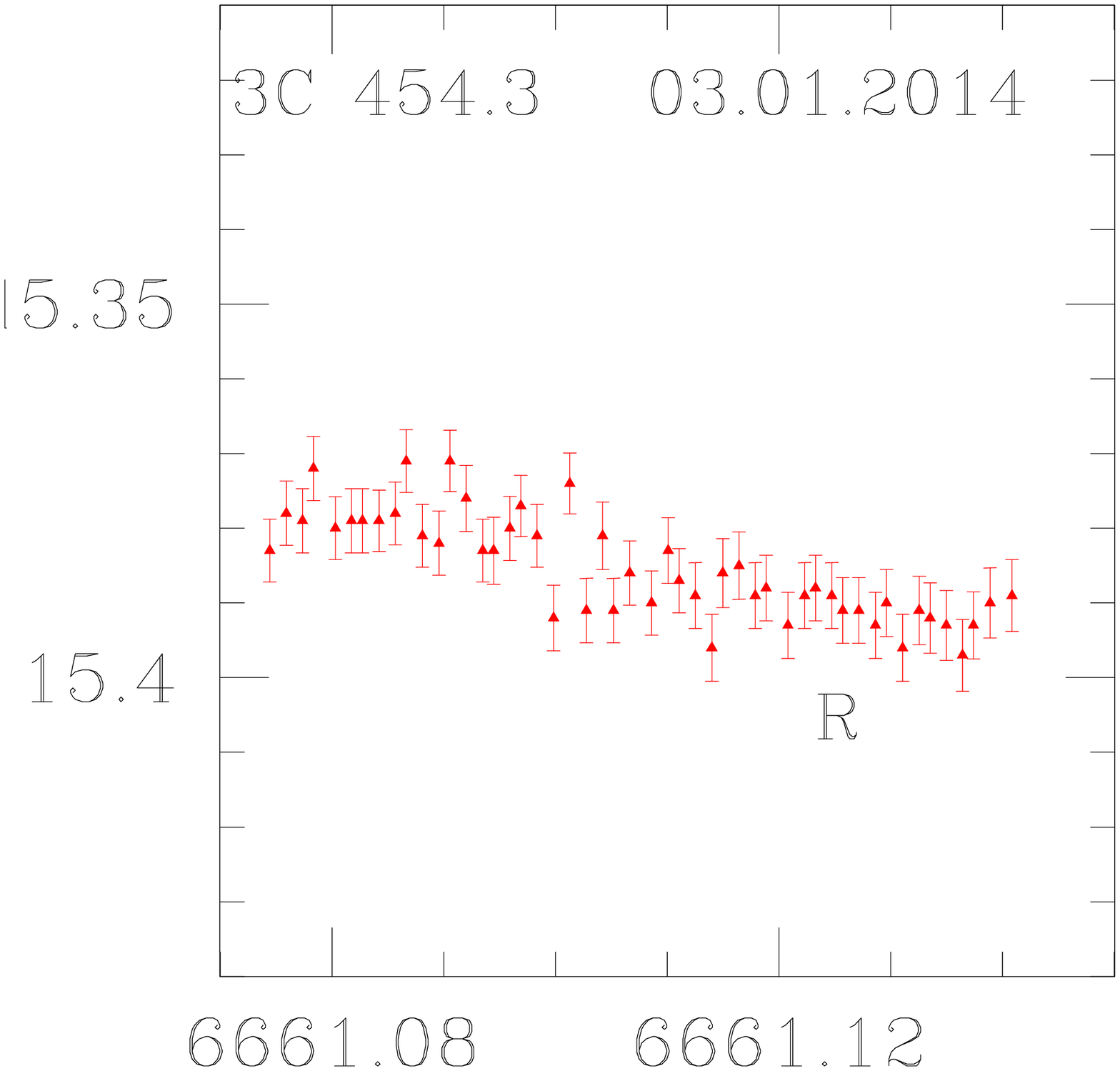 ,height=1.567in,width=1.59in,angle=0}
\epsfig{figure= 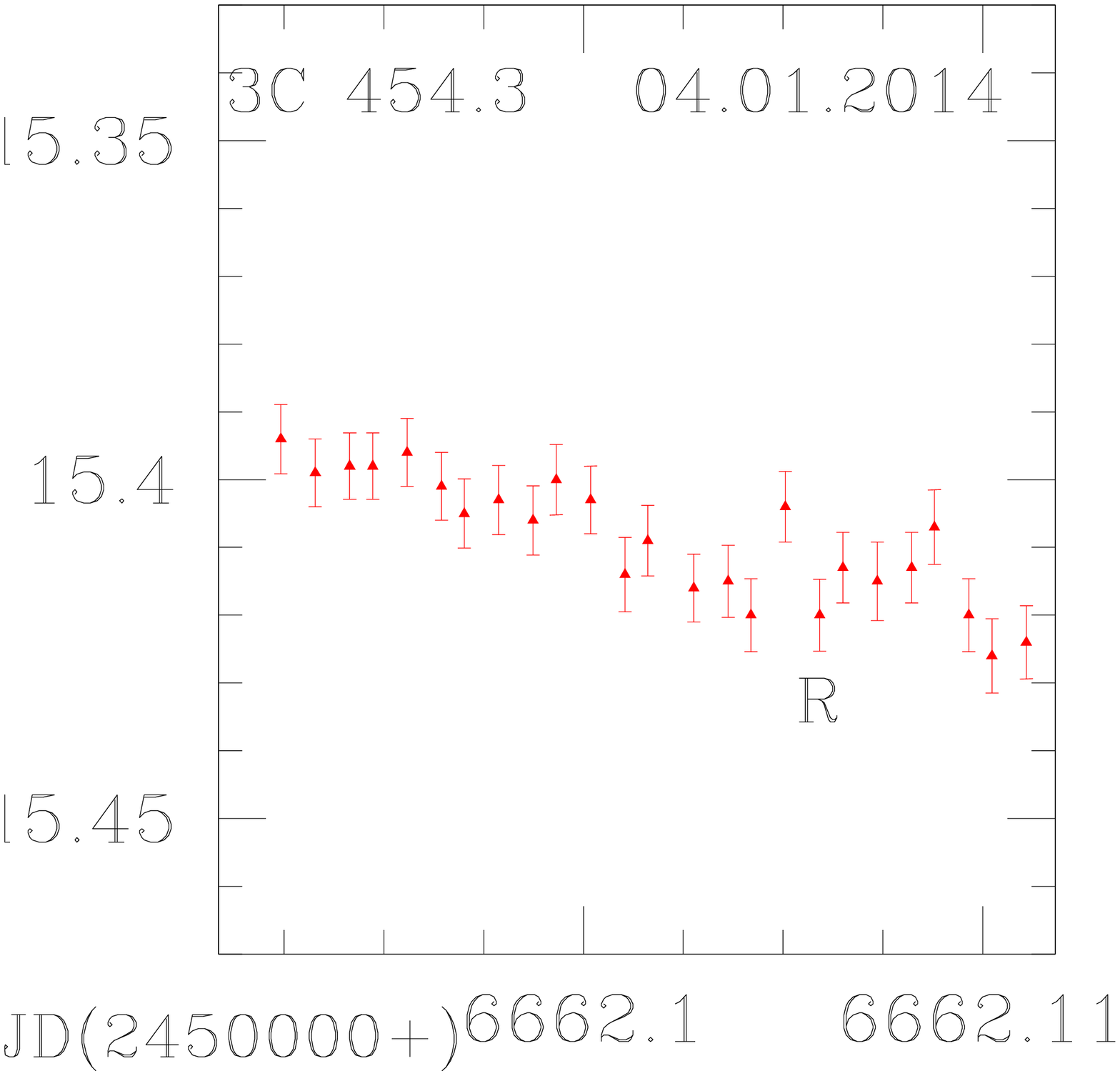 ,height=1.567in,width=1.59in,angle=0}
  \caption{{\bf Light curves for 3C 454.3; open circles denote B filter LC; filled circles, V filter; filled triangles, R filter; 
  open triangles, I filter}. In each plot, X and Y axis are the JD and magnitude, respectively. Source name and observation 
  date are indicated in each panel.}
\label{idvfig}
\end{figure*}

\begin{figure*} 
\epsfig{figure=  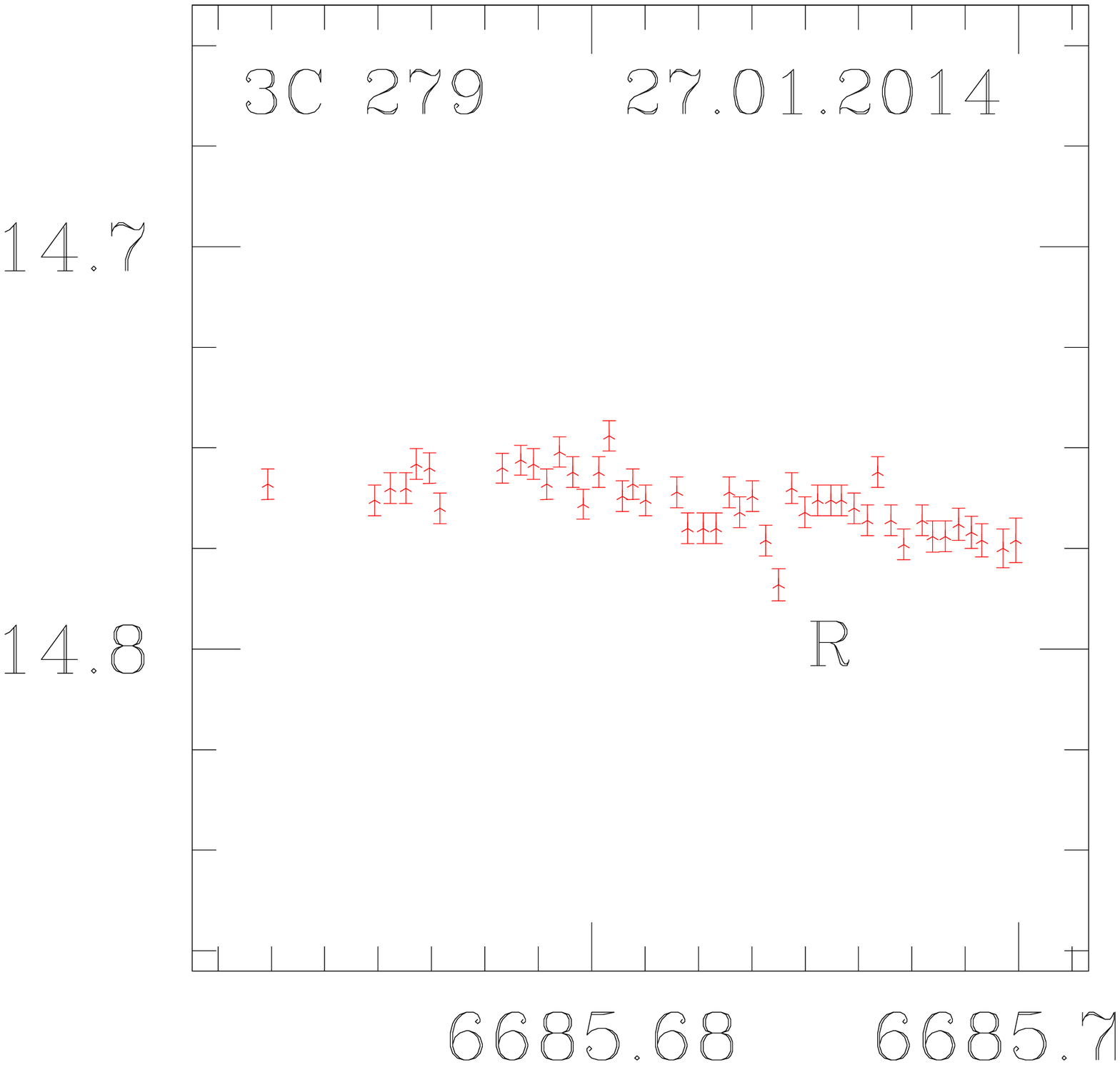,height=1.567in,width=1.59in,angle=0}
\epsfig{figure=  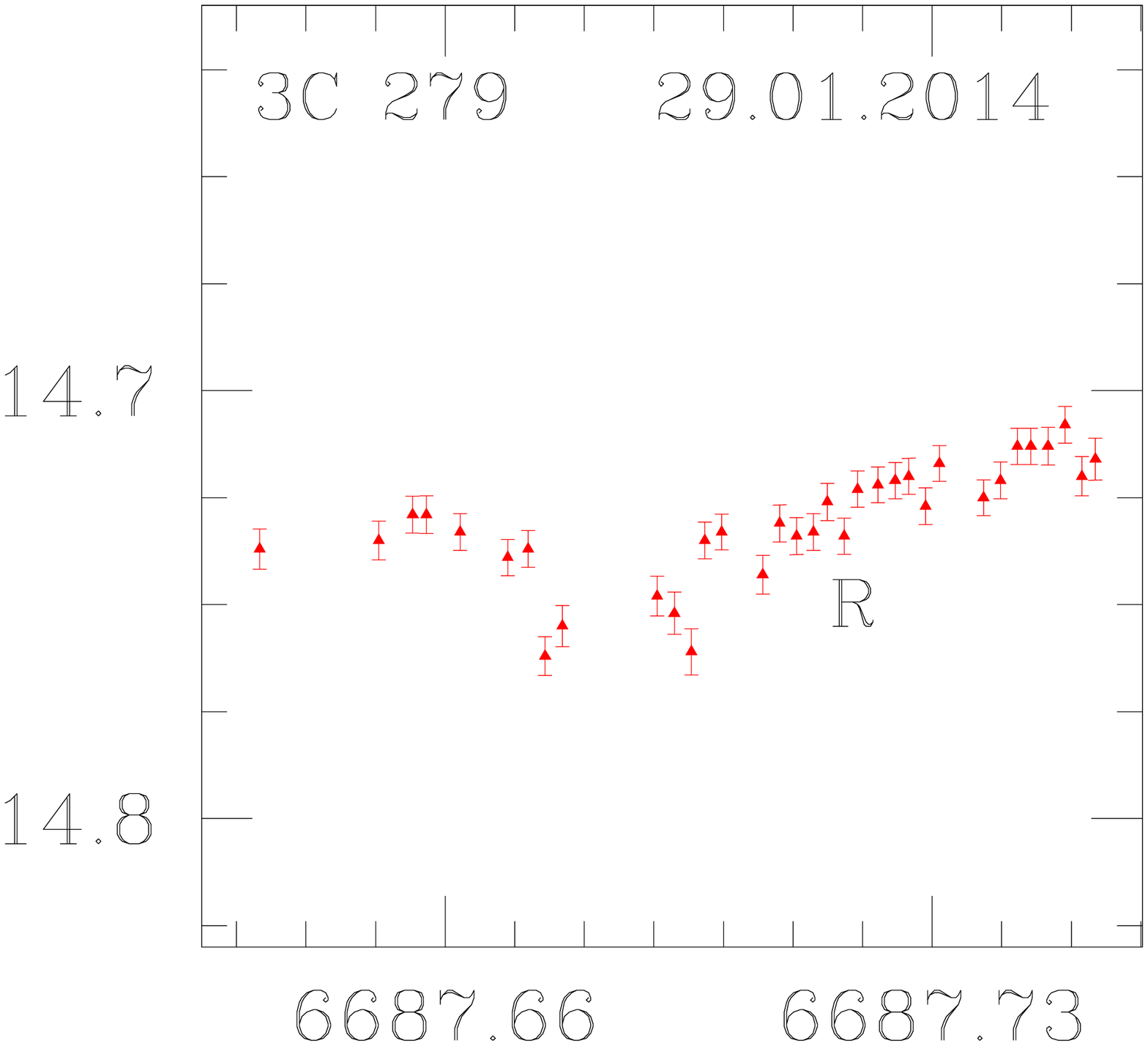,height=1.567in,width=1.59in,angle=0}
\epsfig{figure=  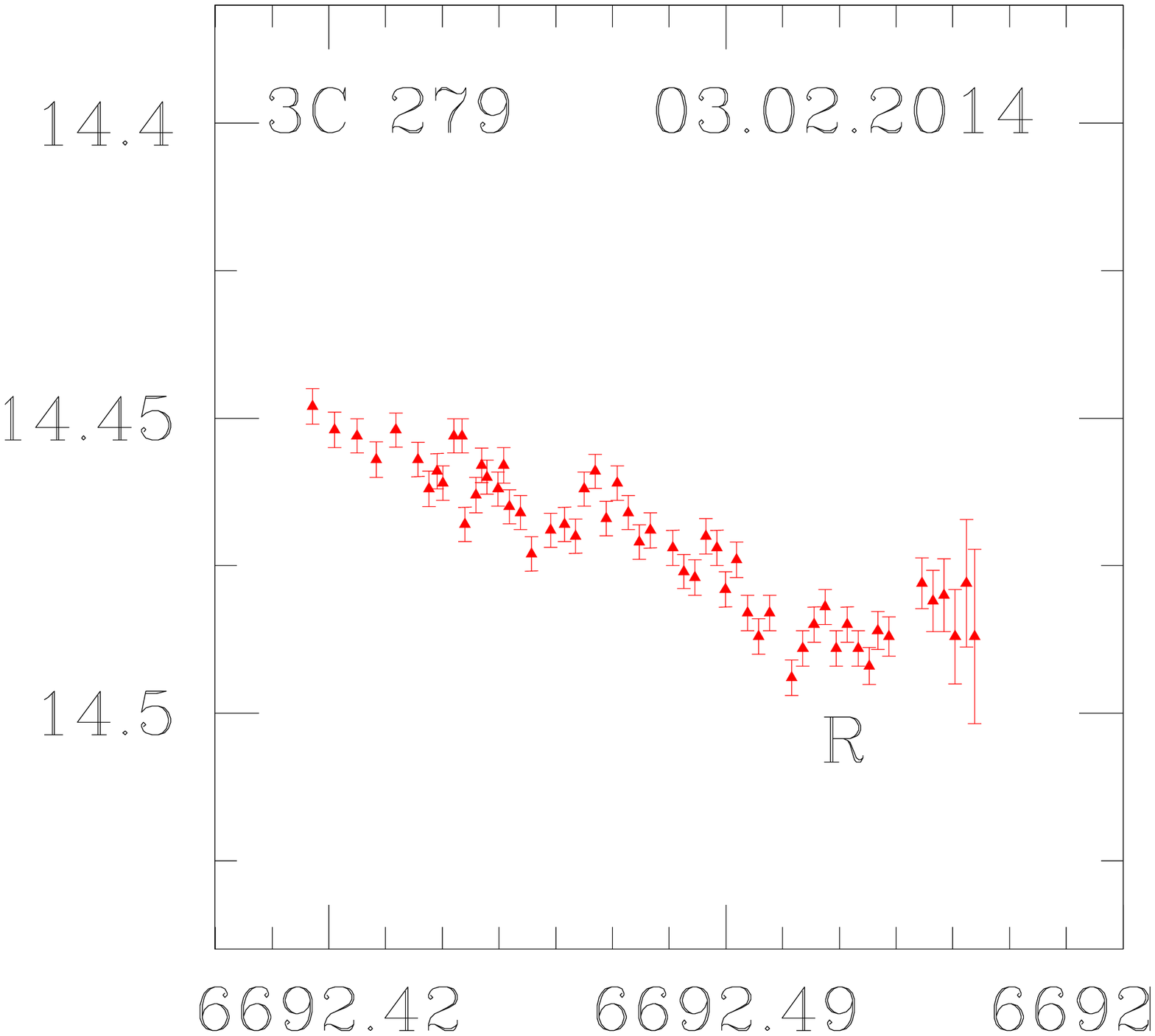,height=1.567in,width=1.59in,angle=0}
\epsfig{figure= 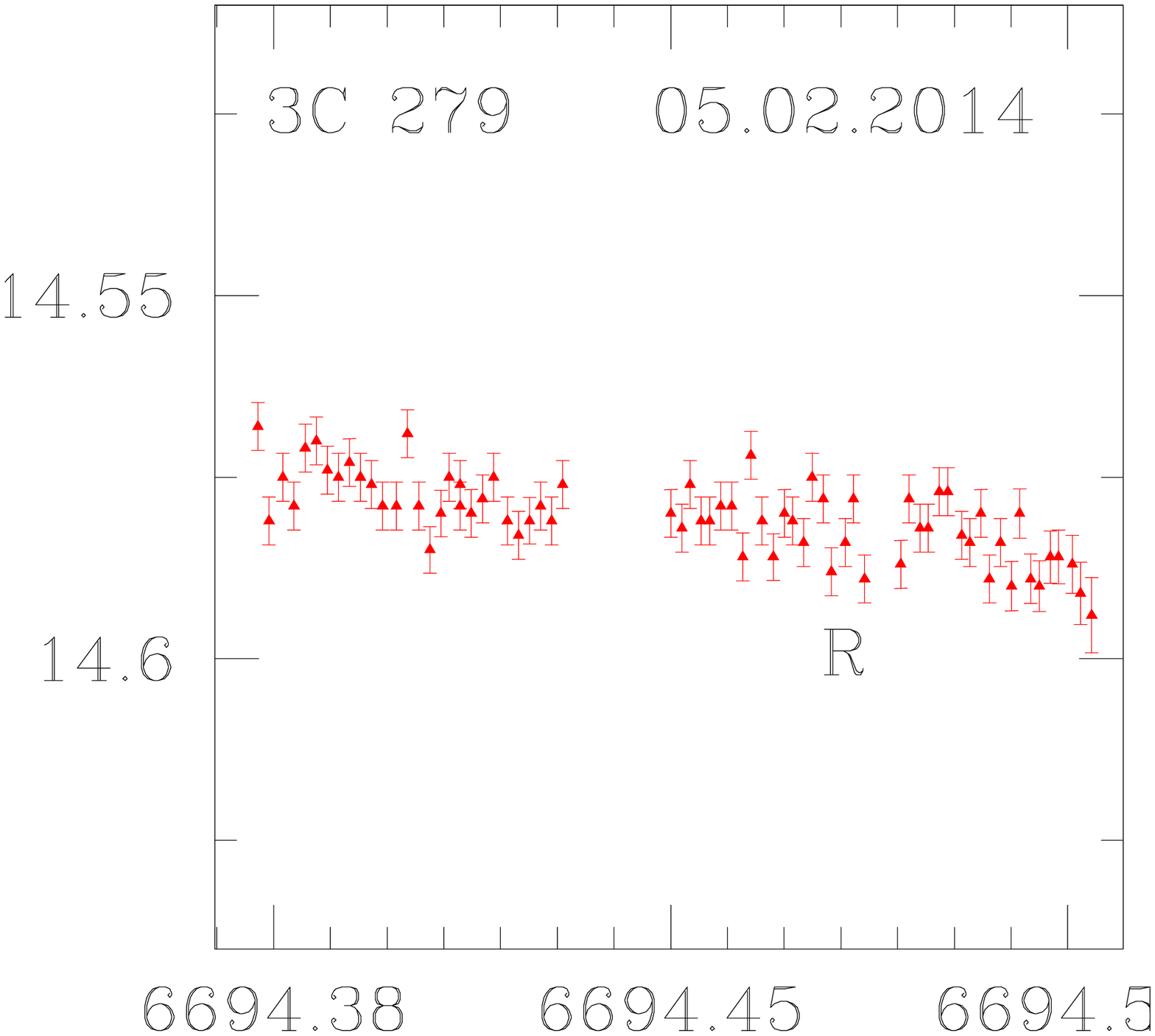,height=1.567in,width=1.59in,angle=0}
 \epsfig{figure=  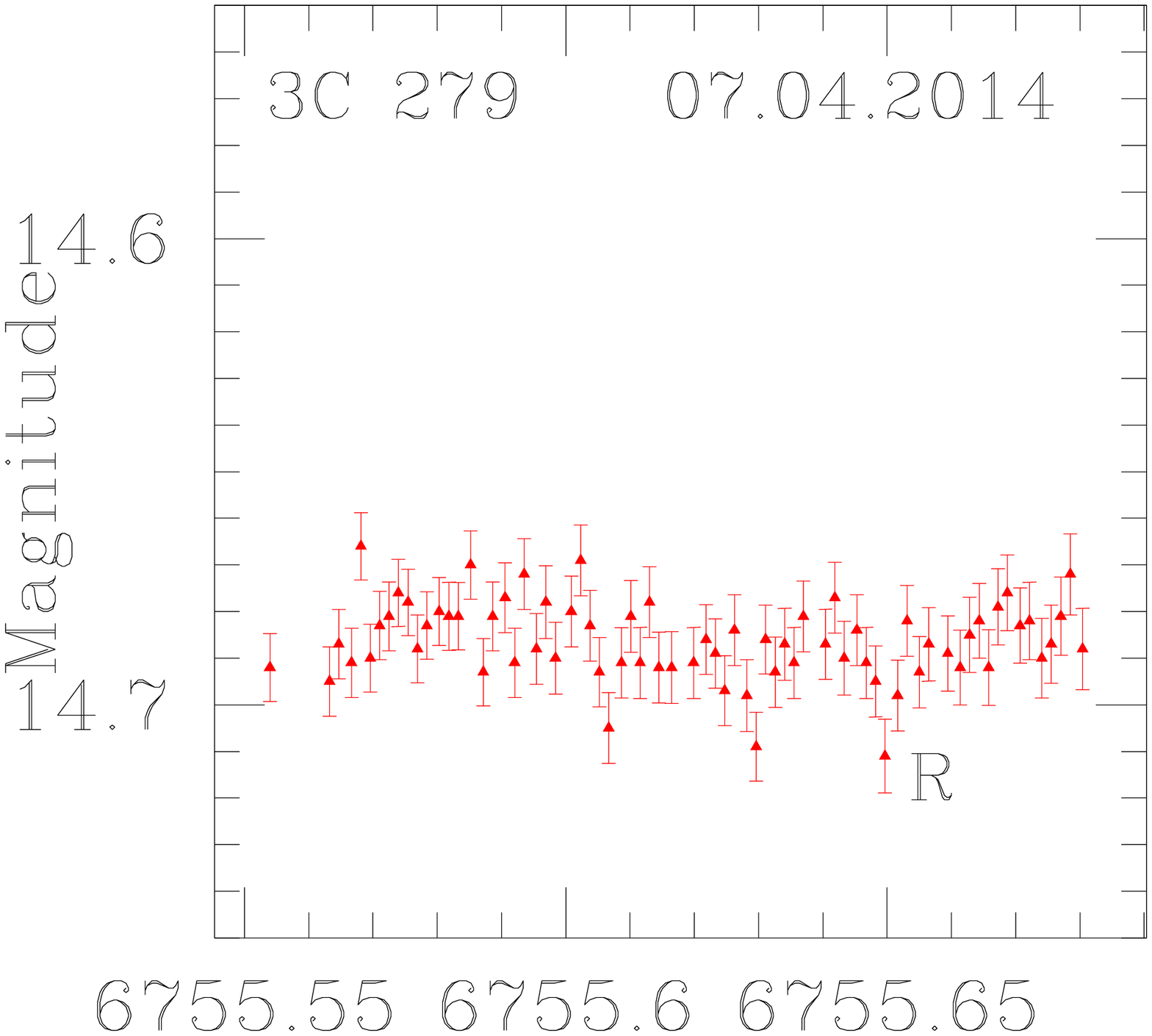,height=1.567in,width=1.59in,angle=0}
 \epsfig{figure=  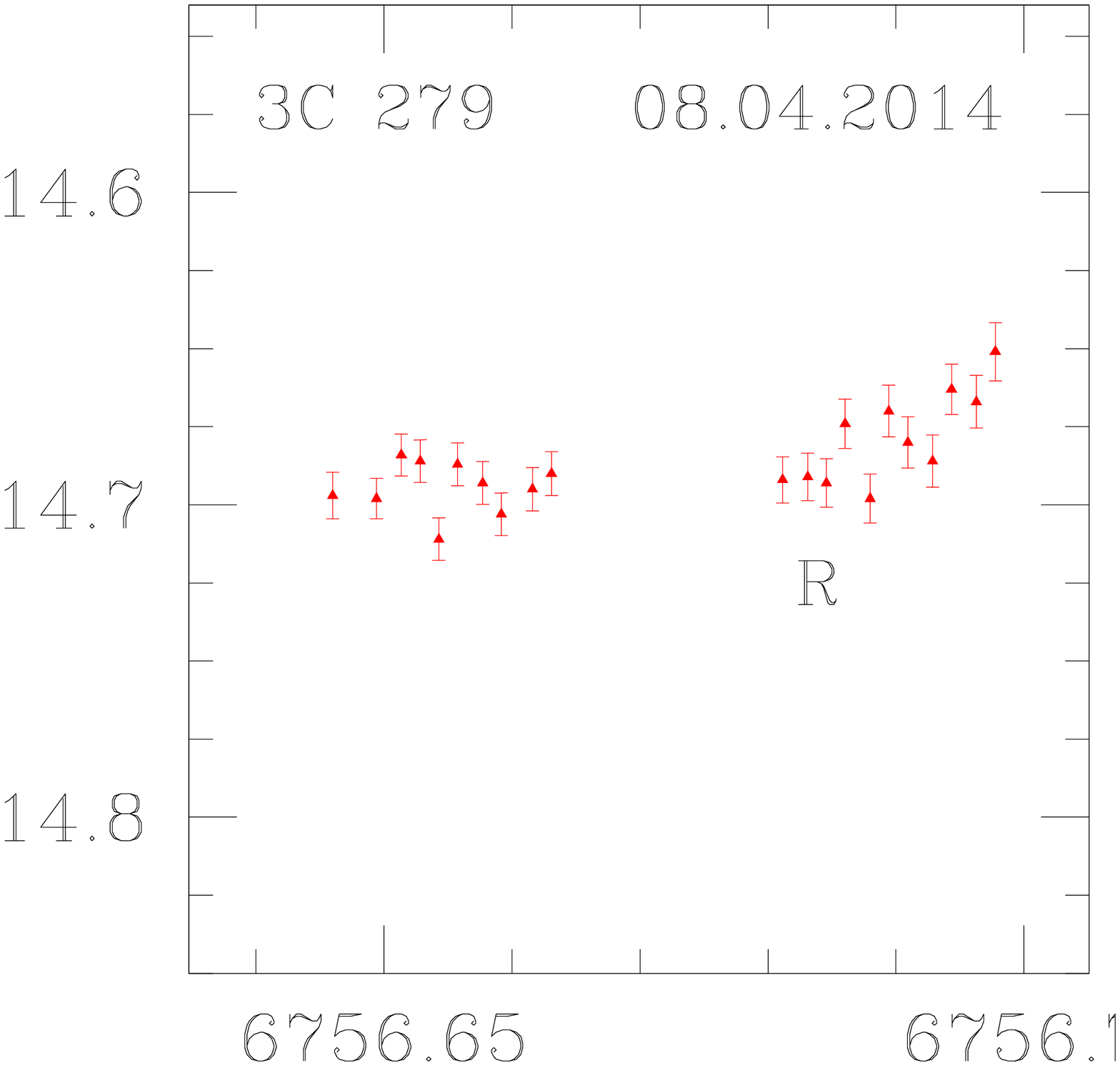,height=1.567in,width=1.59in,angle=0}
 \epsfig{figure=  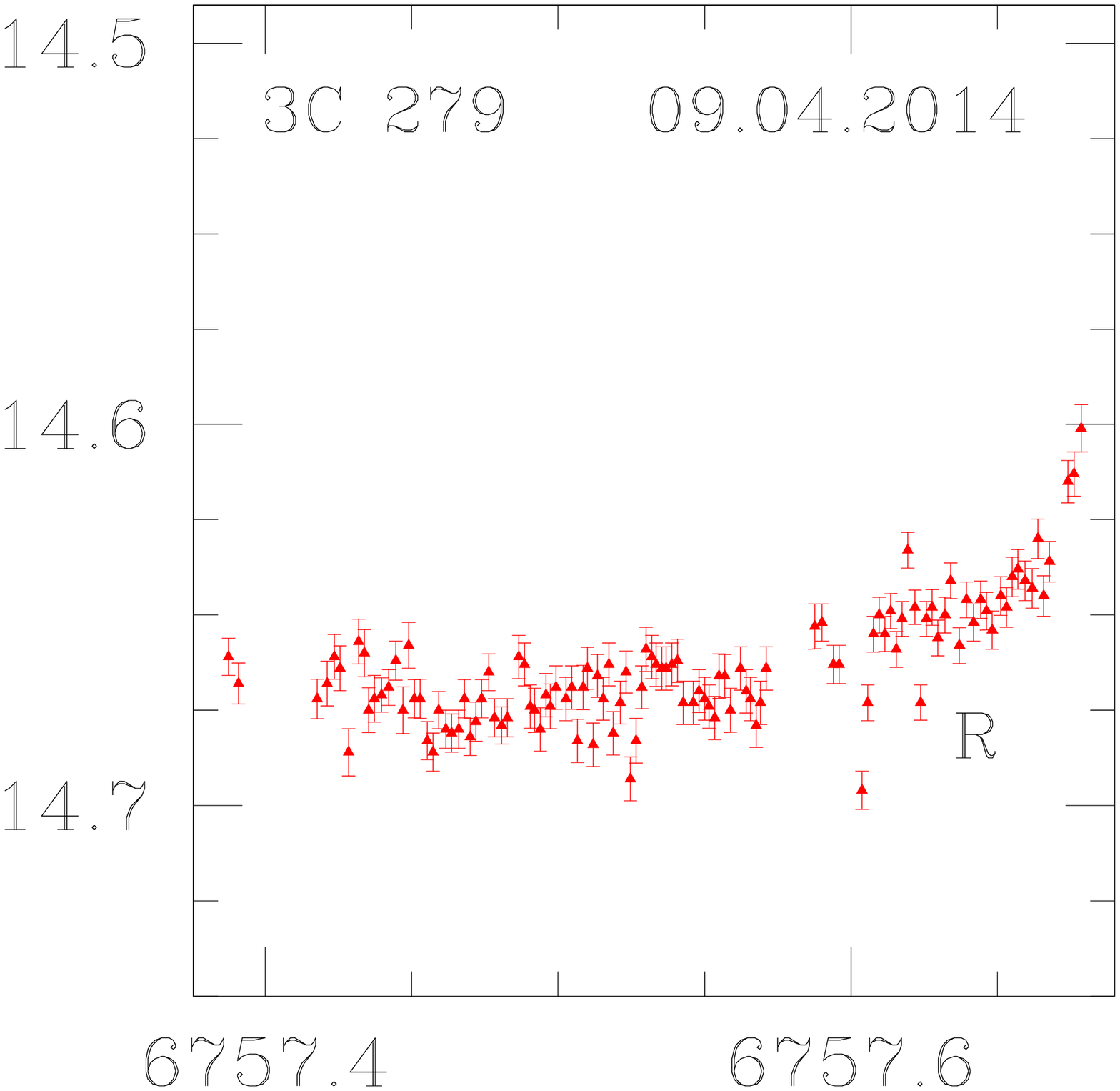,height=1.567in,width=1.59in,angle=0}
\epsfig{figure=  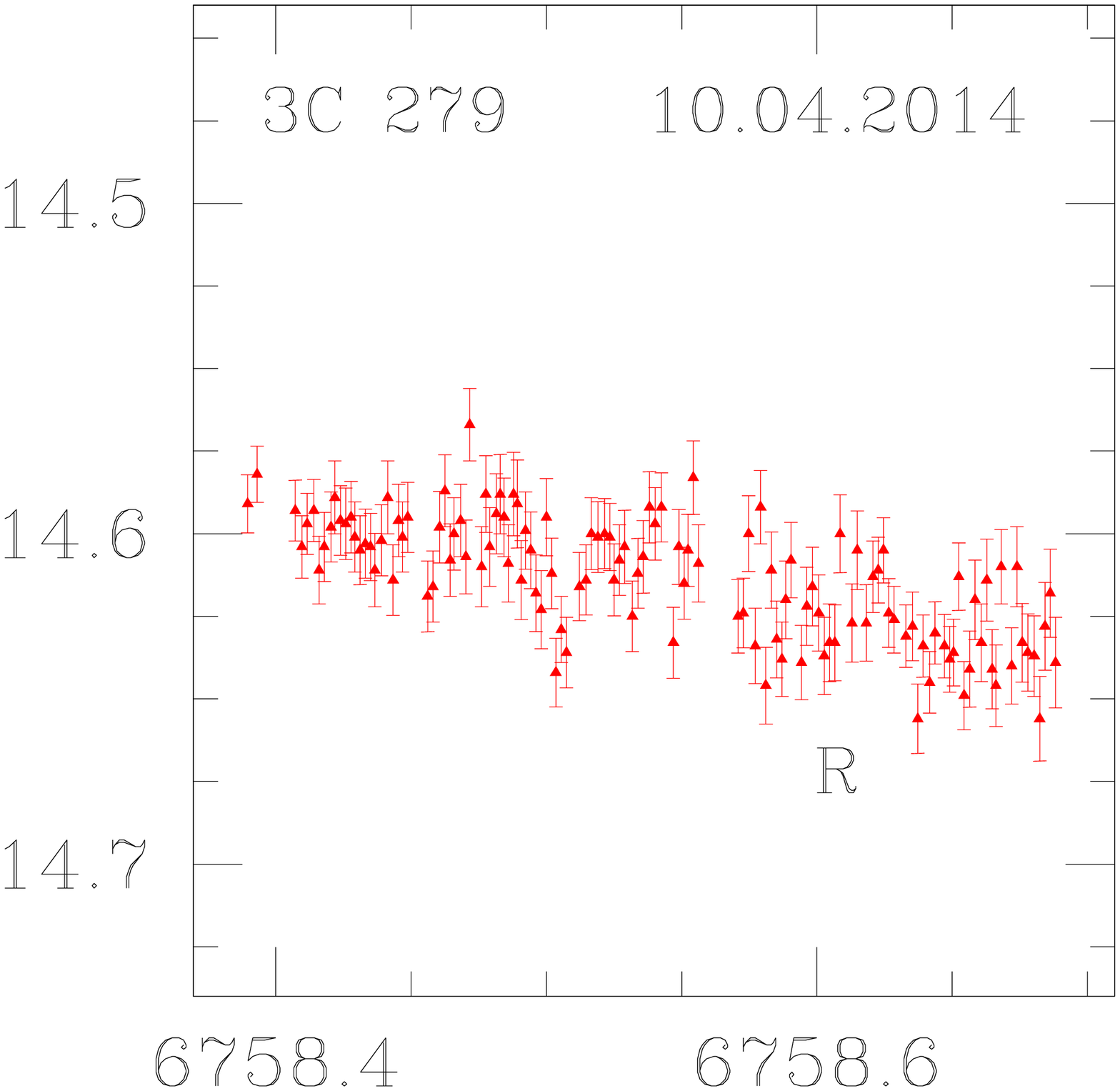,height=1.567in,width=1.59in,angle=0}
\epsfig{figure=  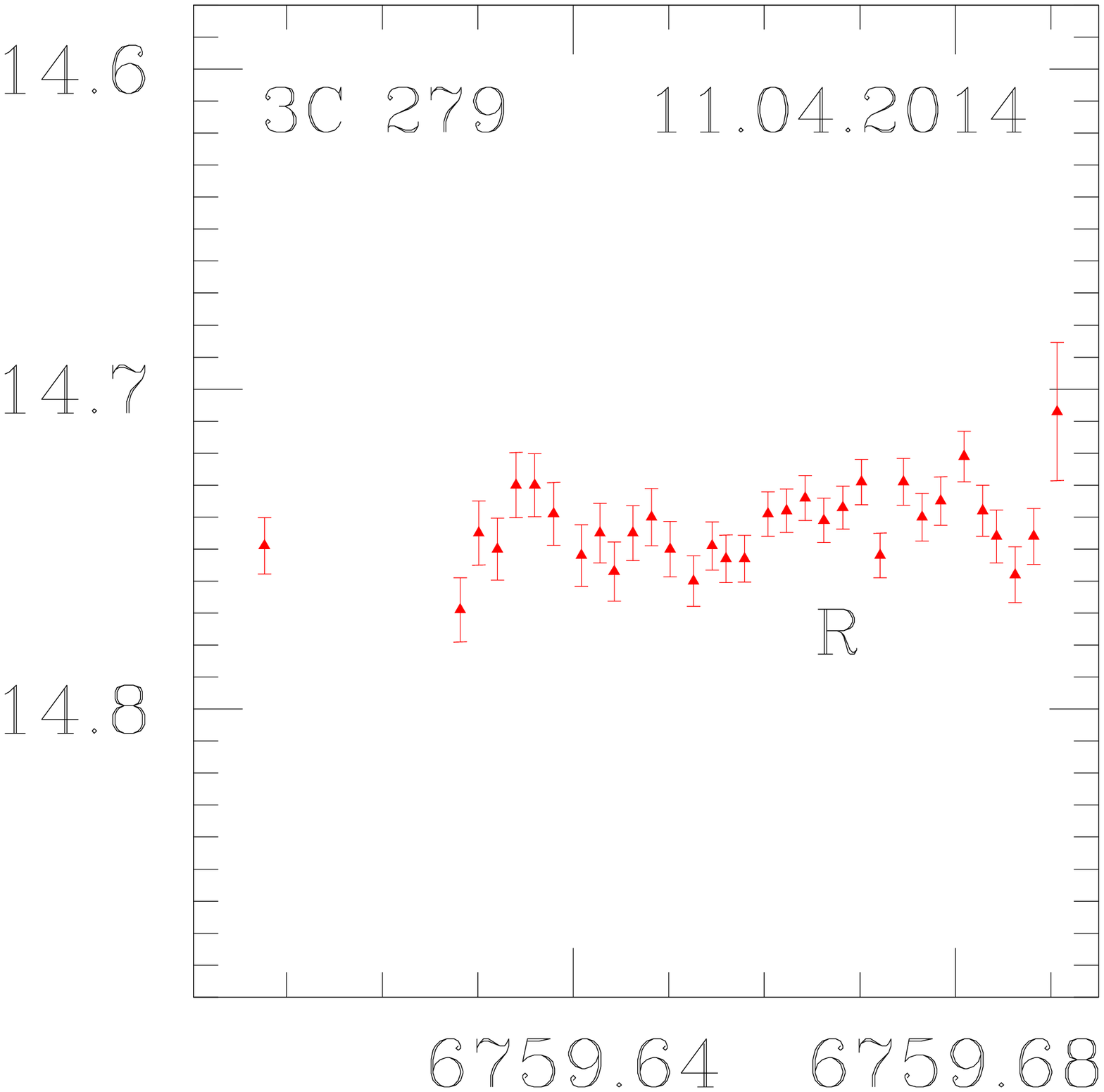,height=1.567in,width=1.59in,angle=0}
\epsfig{figure=  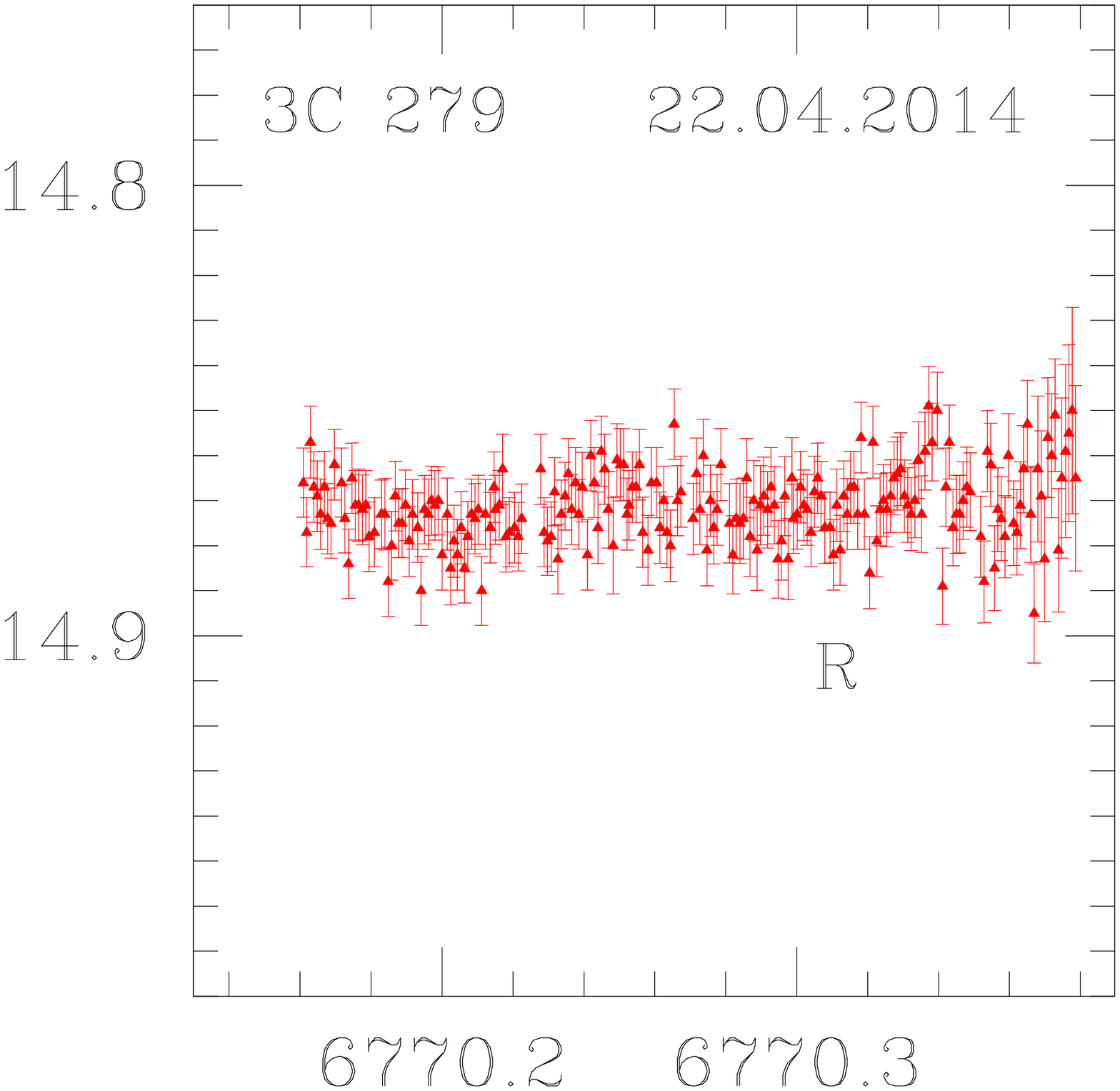,height=1.567in,width=1.59in,angle=0}
\epsfig{figure=  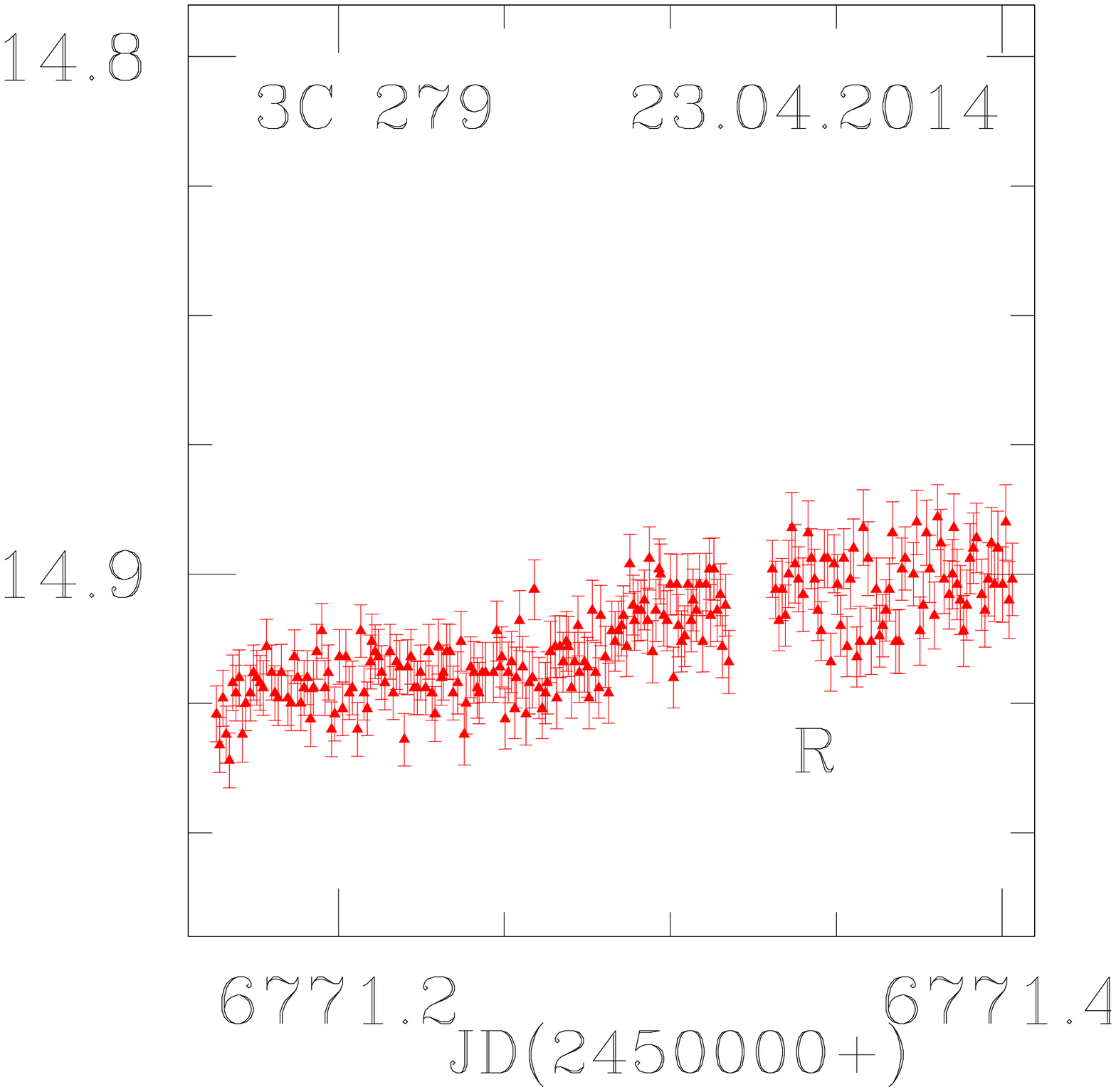,height=1.567in,width=1.59in,angle=0}
\epsfig{figure=  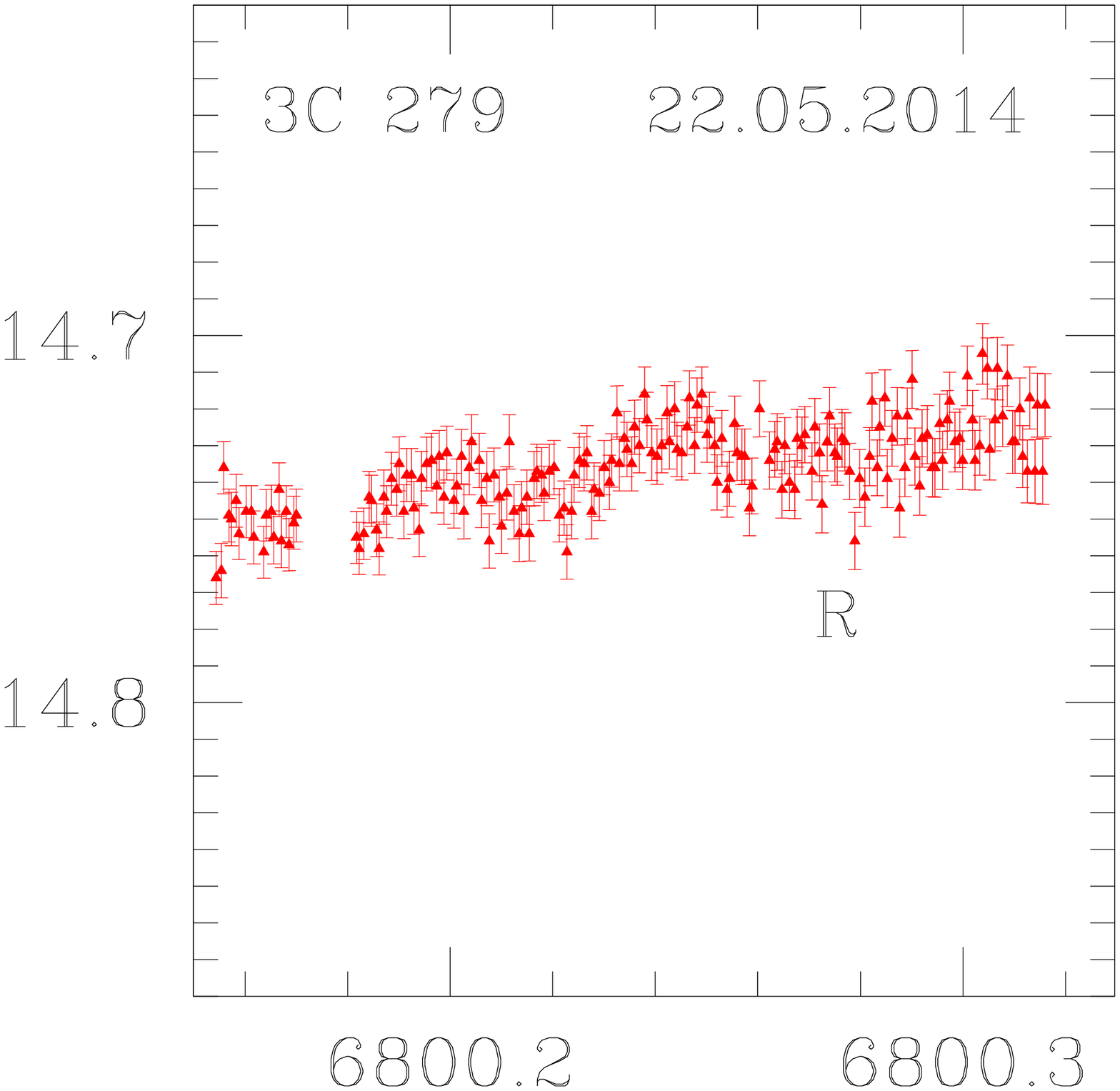,height=1.567in,width=1.59in,angle=0}
  \caption{IDV LCs for 3C 279 with details same as in figure 1.}
\label{idvfig}
\end{figure*}

\begin{figure*}
 \epsfig{figure= 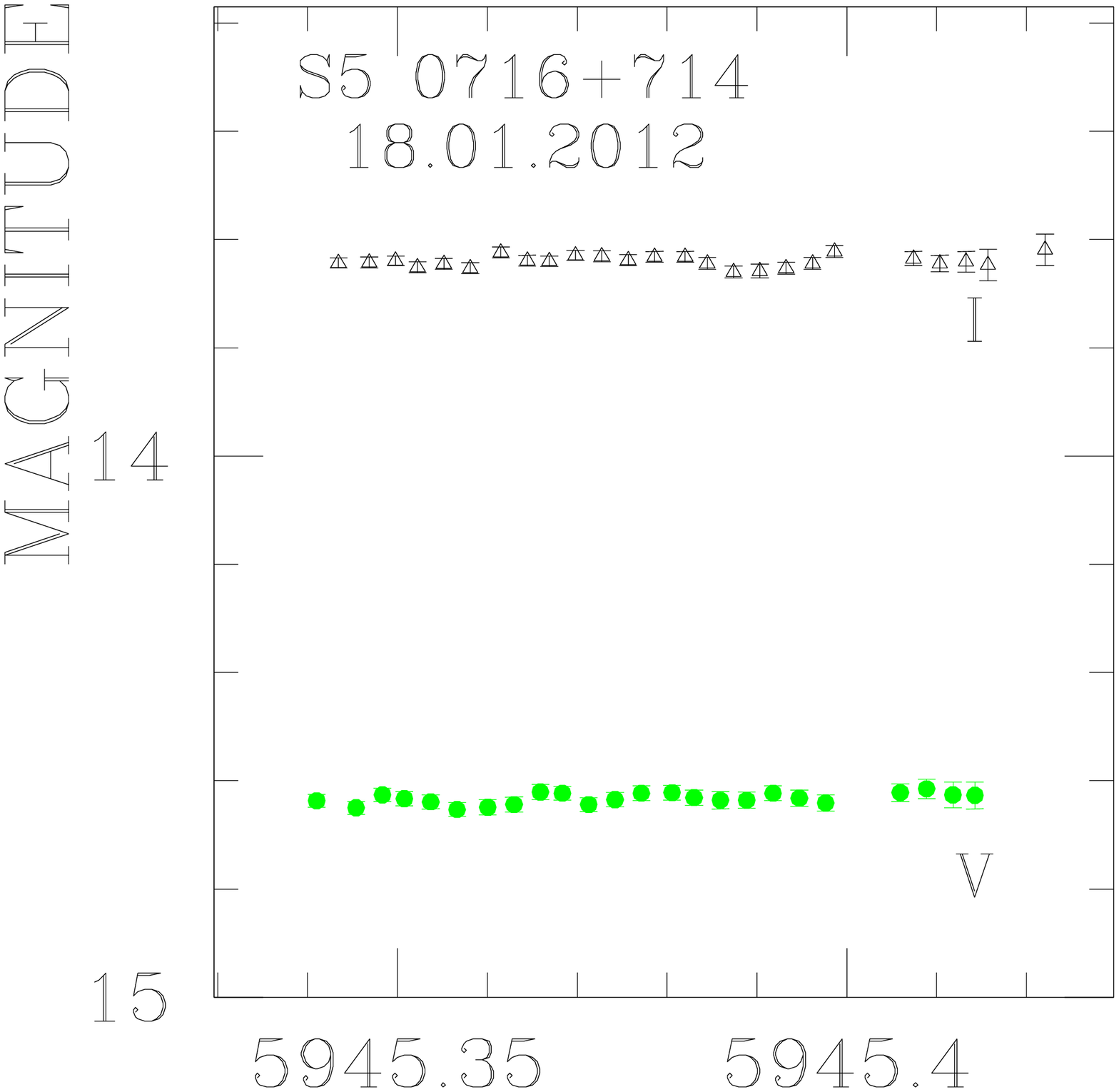,width=1.59in,height=1.567in,angle=0}
 \epsfig{figure= 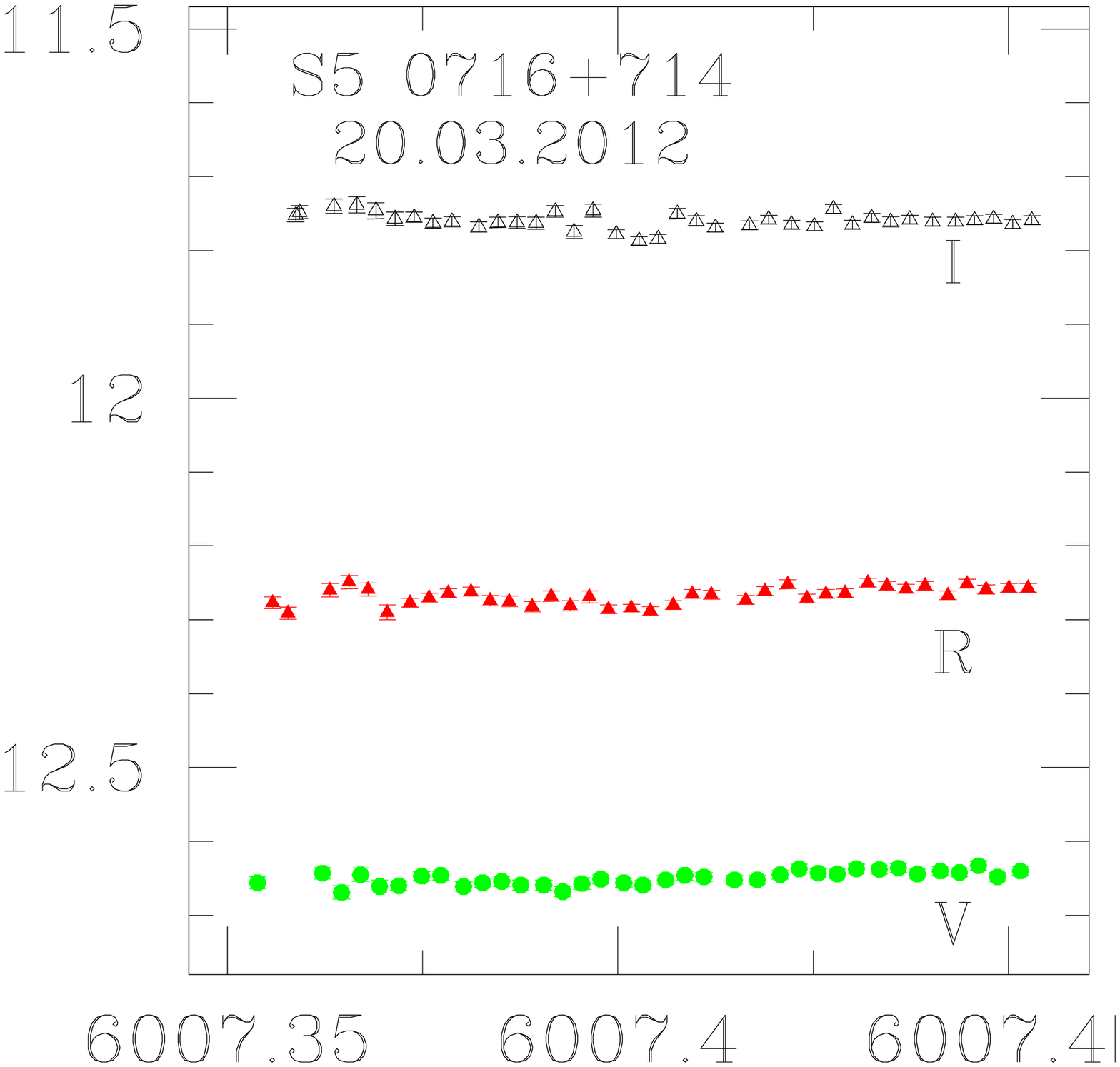,width=1.59in,height=1.567in,angle=0}
 \epsfig{figure=  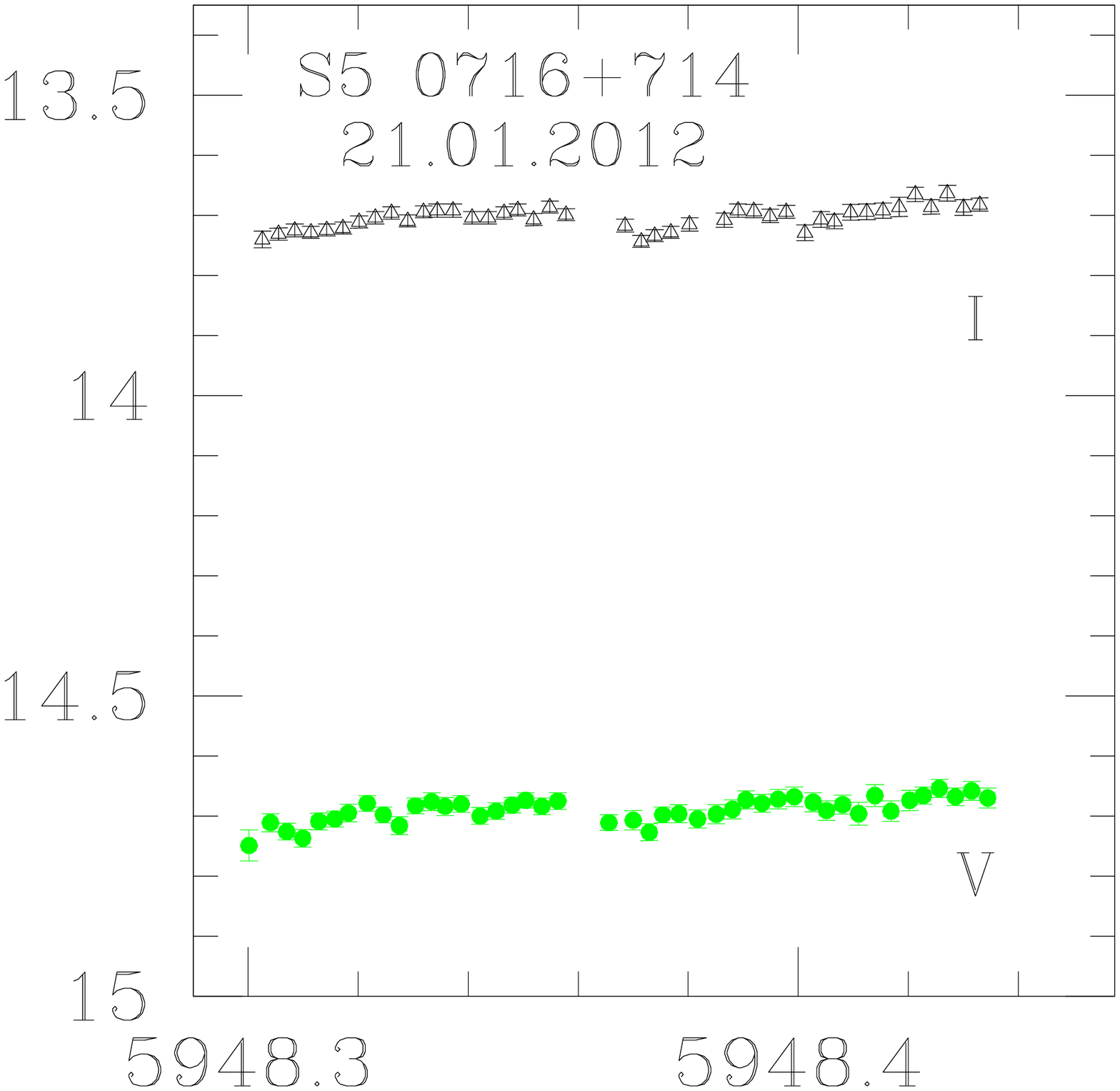,height=1.567in,width=1.59in,angle=0}
\epsfig{figure=  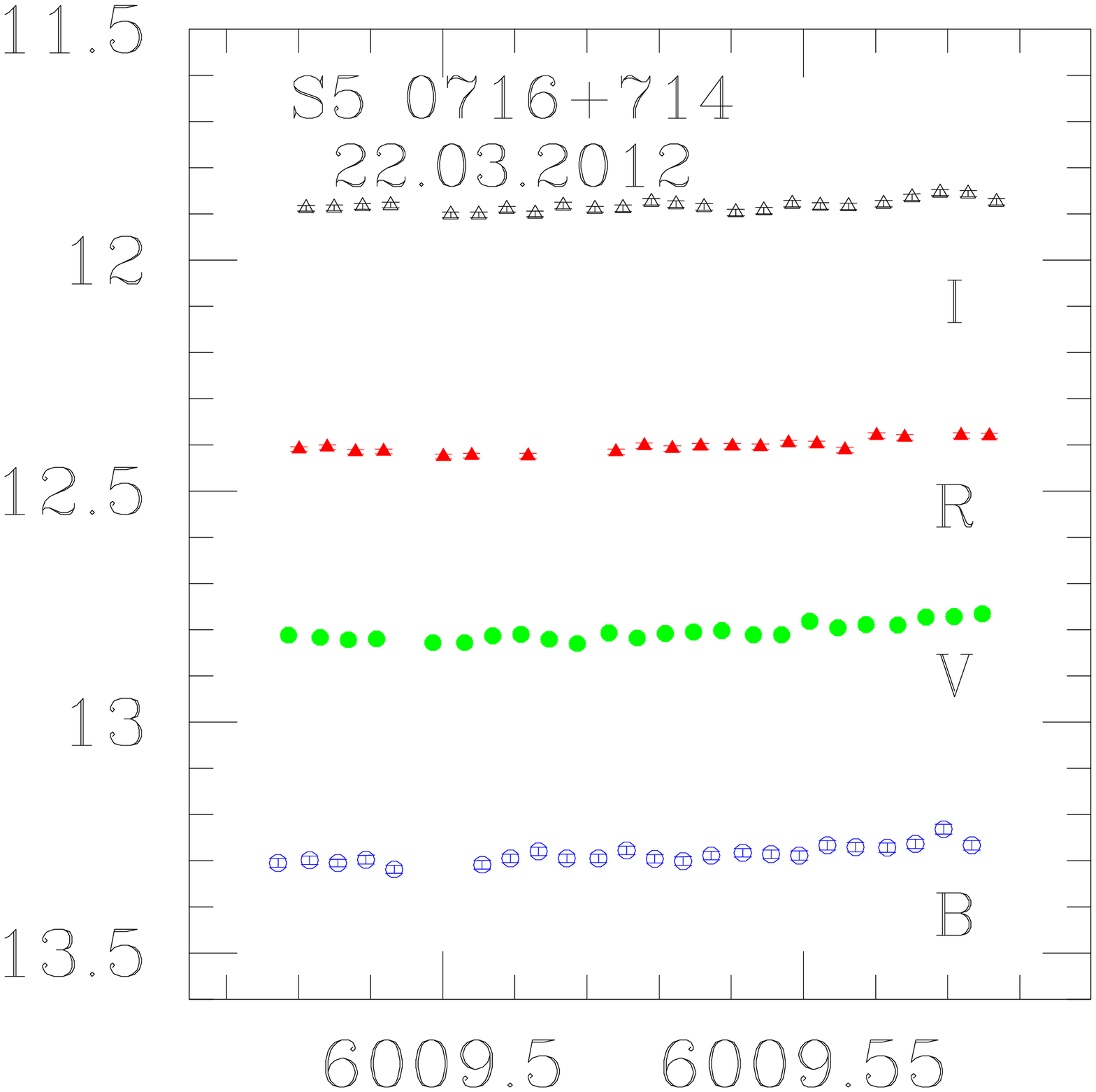,height=1.567in,width=1.59in,angle=0}
\epsfig{figure=  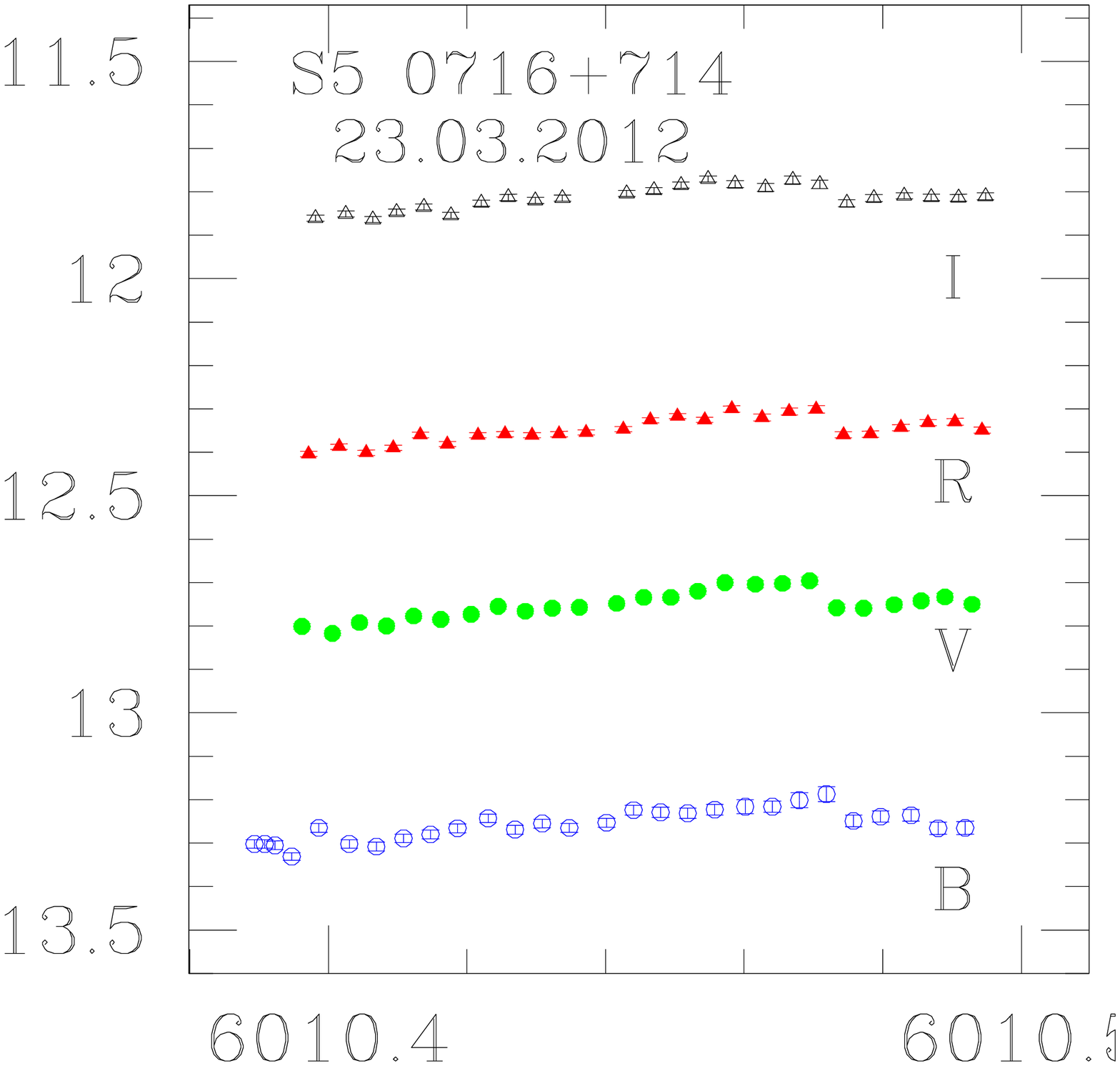,height=1.567in,width=1.59in,angle=0}
\epsfig{figure= 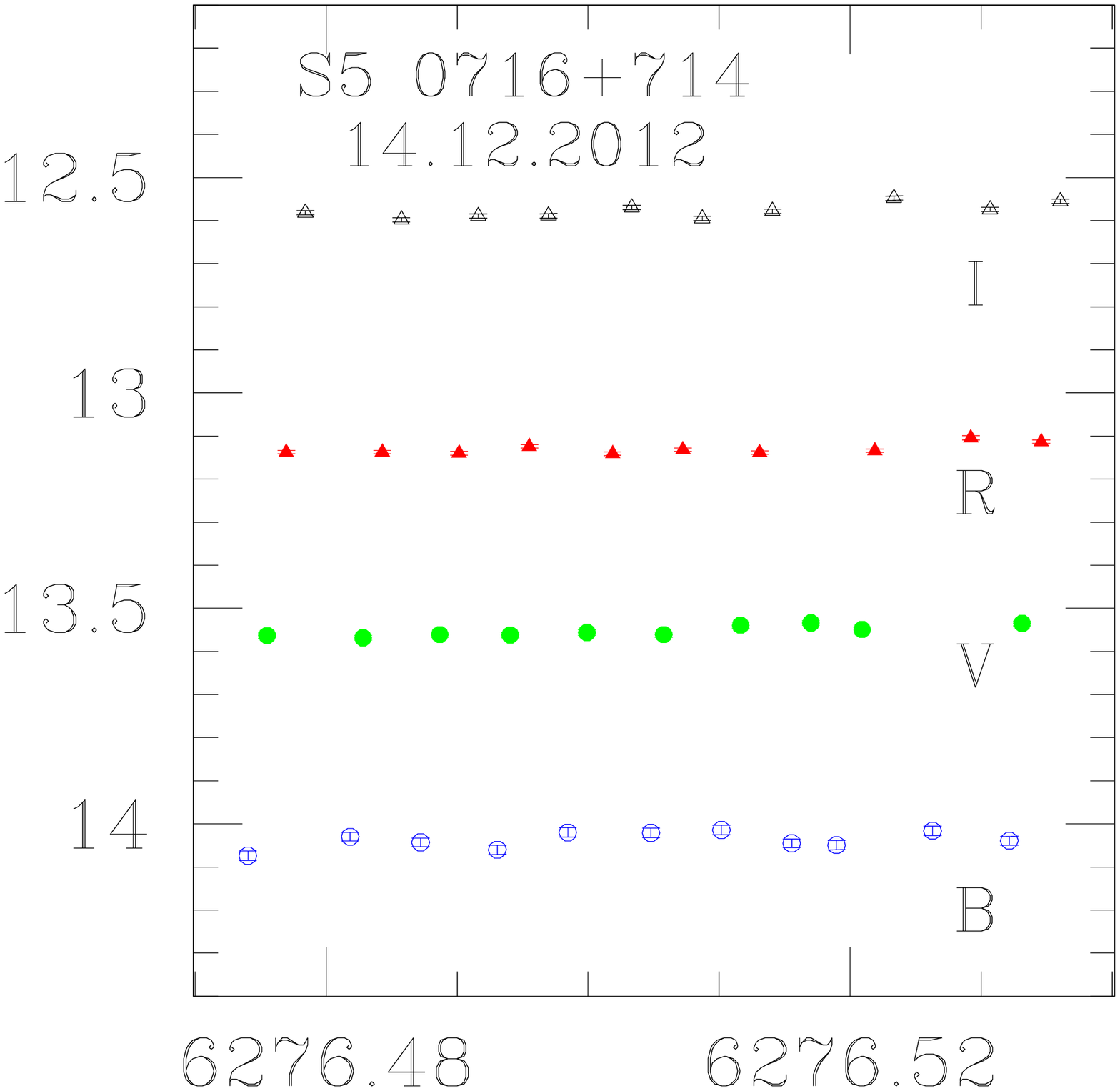,width=1.59in,height=1.567in,angle=0}
\epsfig{figure= 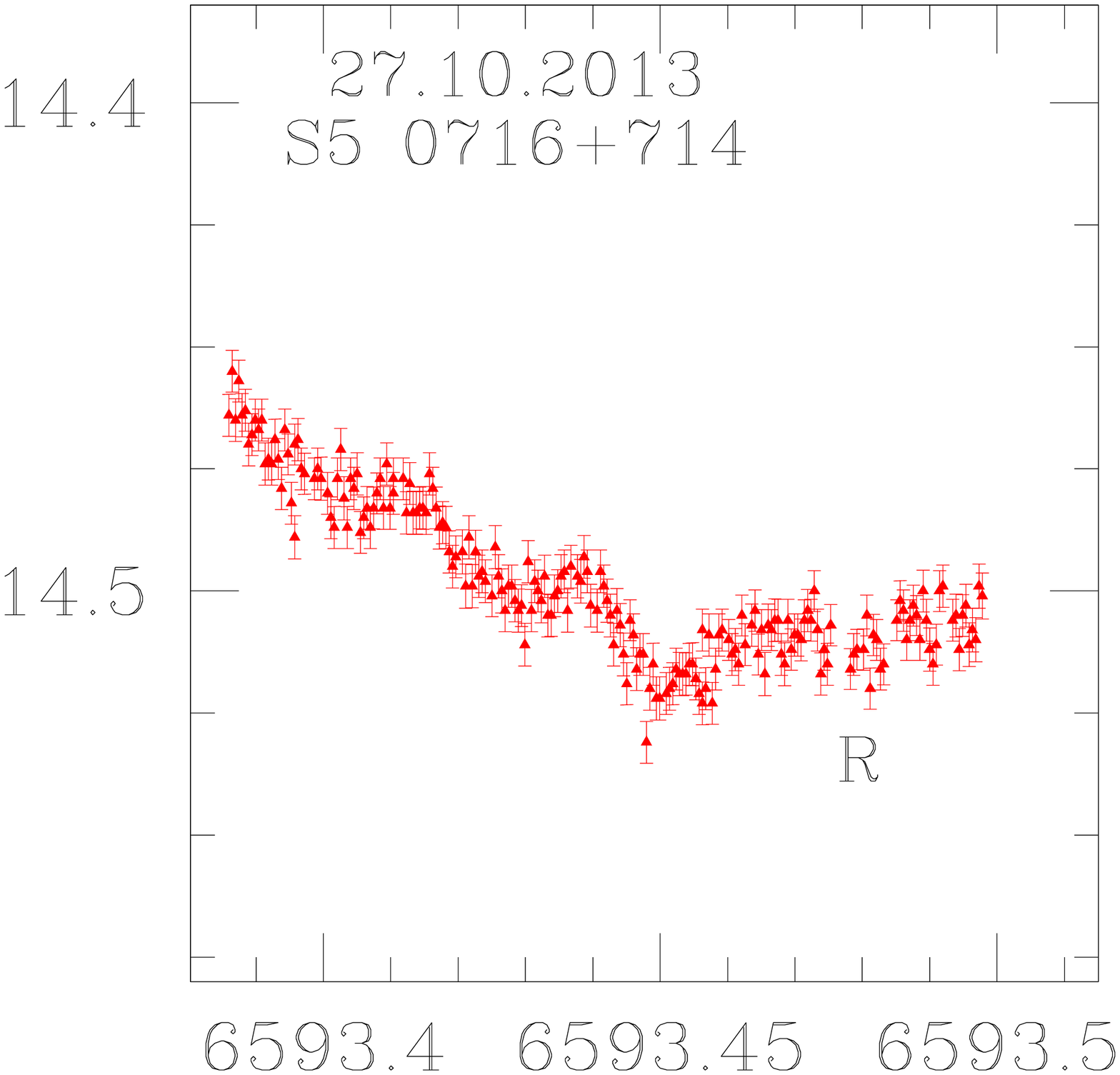,width=1.59in,height=1.567in,angle=0}
 \epsfig{figure=  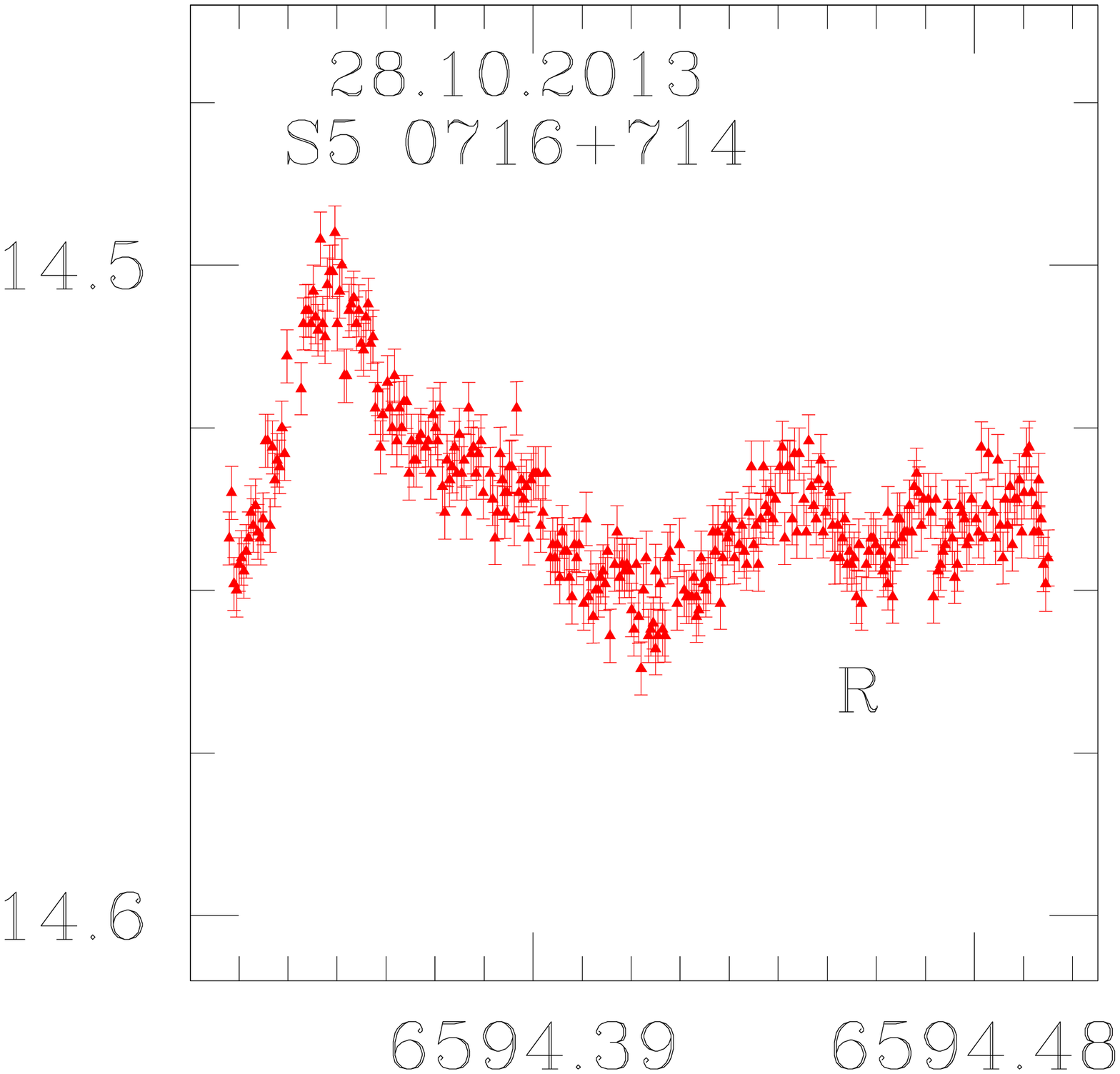,height=1.567in,width=1.59in,angle=0}
  \epsfig{figure=  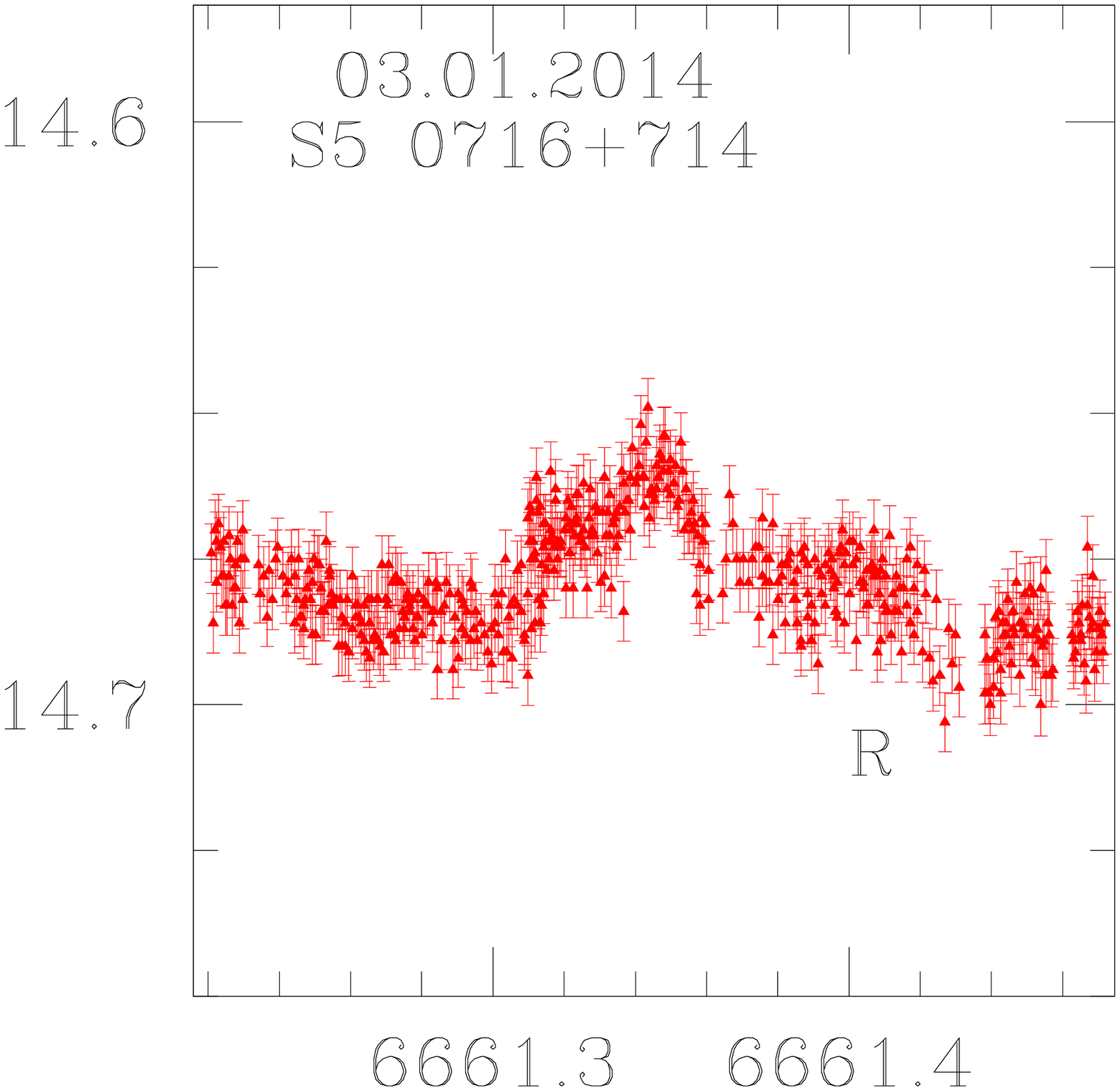,height=1.567in,width=1.59in,angle=0}
     \epsfig{figure=  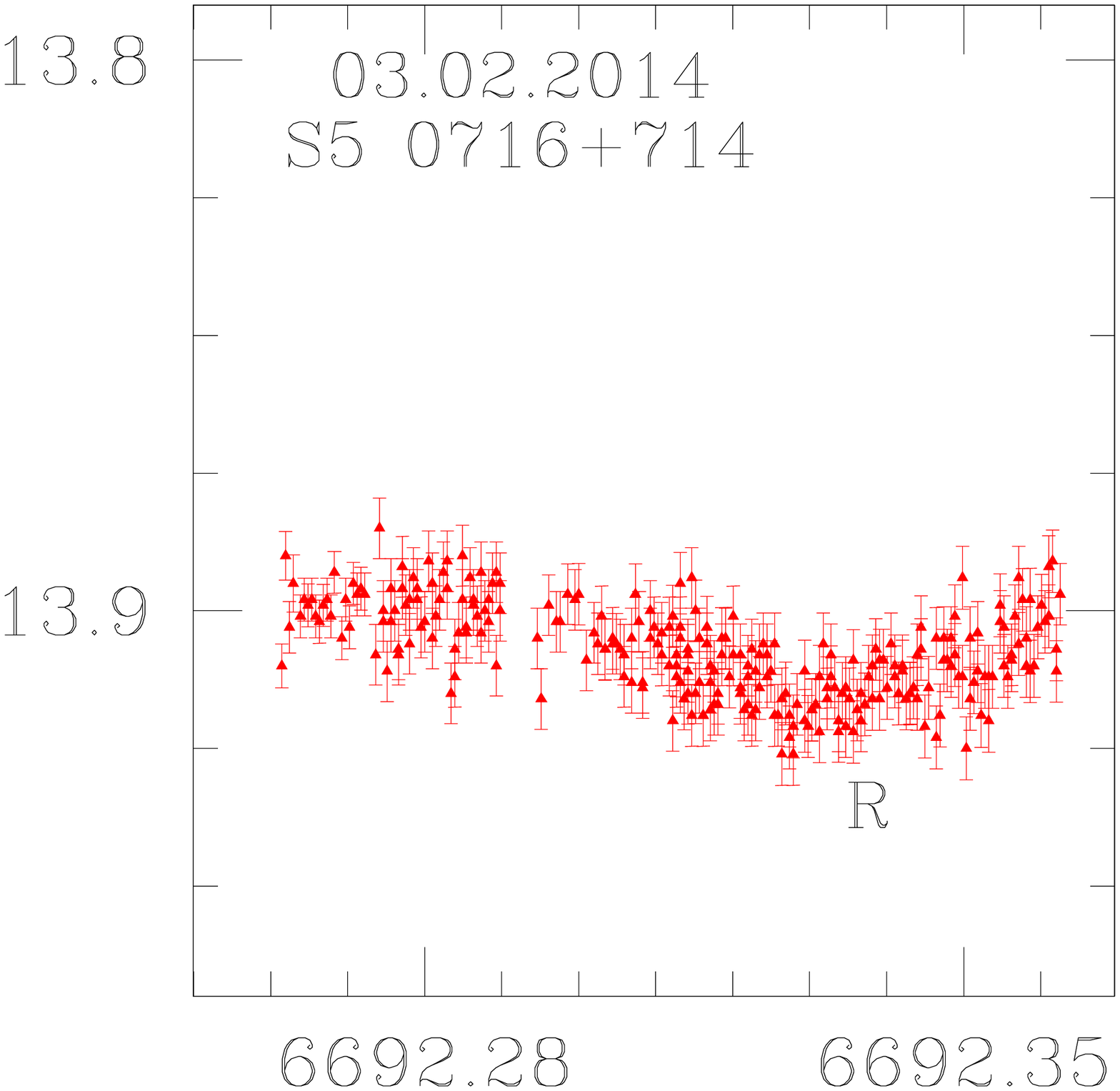,height=1.567in,width=1.59in,angle=0}
 \epsfig{figure=  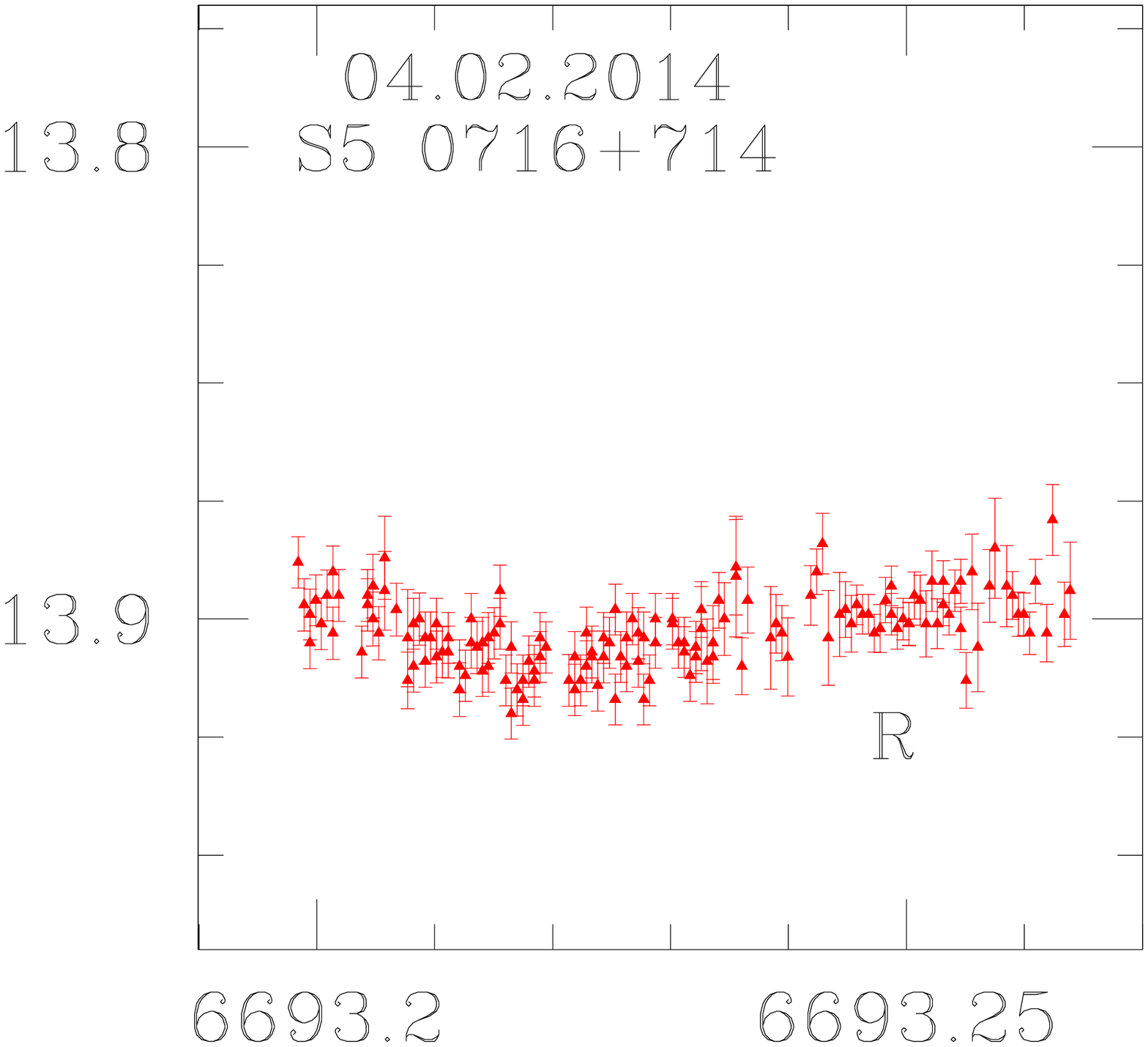,height=1.567in,width=1.59in,angle=0}
\epsfig{figure=  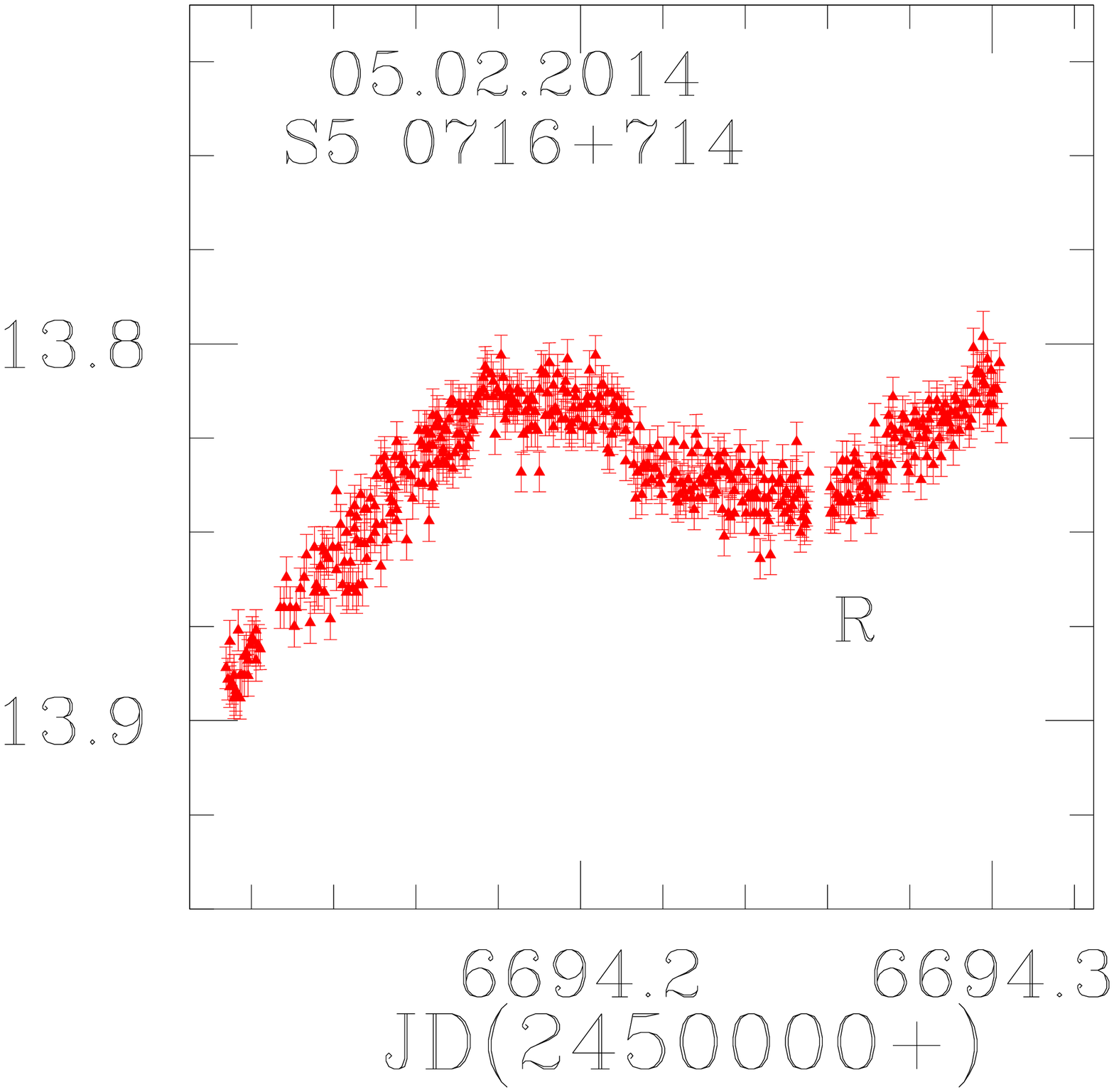,height=1.567in,width=1.59in,angle=0}
\caption{As in Figure 1 for S5 0716$+$714.}
\label{idvfigcont}
\end{figure*}
\section{Analysis Techniques}

\subsection{\bf Variability detection criterion}

The IDV of blazars was examined employing both the popularly used C-statistic and the more reliable F-test.

\subsubsection{\bf C-Test}
The C-statistic was introduced by Romero, Cellone, \& Combi (1999), using the variability detection parameter, C,  defined as 
the average of $C_{1}$ and $C_{2}$ with:
\begin{equation}
C_{1} = \frac {\sigma(BL-S_{A})}{\sigma(S_{A}-S_{B})}~
,~
C_{2} = \frac {\sigma(BL-S_{B})}{\sigma(S_{A}-S_{B})}.
\end{equation}
Here (BL$-$S$_{A}$), (BL$-$S$_{B}$), and (S$_{A}$$-$S$_{B}$) are the differential instrumental magnitudes of 
the blazar and standard star A (S$_{A}$), the blazar and standard star B (S$_{B}$), and S$_{A}$ vs.\ S$_{B}$
determined using aperture photometry of the source and comparison stars, whereas $\sigma$(BL$-$S$_{A}$), 
$\sigma$(BL$-$S$_{B}$) and $\sigma$(S$_{A}$$-$S$_{B}$) are observational scatters of the differential instrumental 
magnitudes of the blazar$-$S$_{A}$, blazar$-$S$_{B}$, and S$_{A}$$-$S$_{B}$, respectively. We observed three or 
more comparison stars of which we selected those two standard stars for which $\sigma$(S$_{A}$$-$S$_{B}$) was
minimal. A value of $C \geq 2.57$ implies that the source is variable at a nominal confidence level of $> 99$\%.
\noindent
\subsubsection{\bf F-Test}

As pointed out by de Diego (2010) the C-test is not a true statistic and is usually too conservative in quantifying 
variability. The F-test is a more powerful and properly distributed statistic that can be used to check the
presence of microvariability or IDV. The F values compare two sample variances and are calculated as

\begin{equation}
 \label{eq.ftest}
 F_1=\frac{Var(BL-S_{A})}{Var(S_{A}-S_{B})}, \nonumber \\ 
 F_2=\frac{Var(BL-S_{B})}{ Var(S_{A}-S_{B})}.
\end{equation}
Here (BL$-$S$_{A}$), (BL$-$S$_{B}$), and (S$_{A}$$-$S$_{B}$) are the differential instrumental magnitudes of blazar and standard A, blazar 
and standard B, and standard A and standard B, respectively, while Var(BL$-$S$_{A}$), Var(BL$-$S$_{B}$), and Var(S$_{A}$$-$S$_{B}$) are 
the variances of differential instrumental magnitudes.

The F value is compared with a critical value, $F^{(\alpha)}_{\nu_{bl},\nu_*}$, where $\nu_{bl}$ and $\nu_*$ 
respectively denote the number of degrees of freedom calculated as the number 
of measurements, $N$, minus 1 ($\nu = N - 1$), while $\alpha$ corresponds to the significance level, which we 
have set as 0.1 and 1 percent (i.e $3 \sigma$ and $2.6 \sigma$) in our analysis. We take the average of $F_1$ and 
$F_2$ to find a mean observational F value and compare it with the critical F value. If the mean F value is 
larger than the critical value, the null hypothesis (i.e., that of no variability) is discarded.  The modest number of measurements
we could make in most bands on most nights made the use of the ANOVA technique (de Deigo 2010) in-feasible.
In order to claim the presence or absence of variability during the observations, we have used the F-test, at 0.99 confidence level. 

\noindent
\subsubsection{\bf Percentage amplitude variation}

The percentage variation on a given night is calculated by using the variability amplitude parameter $A$,
introduced by Heidt \& Wagner (1996), and defined as
\begin{eqnarray}
A = 100\times \sqrt{{(A_{max}-A_{min}})^2 - 2\sigma^2}(\%) ,
\end{eqnarray}
where $A_{max}$ and $A_{min}$ are the maximum and minimum values in the calibrated LCs of the blazar, and $\sigma$
is the average measurement error. The calculated C- and F- statistics and the variability amplitude parameters, $A$, 
are presented in Tables 2 whose full version is available in the supplementary material to this article.  Nightly LCs and colour indices are listed as variable if $F > F_c(0.99)$, though rarely 
do they satisfy the $F > F_c(0.999)$ or the $C > 2.57$ criteria.

\begin{table*}
\caption{ Results of IDV observations of the blazars. The full
version of this table is available in the online version of this manuscript in the Supplementary Materials section. }
\textwidth=7.0in
\textheight=10.0in
\vspace*{0.2in}
\noindent

\begin{tabular}{cccccccccccc} \hline \nonumber

Source & Date       & Band   &N    & $\sigma_{(BL - S_{A})}$     &  $\sigma_{(BL - S_{B})}$         &  $\sigma_{(S_{A} - S_{B})}$       & F-test  &   Variable    & C &A\% \\
     &      &        &     &                             &                                  &                               &$F_{1},F_{2},F,F_{c}(0.99),F_{c}(0.999)$ & & & \\\hline 

3C 454.3 & 29.10.2013   & B     & 23 & 0.0772 & 0.0784 & 0.0310 & 6.2, 6.4, 6.28, 2.79, 3.98 & V & 2.5065 & 26.98  \\
         &    & V     & 24 & 0.0200 & 0.0347 & 0.0278 & 0.5, 1.6, 1.04, 2.72, 3.85 & NV & 0.9840 & --  \\
         &    & R     & 24 & 0.1035 & 0.0222 & 0.1080 & 0.9, 0.04, 0.47, 2.72, 3.85 & NV & 0.5819 & --  \\
         &    & I     & 24 & 0.0207 & 0.0249 & 0.0179 & 1.3, 1.9, 1.63, 2.72, 3.85 & NV & 1.2723 & --  \\
         &    & (B-V) & 23 & 0.0701 & 0.0627 & 0.0353 & 3.9, 3.1, 3.55, 2.79, 3.98 & V & 1.8807 & 23.62 \\
         &    & (B-I) & 23 & 0.0750 & 0.0674 & 0.0321 & 5.5, 4.4, 4.93, 2.79, 3.98  & V & 2.2177 & 23.25 \\
         &    & (V-R) & 24 & 0.1072 & 0.0334 & 0.1050 & 1.0, 0.1, 0.57, 2.72, 3.85 & NV & 0.6695 & -- \\
       &    & (R-I) & 24 & 0.1098 & 0.0306 & 0.1136 & 0.9, 0.1, 0.50, 2.72, 3.85 & NV & 0.6176 & -- \\ \hline
     \end{tabular} \\
\noindent
V : Variable, NV : Non-Variable     \\
\end{table*}

\subsection{\bf Structure Function analysis}

The first order Structure Function (SF), introduced by Simonetti, Cordes, 
\& Heeschen (1985), is applied to each of the non-uniformly
sampled LCs. The SF is designed to search for characteristic variability time scales and possible periodicities.
For details about SF as we have employed it, see Gaur et al.\ (2010).

The first order SF initially increases with larger time lags but can indicate the following behaviours at
larger lags:
(1) a continuing rise with lag indicates that any characteristic timescale of variability is longer than 
the length of the data set, with the longest time lag giving the minimum value of the time scale of uncorrelated 
data points;
(2) uncorrelated data produce a ``white noise'' behaviour, characterized by a constant slope (Ciprini et al.\ 2003);
(3) the presence of a plateau indicates a characteristic time scale of the variability; 
(4) if the SF curve falls, producing a dip, after reaching a plateau, then the time lag at that local minimum 
value of the SF may mark the presence of a periodic cycle and this possibility is strengthened if there is a 
lag corresponding to differences between two or more minima; however, a single dip is considered spurious if 
observed at time lag close to the length of data run; 
(5) a very irregular SF is observed for variations dominated by one or few large outbursts or variations that 
are more or less periodic.

As pointed by Emmanoulopoulos, McHardy, \& Uttley (2010), the SF method sometimes leads to spurious results and thus incorrect 
claims of periodicities or timescales. Hence we have also examined the data for timescales and possible 
periodicities by the Discrete Correlation Function (DCF) method.

\subsection{\bf Discrete correlation function (DCF)}

The DCF technique, first presented by Edelson \& Krolik (1988), is similar to the classical cross-correlation (CCF) 
with an added advantage of being able to be used for the analysis of unevenly sampled data without 
interpolating data points, thus giving a meaningful estimation of the errors. Hufnagel \& Bregman (1992) further 
generalized the method to include a better error estimate.
In general, a DCF value $> 0$ 
implies the two data signals are correlated, while the two anti-correlated datasets have a DCF $<$ 0, and a DCF 
value equal to 0 implies no correlation exists between the two data trains. For quantitative details about the DCF see
Agarwal \& Gupta (2015), and references therein.

\section{\bf Results}

\subsection{3C 454.3}

3C 454.3 (PKS 2251+158) ($\alpha_{2000.0}$ = 22h 53m 57.75s $\delta_{2000.0}$ = $+16^{\circ} 08^{'} 53.56^{\prime \prime}$) 
is a well known FSRQ at a redshift of $z = 0.859$.  It has displayed quite high optical
activity since 2001 that has been suggested to be, at least in part, due to changes in the viewing 
angle caused by a helical jet geometry (e.g., Villata \& Raiteri 1999; Raiteri et al.\ 2008).
A significant increase in activity at optical and radio wavelengths in conjunction with multiple $\gamma$-ray flaring
events detected with the AGILE and Fermi $\gamma$-ray satellites, renders 3C 454.3 well suited for 
multi-wavelength campaigns (e.g., Pacciani et al.\ 2010; Fuhrmann et al.\ 2006).\\

\begin{table}
\caption{ Results for IDV studies displaying magnitude changes in each band.  }
\textwidth=6.0in
\textheight=9.0in
\vspace*{0.2in}
\noindent
\begin{tabular}{p{1.5cm}p{0.3cm}p{0.5cm}p{1cm}p{0.6cm}p{1cm}p{0.4cm}} \hline

Source      & Band & Faintest  & ~~~Date  & Brightest & ~~~Date  & $\Delta m$\\ 
            &      & ~~Mag       & ~~~(Max) &~~Mag  & ~~~(Min) &           \\ \hline 

3C 4543.3   & B    & 16.78  & 04.11.2013 & 14.45 & 25.09.2013 & 2.33 \\
            & V    & 15.84  & 04.11.2013 & 13.86 & 25.09.2013 & 1.98 \\
            & R    & 15.40  & 04.01.2014 & 13.14 & 24.06.2014 & 2.26 \\
            & I    & 15.00  & 04.01.2014 & 12.74 & 25.09.2013 & 2.26 \\
3C 279      & R    & 14.56  & 10.04.2014 & 14.45 & 03.02.2014 & 0.11 \\
S5 0716+714 & B    & 14.07  & 14.12.2012 & 13.19 & 23.03.2012 & 0.88 \\
            & V    & 14.75  & 21.01.2012 & 12.63 & 20.03.2012 & 2.12 \\
            & R    & 14.82  & 04.01.2014 & 12.25 & 20.03.2012 & 2.69 \\
            & I    & 13.74  & 21.01.2012 & 11.74 & 20.03.2012 & 2.00 \\
 \hline
\end{tabular}  \\
Column 1 is the Source name, column 2 indicates the band in which observations were taken,
column 3 represents the maximum magnitude attained by the source in the filter mentioned in previous column on a particular date which is given
in column 4, followed by minimum magnitude value and respective date in column 5 and column 6 and finally
the net magnitude change during IDV observations is given in column 7.
\end{table}  

In May 2005, a dramatic outburst was recorded in the radio to X-ray energy bands (e.g., Vercellone et al.\ 2010),
reaching $R \approx 12$~mag before decaying to $R = 15.8$~mag in $\sim$75d (Villata et al.\ 2006).
Based on the continuous monitoring of the object with the Whole Earth Blazar Telescope (WEBT) a quiescent state
was reported during spring 2006 -- 2007, which revealed the ``big blue bump'' in the optical spectrum
(a feature more typical of radio quiet AGN), indicating thermal emissions from the accretion disk,
as well as a ``little blue bump'' due to line emission from the broad-line region of the AGN (Raiteri et al.\ 2007). 
After this faint state the object again underwent multi-frequency flaring activity in July 2007 
(Vercellone et al.\ 2008) followed by multiple flares during Nov -- Dec 2007,
reaching optical levels of  $R = 12.69$~mag  (Zhai, Zheng, \& Wei 2011). 
The 2010 Oct -- Dec outburst extended over all wavelengths with $\gamma$-ray flux peaking at 
$85 \times 10^{-6}$ph~cm$^{-2}$~s$^{-1}$ and optical emissions increasing simultaneously (Vercellone et al.\ 2011) as 
observed at several observatories (Larionov, Villata, \& Raiteri 2010; Semkov et al.\ 2010; Raiteri et al.\ 2011),
along with increases in X-ray and radio fluxes (Jorstad et al.\ 2013; Wehrle et al.\ 2012). 

Extensive flux variability
observations at different frequencies over many years encouraged detailed modeling and analysis 
of SEDs which helped to interpret the physical conditions of the jet and the emission mechanism 
responsible for the emitted radiation (e.g., Finke \& Dermer 2010; B{\"o}ttcher et al.\ 2013). 

\begin{figure}
\centering
\includegraphics[width=3.5in,height=4.5in]{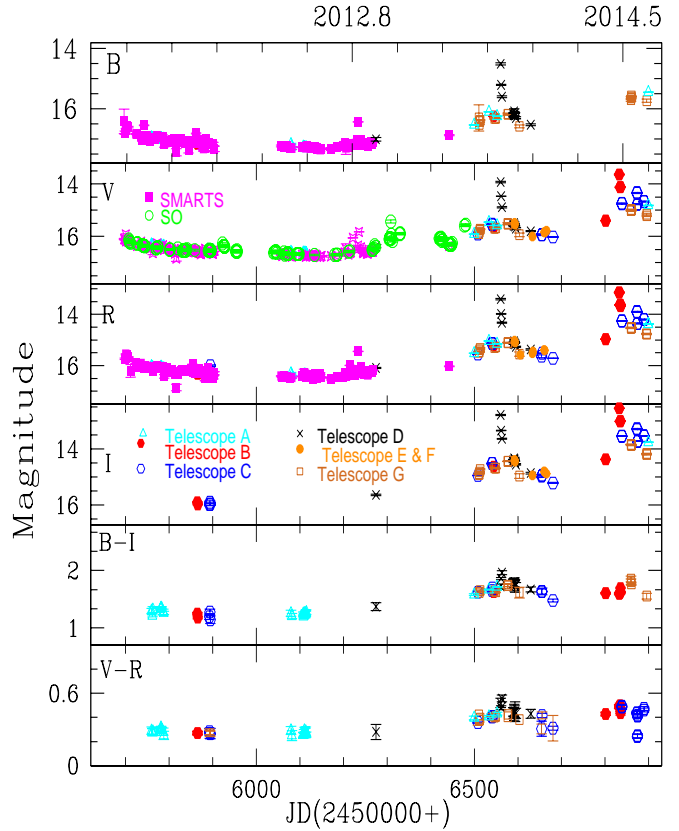}
\caption{Short/Long-term variability LCs and colour indices of 3C 454.3 in the B, V, R and I bands 
and (B-I) and (V-R) colours. Different colours denote data from different observatories: Cyan = telescope A, 
Red = telescope B, Blue = telescope C, Black = telescope D, Dark orange = telescopes E \& F, Chocolate =  telescope G, 
Magenta = SMARTS, Green = SO. }
\end{figure}

\subsubsection{\bf Intra$-$Day Variability}

Our observations of 3C 454.3 were carried out on 13 nights spanning the period between 2013 September and 2014
January, to search for variability on intra-day timescales. The light curves of the blazar  
are displayed in Fig.\ 1.
Photometric observations in the B, V, R, and I bands were carried out 
essentially continuously and quasi-simultaneously on 2013 September 25, 28, October 24, 25, 26, 27, 28, 29 
and November 04 for a span of $\sim 4$~hrs each. Also photometric observations only in the  R band were carried 
out on 2013 Oct 27, 28, Dec 7, 8, and 2014 Jan 3, 4 for a span of $\sim 4$~hrs each. We have labeled the plots 
with the blazar name 
and the date of monitoring. In order to claim the presence or absence of variability during the observations, 
we have used the F-test, at 0.99 confidence level, with results summarized in Tables 2 whose complete version 
is presented in the supplementary materials online. Also, the details of the brightness changes in the source
during the INOV observations are listed in table 3.

{\bf B pass-band:} The target was observed in the B pass-band on 8 nights for $\sim 4$~hrs on an average.
The F-test indicates that the source was variable in the B band on 3 nights: 2013 Oct 25, 28 and 29. During these 8 nights the B-band
brightness decayed to $B = 16.78$~mag which is $\sim 3.01$~mag fainter than the brightest B magnitude level of 13.77 reported by Zhai, Zheng, 
\& Wei (2011). Hence the source might be best characterized as being in a post-outburst state during our 
observations of 2013. Recently, 3C 454.3 showed drastic increase in brightness reaching B band magnitude of 
14.14 on 2014 June 24. It decreased in brightness after then.

{\bf V pass-band:} Our V filter observations of 3C 454.3 lasted for a total of 9 nights over spans of 
$\sim 4$~hrs on each night. C test results reveal the lack of micro-variability in the source in the V band.
The F-test results also support the absence of micro-variability.

{\bf R pass-band:} We monitored the blazar in R pass-band for 13 nights from 2013 Sept 25 through 2014 Jan 04, 
with duration ranging from 3 to 4 hrs. We detected noticeable variability on 4 out of these 13 nights: 2013 Oct 
28, Dec 07, 08 and 2014 Jan 04. The C values for above dates
indicate that the source is variable on 2013 Dec 07 with amplitude of variability reaching 9.16~\% as the source 
changed in brightness from 15.525~mag to 15.520~mag in $\sim 3.1$~hrs with a prominent dip observed at UT 14.57 
hrs when it faded to 15.58~mag. On this night, the maximum variation noticed was $\Delta R \sim 0.1$ with the 
brightest level of 15.49 mag and faintest level at 15.58 mag.
Analysis using the F-test indicates the presence of micro-variability on the remainder of the observation dates 
mentioned above. During the span of 13 nights, we found that the target faded to $R = 15.40$~mag on 2014 Jan 14,
which is $\sim 3.4$~mag fainter than the
brightest magnitude of $R \sim 12$~mag, as reported by Villata et al.\ (2006), but is still brighter by 
$\sim 1.6$~mag than the faintest level of 17 mag. So, as noted for the B-band, the FSRQ appears to be in a 
post-outburst stage during the later months of 2013. But during the flaring state of May 2014 it reached the 
brightest R Band magnitude of 13.143 on 2014 June 24.

{\bf I pass-band:} We carried out monitoring observations of 3C 454.3 in the I band for 9 nights between
2013 Sept 25 and Nov 11. The C values reveal the lack of micro-variability in the source in the I band.
The F-test results also  support the absence of micro-variability in I.
\begin{figure}
\centering
\includegraphics[width=3.5in,height=4in]{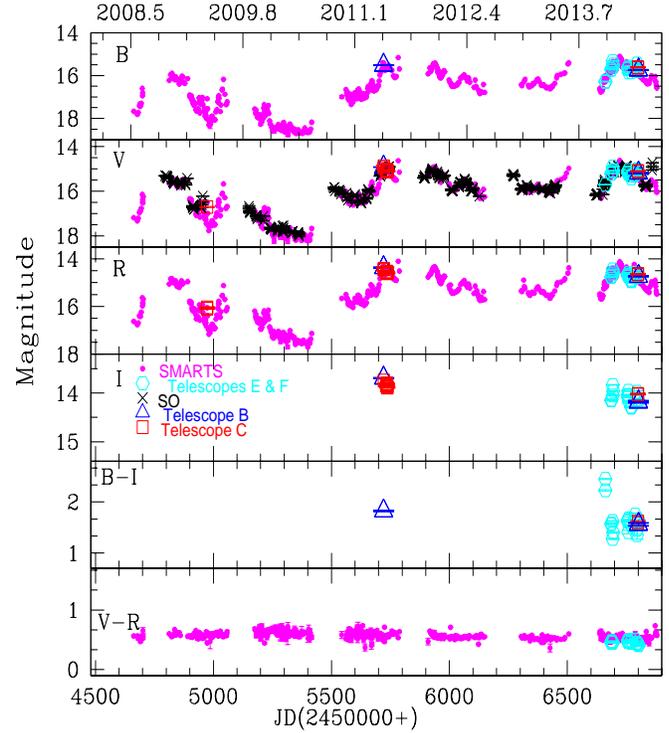}
\caption{Short/Long-term variability LCs and colour indices of 3C 279 in the B, V, R and I bands 
and (B-I) and (V-R) colours. Different colours denote data from different observatories: magenta, SMARTS; cyan, telescopes E \& F;
blue, telescope B; red, telescope C; black = Seward. }
\label{stvfigure2}
\end{figure}

\begin{figure}
\centering
\includegraphics[width=3.3in,height=3.8in]{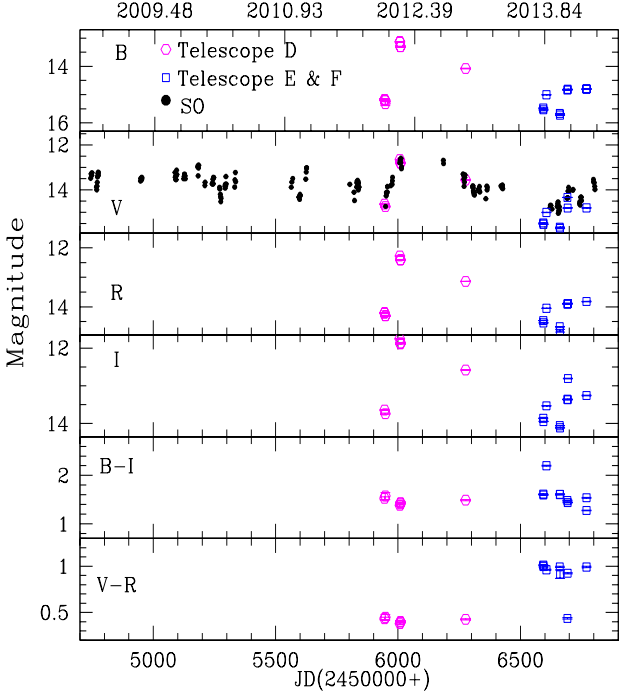}
\caption{Short/Long-term variability LCs and colour indices of S5 0716+714 in the B, V, R and I bands 
and (B-I) and (V-R) colours. Different colours denote data from different observatories: magenta, D; blue, telescopes E \& F;
black, SO. }
\end{figure}
\subsubsection{\bf Short-term and Long-term flux variability}

The {\bf short-term/long-term variability (STV/LTV)} LCs of 3C 454.3 in the B, V, R, and I pass-bands are shown in Fig.\ 4
along with the colour variations (B-I) and (V-R).  For a particular date the magnitude is taken to be the mean magnitude 
of all image frames in a specific pass-band and the time is taken as the Julian Date (JD) value at UT = 00 hrs on 
the same date. For calculating the amplitude of variability using Eqn.\ (3), we considered every data point 
available to us. Also, the details of the magnitude changes during the whole observation period are listed in table 4.

{\bf B passband:} 
The top panel of the figure shows the {\bf STV/LTV} LC of 3C 454.3 in the B band; it includes our data along 
with those provided by SMARTS.  We have monitored the source in B band for STV studies from JD 2455620 to 2456930
using the seven different telescopes described in Section \ref{observations} and indicated in the figure, 
which shows that after constantly decreasing in brightness, the source attained the faintest level
on JD 2454986.5, followed by a brightening trend till JD 2455510.5. 
Since then the target decreased in brightness, displaying a maximum magnitude variation of $\sim 3.181$. The same 
trends were noted by other authors (e.g., Zhai, Zheng, \& Wei 2011). The source showed increase in brightness during mid year 2014.
During the total time span, the amplitude of variability in the B band was calculated to be 318~\%.       

{\bf V passband:} 
The corresponding {\bf STV/LTV} LC in the V band is 
generated using our data along with that provided by SMARTS and SO, covering the time period between 
JD~2454662.5 and 2456930. After constantly decreasing
in brightness since JD 2455816.5, it recently entered a flaring stage reaching a minimum value of V band magnitude on 2014 June 23.
We found the the amplitude of variability in V band to be $\sim 306$~\%.

{\bf R passband:}
The {\bf STV/LTV} LC in the R band is displayed in the third panel from top in  Fig. 4, which includes observations 
from SMARTS in addition to our data, and runs from JD 2455759.5 to JD 2456930. The object attained a large
flux state of $R = 13.170$~mag on 2455519.5. Zhai, Zheng, \& Wei (2011) have also reported the B and R band magnitudes of the
source to be $\sim 14$ and $\sim 13$, respectively, in 2010 Nov, when the source was in a remarkably bright state. 
Since then the brightness of 3C 454.3 generally has been declining.
It seems to have entered the post outburst stage in 2013. As in the other bands, in mid 2014 there was an increase in the flux
of the target. But we also saw it decreased in brightness reaching a R band magnitude of
14.38 on JD 2456899.5. The amplitude of variability was found to be as large as $\sim 362$~\%.

{\bf I passband:}
The third panel from the bottom in above mentioned figure represents the I band LC during observations made between  
JD~2455759.5 and 2456930. We 
calculated the variability amplitude to be 320~\% using equation 3.

\subsubsection{\bf Colour variability on diverse timescales}

We have investigated the variations of the (B-V), (B-I), (V-R) and (R-I) colours on intra-day timescales and 
presented the results in Table 2. Performing C- and F-tests on each night's data sets allowed us to quantify 
that the source rarely showed significant colour variations: according to the F-test, colour variability was
detected on 3 nights, namely 2013 Oct 25, 28, and 29.

In Figure 4, we have shown the (B-I) and (V-R) colour indices as functions of time in the second and 
first panels from the bottom, respectively. The plots indicate moderate colour variations in both of these indices. 
The maximum variation seen in the source in (V-R) is 0.308 mag found between 0.249 mag at JD 2456081.5 and 
0.557 mag at JD 2456563.5.  In the (B-I) colour-index the maximum variation noticed during our observation 
span was found to be 0.820 between the colour index values of 1.963 mag at JD~2456563.5 and 1.143 mag at 
JD~2455894.5.

\subsection{3C 279}

{\bf The FSRQ 3C 279 ($\alpha_{2000.0}$ = 12h 56m 11.17s, $\delta_{2000.0}$ = $-05^{\circ} 47^{'} 21.5^{\prime \prime}$) 
at a red-shift of 0.536 (Burbidge \& Rosenburg 1965) displays violent flux variability in all wavelength 
bands. }
It has been intensively studied through several simultaneous multi-wavelength campaigns (e.g.\ Hartman et al.\ 1996; 
Wehrle et al.\ 1998; B{\"o}ttcher et al.\ 2007). The source has been regularly variable in optical bands with very 
large optical variation of $\Delta B > 6.7$~mags (ranging between 11.3 and 18.0; Netzer et al.\ 1994) from the archival plates
of the Harvard collection as reported long ago by Eachus \& Liller (1975). 
A large amplitude STV of $\Delta R = 1.5$~mags was reported in 42 days (Gupta et al.\ 2008b). 

During a 2006 WEBT (Whole Earth Blazar Telescope)
campaign  a dramatic flare was reported at radio, NIR and optical frequencies with R $\sim 14.0-14.5$~mag. Quasi-exponential
decays of B,V, R, and I fluxes of $\sim 1$~mag on a timescale of $\sim 12.8$d, between JD 2453743 - JD 2453760,
was displayed by the source (B{\"o}ttcher et al.\ 2007), which prompted B{\"o}ttcher \& Principe (2009) to propose this being a signature
of deceleration of a synchrotron emitting jet component.


The SED of 3C 279 has two broad peaks believed to be associated with synchrotron and Inverse Compton (IC) emission. The ``Compton hump''
can be explained by the synchrotron self Compton (SSC) process through the cooling of a relativistic electron distribtion. But, since 
it requires extremely low magnetic field (B{\"o}ttcher, Reimer, \& Marscher 2009) and high radiation energy density in comparison to 
the magnetic energy density (Maraschi, Ghisellini, \& Celotti 1992), thus poses problems in explaining the VHE emission. So another 
explanation is the external Compton mechanism where the target photons from the accretion
disc, or broad line region (BLR) clouds, or dusty torus are strongly deboosted. 

\subsubsection{\bf Intra$-$Day Variability}

{\bf R passband:}
We carried out our photometric observations of 3C 279 
during the period Jan 2014 to May 2014 for IDV studies in the R passband, for
a total of 12 nights. The extracted LCs are displayed in Figs.\ 2.
The details of magnitude changes on intra-day basis during the 12 nights are given in table 3.
Gupta et al.\ (2008b) reported the source to be in the outburst state with R $\sim$ 12.6 mags. During our last observations the FSRQ seems to be in
somewhat of a post outburst state
with R band magnitude of 14.75 as on 2014 May 24. It was found to be variable with continuous,
but small decreases, in magnitude until it again flared in May 2014.
So 3C 279 was found to be highly variable during the IDV observations and seems to be in an active, if not outburst, state.

\begin{table}
\caption{ Results for STV/LTV studies displaying magnitude changes in each band.  }
\textwidth=6.0in
\textheight=9.0in
\vspace*{0.2in}
\noindent
\begin{tabular}{p{1.5cm}p{0.3cm}p{0.5cm}p{1cm}p{0.6cm}p{1cm}p{0.4cm}} \hline

Source      & Band & Faintest  & ~~~JD  & Brightest & ~~~JD  & $\Delta m$\\ 
            &      & ~~Mag       & ~~~(Max) &~~Mag  & ~~~(Min) &           \\ \hline 

3C 4543.3   & B    & 17.47  & 2454986.5 & 14.14 & 2456832.5 & 3.33 \\
            & V    & 16.84  & 2455816.5 & 13.63 & 2456832.5 & 3.21 \\
            & R    & 16.85  & 2455816.5 & 13.14 & 2456832.5 & 3.71 \\
            & I    & 16.07  & 2456108.5 & 12.55 & 2456832.5 & 3.52 \\
3C 279      & B    & 18.77  & 2455316.7 & 15.10 & 2456721.8 & 3.67 \\
            & V    & 18.25  & 2455314.6 & 14.62 & 2456721.8 & 3.63 \\
            & R    & 17.70  & 2455400.5 & 14.12 & 2456721.5 & 3.56 \\
            & I    & 14.30  & 2456771.1 & 13.84 & 2456692.4 & 0.46 \\
S5 0716+714 & B    & 15.72  & 2456662.1 & 13.13 & 2456007.3 & 2.59 \\
            & V    & 15.70  & 2456662.1 & 12.60 & 2456011.7 & 3.12 \\
            & R    & 14.70  & 2456661.4 & 12.28 & 2456007.4 & 2.42 \\
            & I    & 14.11  & 2456662.1 & 11.75 & 2456007.4 & 2.36 \\
 \hline
\end{tabular}  \\
Column 1 is the Source name, column 2 indicates the band in which observations were taken,
column 3 represents the maximum magnitude attained by the source in the filter mentioned in previous column on a particular JD which is given
in column 4, followed by minimum magnitude value and respective JD in column 5 and column 6 and finally
the net magnitude change during IDV observations is given in column 7.
\end{table}

\begin{figure*}
\epsfig{figure=  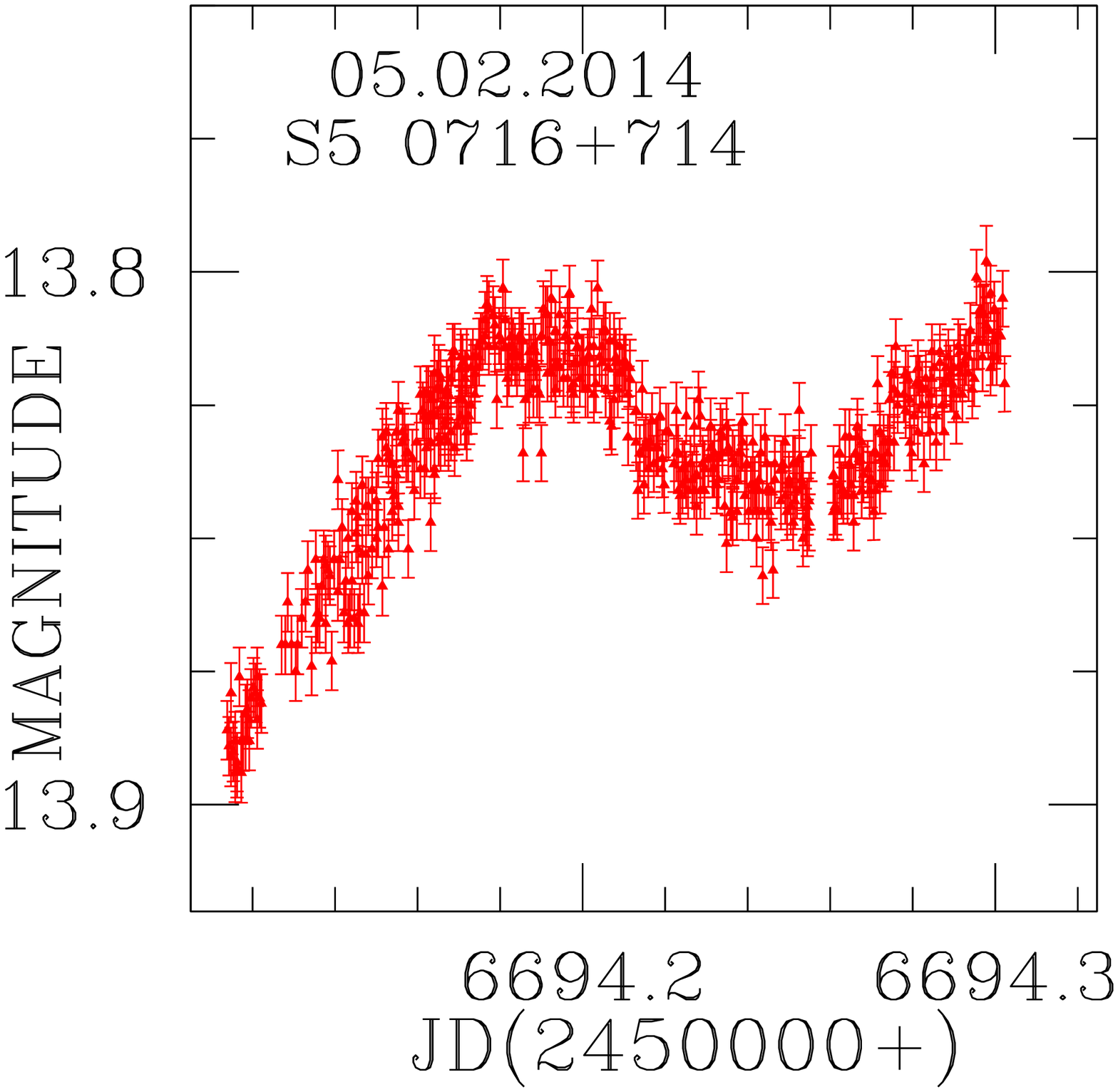,height=1.8in,width=2in,angle=0}
\epsfig{figure=  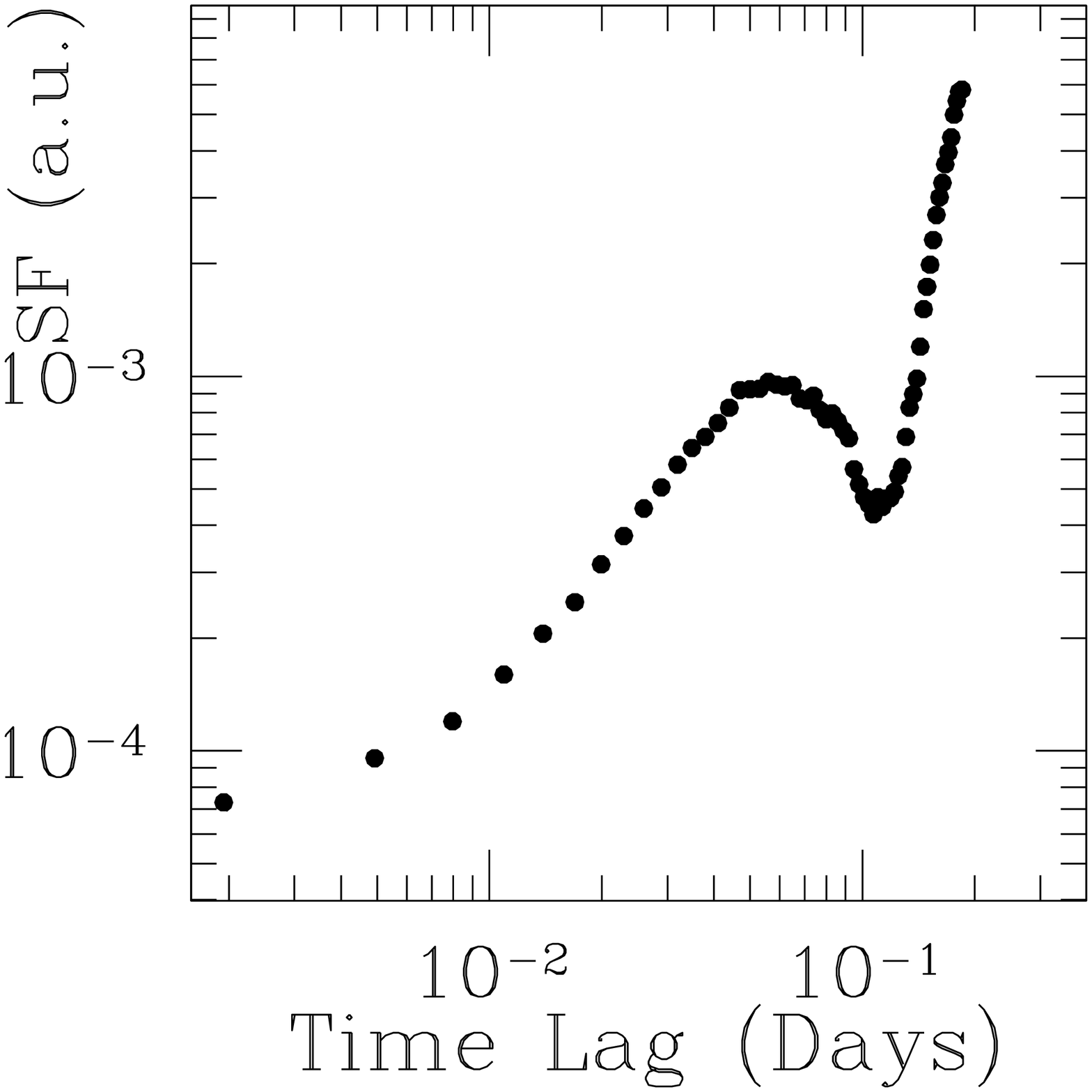,height=1.8in,width=2in,angle=0}
\epsfig{figure=  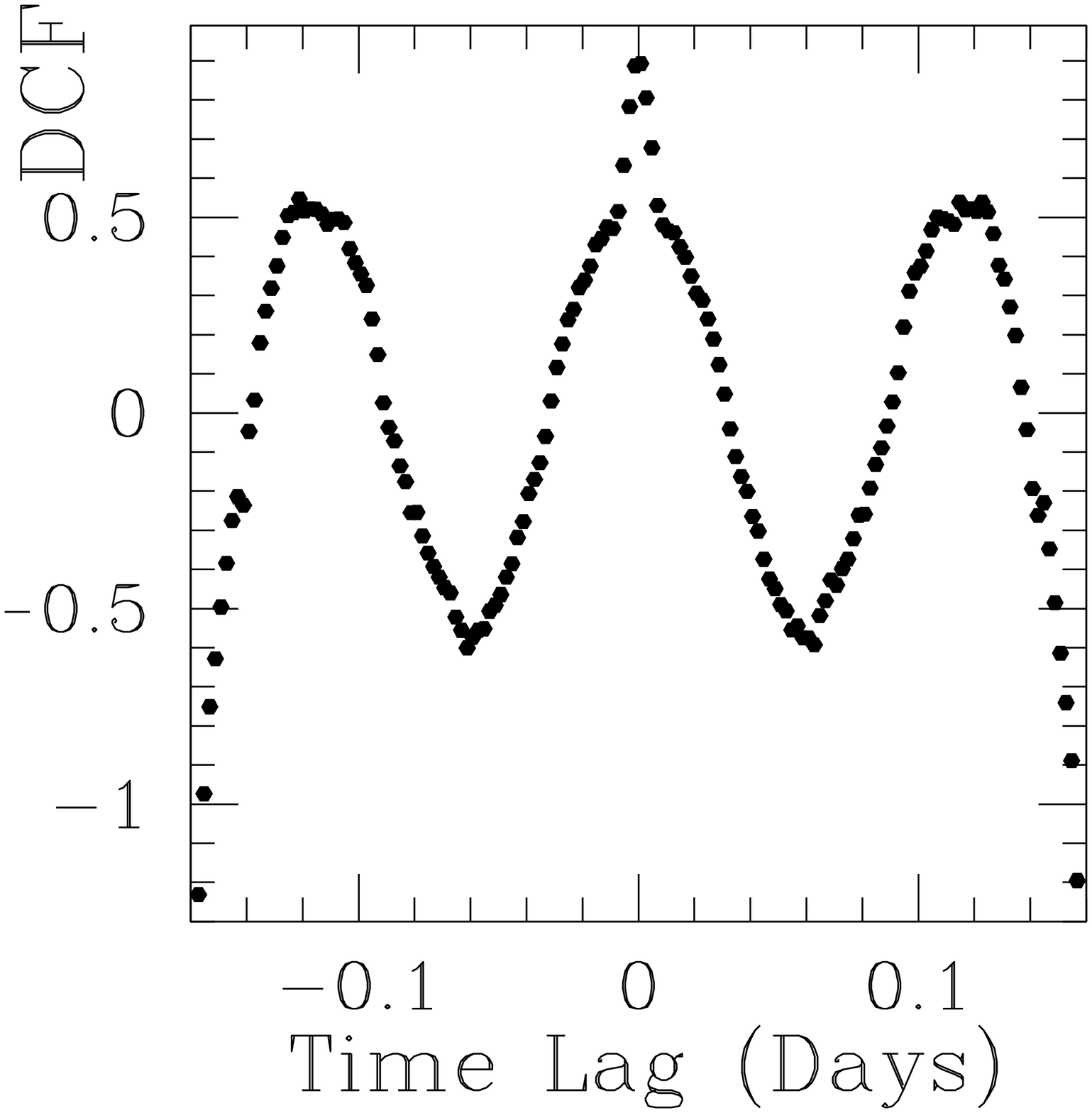,height=1.8in,width=2in,angle=0}
\caption{R-band optical IDV LC of the blazar S5 0716+714 and their respective SFs and DCFs. The remainder of these observations and analyses are presented as
online-only material.}
\end{figure*}
\subsubsection{\bf Short-term and Long-term flux variability}

The STV/LTV LCs of 3C 279 are displayed in Fig 5 in B, V, R, and I filters which includes our B, V, R, and I data points along with
the SMARTS (in B, V, R) and SO (in V passband only) data points covering the time interval from
JD 2454480  to JD 2456900.
The colour variation (B-I) \& (V-R) are also displayed in Fig 5 lower 2 panels.
Also, the details of the magnitude changes during the whole observation period are listed in table 4 for each passband.

{\bf B passband:}
STV/LTV LCs of the source in B band includes our data along with those provided
by SMARTS and is plotted in the top most panel of Fig 5. The source has recently been found to be in a
flaring state, with the brightest B magnitude in last 5 years noted as 15.101 on 2014 March 05. Thus it is still 3.8 mags
fainter than the 1936-1937 outburst, which is the classic outburst used for comparison, when it reached B $\sim$ 11.3.
The target has been highly variable on yearly basis with LTV amplitude to be 338\%, during the $\sim$ 5 years period.

{\bf V passband:}
The photometric observations of 3C 279 in V band includes our data
from seven telescopes mentioned in Table 1 along with SMARTS data sets and supplemented with the data from SO.
The STV/LTV LCs for V filter are shown in Fig.\ 5  where different colours refer to different
telescopes, as listed in the caption. Recently, the source was noted to show a rapid brightness increase
reaching the brightest level on 2014 March 5. The STV/LTV amplitude in V passband was calculated to be 338\%. The source
was found to undergo small decrease in brightness with V = 15.15 on 2014 on 2014 May 24 (JD 2456802.175).

{\bf R passband:}
The STV/LTV R band LC plotted in Fig.\ 5 is comprised of data from 7 telescopes along
with data sets obtained from SMARTS. The source was found to be in high state during 2014.
We calculated the amplitude of variation to be 263.6\%. The source was highly variable
during last 5 years and seems to be in an active stage presently.

{\bf I passband:}
3C 279 was observed in I band for STV/LTV purpose using the 7 telescopes mentioned in Table 1.
There were no data points from SMARTS in the I passband.
The last data point taken by us in this band was on 2014 May 24 when the source was found  to have I = 14.2; this
still  corresponds to a  bright state. The amplitude of variability calculated for the source during our shorter span of  observations reached only 43.2\%.

\begin{table*}
\caption{Fits to colour-magnitude dependencies and colour-magnitude correlation coefficients}

\noindent
\begin{tabular}{p{1.4cm}p{1.4cm}p{1.4cm}p{1.4cm}p{1.4cm}p{1.4cm}p{1.4cm}p{1.4cm}p{1.4cm}p{1.4cm}} \hline
Source &Date of   &\multicolumn {2} {c} {B$-$V vs V}  &\multicolumn {2} {c} {V$-$R vs V}  &\multicolumn {2} {c} {R$-$I vs V} &\multicolumn {2} {c} {B$-$I vs V} \\
&observation               &      $m^a$      & $c^a$               &     $m$      &      $c$         &  $m $       &   $c $           &   $m$      &    $c$           \\
        &       &    $r^a$       &$p^a$               &$r$           &$p$               & $r$        &$p$               &$r$        &$p$               \\\hline
3C 454.3 &25.09.2013     & $-$0.043     &~~1.487            & ~~0.322       &$-$3.969         & ~~0.102      &  $-$0.203          & ~~0.302     &    $-$2.473         \\
&               &$-$0.045     &~~0.833          & ~~0.496     & ~~0.014          & ~~0.163     & ~~0.445        & ~~0.288    & ~~0.173          \\
&28.09.2013     &~~~~*       & ~~~~*            &~~0.613       &   $-$8.568         &~~0.099      & $-$0.779        & ~~~~*     & ~~~~*            \\
&               &~~~~*      &~~~~*          &~~0.577     & ~~0.001         & ~~0.010     & ~~0.613         & ~~~~*    & ~~~~*          \\
&24.10.2013     & $-$1.004    &~16.562          &~~0.443     & $-$6.442       & $-$0.117   & ~~3.133           & $-$1.049  & ~16.953           \\
 &              & $-$0.462    &~~0.030          &~~0.594      & ~~0.003         &$-$0.145   & ~~0.509         &$-$0.488  & ~~0.040          \\
&25.10.2013     & $-$0.598    &~10.260           &~~0.452      & $-$6.611       & $-$0.063      & ~~2.291        & ~~0.077     & ~~0.576         \\
 &              & $-$0.301   &~~0.173          & ~~0.603    & ~~0.002          &  $-$0.15    & ~~0.580         & ~~0.039    &~~0.862         \\
               
&26.10.2013     & ~~0.156     & $-$1.818       &~~0.060       & $-$0.482        & ~~0.082      & ~~0.292        &~~0.414         & $-$4.425               \\
  &             &  ~~0.132    &~~0.559          &~~0.150     & ~~0.484         & ~~0.258     & ~~0.234         &  ~~0.367       &  ~~0.102              \\
            
&27.10.2013         & $-$0.393     &~~7.231            &~~0.256    & $-$2.438           & ~~0.078   & $-$0.488           & $-$0.153  & ~~4.170            \\
   &               &  $-$0.026         &~~0.106   &~~0.565         &~~0.005   &   ~~0.153       &~~0.497  & $-$0.130     & ~~0.607    \\
                  
&28.10.2013      & ~~0.479     & $-$6.574           & ~~0.799    &  $-$11.158         & $-$0.032      & ~~0.918       & ~~1.020  &$-$14.183           \\
    &           & ~~0.249    &~~0.202           & ~~0.849   & ~~0.00001         & ~~0.052     & ~~0.801          &~~0.530  & ~~0.004          \\
&29.10.2013         & 0.430     &  $-$5.640         &~~0.759       & $-$10.382        & ~~0.025   &~~0.322           & ~~1.167  &  $-$16.576           \\
     &          & ~~0.237    &~~0.277          &~~0.789     & ~~0.000        &~~0.028   &~~0.897          &~~0.598  &~~0.003          \\
&04.11.2013        &$-$0.733           &~12.352               &~~0.840       & $-$13.032        & ~~0.0481      & $-$1.775        & ~~0.171        & $-$1.068               \\
      &         & $-$0.338          &~~0.157               &~~0.763     & ~~0.0001        & ~~0.078     & ~~0.735         & ~~0.086       & ~~0.735               \\
  
   S5 0716 &20.03.2012 & ~~~~* & ~~~~*    &~~0.412   & $-$4.829  & ~~0.602 &$-$7.111  & ~~~~* & ~~~~*  \\
 ~~~+714&           & ~~~~* & ~~~~*    & ~~0.334   & ~~0.046  & ~~0.422 & ~~0.010  & ~~~~* & ~~~~* \\
 &22.03.2012 & $-$0.219 & ~~3.336   & ~~0.337 & $-$3.912  & ~~0.228  & $-$2.255 & ~~0.337 & $-$3.509 \\
 &           & $-$0.300 &~~0.165    & ~~0.579 &~~0.007    & ~~0.466 & ~~0.038   &~~0.490 &0.018 \\
 &23.03.2012 & $-$0.044 & ~~1.121  &~~0.137 &$-$1.350   &~~0.110 & $-$0.722   &~~0.068 &$-$0.030 \\
 &	     &$-$0.107 & ~~0.626   &~~0.476 &~~0.016    &~~0.373  & ~~0.072   &~~0.159 &~~0.468 \\
 &14.12.2012 & $-$0.700     &~~9.972  & ~~0.638  & $-$8.227   & $-$0.0344 & ~~1.019   & $-$0.039 & ~~1.190 \\
 & & $-$0.432  & ~~0.213    & ~~0.673 & ~~0.047      & $-$0.0293 & ~~0.936     & $-$0.022 & ~~0.952\\
\hline
\end{tabular}  \\
$^a$ $m =$ slope and $c =$ intercept of CI against V; $r =$ Pearson coefficient; $p =$ null hypothesis probability\\
\end{table*}
\subsubsection{\bf Colour variability}

We have also examined the colour variability in the source on short term basis and this is displayed in the bottom 2 panels of Fig 5.
The maximum difference in the (V-R) Colour for 3C 279 is 0.85 mag 
(between its colour range 0.29 mag and 1.14 mag) while for (B-I) difference was of 1.173 mag when maximum colour difference was 2.45 and minimum being
1.277.

\subsection{S5 0716+714}

The high declination ($\alpha_{2000.0}$ = 07h 21m 53.4s, $\delta_{2000.0}$ = $+71^{\circ} 20^{'} 36.4^{\prime \prime}$)
BL Lac is one of the most well-studied objects over entire EM spectrum and is an extremely
variable source on diverse timescales, ranging from minutes to years (Heidt \& Wagner 1996; Nesci et al.\ 1998;
Giommi et al.\ 1999; Raiteri et al.\ 1999; Gupta et al.\ 2008a,b).
S5 0716+714 is one of the most active and bright BL Lacs in the optical bands with a featureless optical-UV continuum 
(Biermann et al.\ 1981; Stickel, Kuehr, \& Fried 1993) thus making it very hard to estimate the redshift. 
Nilsson et al.\ (2008) determined the redshift of the source to be 0.31 $\pm$ 0.08 using the marginally detected host galaxy
as the standard candle plus its location close to 3 galaxies of red-shifts 0.26 (Bychkova et al.\ 2006).
In the past, S5 0716+714 has been found to have
duty cycle of 1 (Wagner \& Witzel 1995) implying the source to be always in active state and hence has been extensively
observed on IDV timescales (e.g. Montagni et al.\ 2006; Gupta et al 2008a,b and references therein) covering 5 major optical outbursts
with an interval of $\sim$ 3.0 $\pm$ 0.3 years (e.g. Gupta et al 2008b).

Gupta el al.\ (2009) used wavelet analysis to analyze the archived optical data of the source obtained
by Montagni et al.\ (2006) and found {\bf quasi-periodic oscillation (QPO)} components ranging between $\sim$ 25 \& $\sim$ 73 minutes on several nights. Later, a QPOs of
$\sim$ 15 min was reported by Rani et al.\ (2010b) at optical frequencies. Quirrenbach et al.\ (1991) claimed a possible QPO of
$\sim$ 1day in optical and radio bands simultaneously. 
\subsubsection{\bf Intra$-$Day Variability}

We observed S5 0716+714 on 12 nights between 2012 Jan and 2014 Feb, to search for IDV in multiband fluxes.
The calibrated LCs of the blazar are
displayed in Fig 3. To search and analyze blazar variability, we have employed the C- and F-tests, as discussed in Section 3.1.
The brightness changes on intra-day basis during the 12 nights are given in table 3 in each particular filter.

{\bf B passband:} 
The source was observed in the B filter on 3 nights for $\sim 4$~hrs on average. Following F-Test
we found the source to be variable on March 23 with amplitude of variability reaching a maximum of $\sim $13.5\%.

{\bf V passband:}
Our V passband observations of the target were carried out for 6 nights (2012 Jan 18, 21, March 20, 22, 23 and Dec 14).
Using F test the source was found to be variable on Jan 21 and Mar 23 when their variation exceeded
0.99 significance level with amplitude of variability reaching 11.17\%.

{\bf R passband:}
The BL Lac's monitoring in the R filter for 10 nights span the period from 2012 March to 2014 Feb with $\sim 4$~hrs average on each night.
On 6 of these nights the variations exceeded .99 significance,
with amplitude of variability reaching a maximum of $\sim$ 9.5\% on 2012 March 23. During those nights, the source was found to be in
an active state.

{\bf I passband:}
This source was monitored in I passband for a span of about 6 days occurring on 2012 Jan 18,21, March 20, 22, 23 and Dec 14. We tested
for IDV using both tests. Using the F-test we found that the source showed significance levels above 99\% for microvariability on 2 nights, 
2012 Jan 01 and March 03. During our observations,
the maximum amplitude of variability went up to $\sim 8.5$~\% by 2012 March 23.

\subsubsection{\bf Short-term and Long-term flux variability}

We investigated the BL Lac in B, V, R, and I filters during the period between JD 2454743.95454 and JD 2456804.65226 to study optical properties
corresponding to short term flux-variability. These LC plots are displayed in Fig.\ 6
along with the colour variation LCs of (B-I) and (V-R) in different panels.
Also, the details of the magnitude changes during the whole observation period are listed in table 4 for each passband.

{\bf B passband:} The {\bf STV/LTV} LC of S5 0716+714 in the B passband is displayed in the upper panel of the  figure where we used the data from the seven telescopes
mentioned in Table 1, observed between the above mentioned period. We calculated variability amplitude of $\sim$ 225\% in B band.

{\bf V passband:}
The {\bf STV/LTV} LC of the source in the V passband is displayed in the second panel from the top of the above mentioned figure with our data points using the 7 telescopes
mentioned in Table 1 and the data from SO. 
We calculated the variability amplitude
using equation 3 and found that the source varied by 306\%.

{\bf R passband:}
The LC for {\bf short/long} term studies of the source in R band is generated using the data taken from the
seven optical telescopes mentioned in Table 1.
The variability amplitude was calculated to be $\sim$ 240\% in R band.

{\bf I passband:}
The corresponding LC for I band is displayed in the third panel from the bottom of the above figure where data used is taken from the
seven optical telescopes mentioned in Table 1.
The maximum variation noticed in the target is 2.361 with {\bf STV/LTV} amplitude of $\sim$ 221\% in I band.

\subsubsection{\bf Colour variability}

The maximum variation noticed in the target for (B-I) is 0.817 (between its
colour range 1.382 mag on JD 2456007.3 and 2.199 mag on JD 2456694.3) while for (V-R) is 0.634 (between its
colour range 0.379 mag on JD 2456007.3 and 1.013 mag on JD 2456593.4).

\begin{figure}
\epsfig{figure= 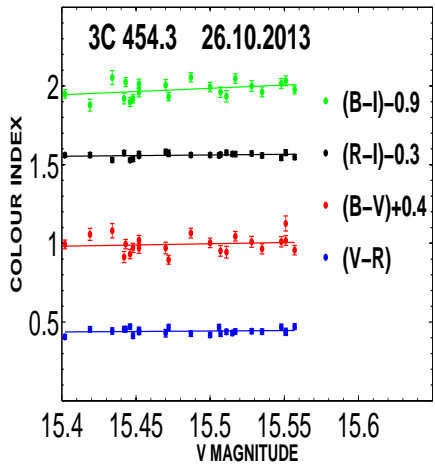 ,height=2.6in,width=2.5in,angle=0}
\caption{Colour magnitude plots on intra-day timescales for 3C 454.3.  
The V magnitudes are given on the X-axis and the various labeled colour indices are plotted 
against them for each labeled  date of observation. Colour magnitude plots for remaining observation dates are available online.
}
\label{colormagfigure}
\end{figure}

\section {\bf Cross-correlated variability }

To search for and quantify any possible correlations between the optical fluxes we used the DCF technique, 
as explained in Section 3.3.  We searched for possible time lags, which could be used to probe physical 
conditions in the inner regions of AGNs. The shortest possible timescale of variability
is normally associated with the changes occurring in
jet emitting region, whose time scales are shortened by Doppler beaming..

For 3C 454.3, on most of the nights the DCFs between the light curves in different
optical bands peak at small time lags ranging from 0 hrs to 0.07 hrs which can not be considered significant 
as these time lags are close to the measurement intervals or might be due to photometric and systematic errors 
on individual data points of both LCs. According to White \& Peterson (1994) and Peterson et al.\ (1998), any detection of time 
lags is limited to lags longer than the LC sampling time scale, which prevents us from measuring time lags 
on time scales less than a few minutes. In order to further investigate this issue we performed a SF analysis to see if there was a discernable 
time scale of variability by following the description given in Section 3.2.  This SF approach indicated 
that any time scale of variability is greater than or close to the total length of our observations, and 
therefore not reliable. Also, any nominally positive SF results were not supported by the DCF results. We
therefore conclude that our observations did not reveal any characteristic time scale of variability nor
any time delays among the various optical bandpasses. The small frequency intervals in the optical regime lead to null or small time lags which
implies that the photons in these wavebands {\bf are emitting by the same} physical process and from the same emitting region.

For the FSRQ 3C 279,  the SF for the R band LC
on Jan 29 gives a hint of variability at 72 minutes from both SF and DCF analysis, but it is likely to arise from gaps in the LC.
After that nothing significant is seen on 2014 Feb 3 and 5. For rest of the R band LCs of 3C 279 we do not find any timescale
of variability as the shape of the SF indicates that any timescale of variability that might be present is greater than the length of our
observation. Plots are given in the online-only material.

From the SF evaluation of S5 0716+714 for the LC of 2012 Jan 21 in I and V bands, we get a possible timescale
of variability $\sim$1.2 hrs in both the bands, which is to some extent supported by the DCF also.
On March 23 we observed the source in B, V, R and I.
The SF and DCF for R band LC indicate a variability timescale of 1.44 hrs, while for the B, V, and I bands for the same day we get variability timescales
of 1.92 hrs, 1.44 hrs and 1.44 hrs, respectively. The SF plot for the 2013 Oct 27 LC displays a continuous rising trend that indicates that
any timescale of variability exceeds the length of our observations. 
The nominal timescale of variability for the LC observed on 2014 Jan 03 is found to be 2.4 hrs which is also supported by DCF analysis.
Moving on to 2014 Feb we see from the R band LCs taken on Feb 3
shows a possibility of intra-day variability timescale of 1.44 hrs from the SF analysis which is supported by DCF technique
also. Whereas on Feb 4 and 5 we see a continuous rising trend followed by a dip, thus giving possible variability timescales of 72 min and 86.4 min
from both SF/DCF approaches. The SF and DCF plots for the target are displayed in the supplementary material
while a sample of the same is given in Fig.\ 7.

\section{\bf Colour-magnitude relationship}

In this section we investigate any relationship between the optical variations in the (B-V), (V-R), (R-I) and 
(B-I) colour indices of the source and the brightness variations, which can be used to study colour behaviour 
and variability scenarios.

The colour-magnitude plots for the {\bf 9 nights in case of 3C 454.3 (i.e., 2013 Sept 25, 28, Oct 24, 25, 26, 27, 28, 29, Nov 04) and 6 nights for
S5 0716+714 (i.e., 2012 Jan 18, 21, March 20, 22, 23 and Dec 14)}
are displayed in the Supplementary Materials section while a sample of the same is
displayed in Fig.\ 8.
Colour variation behaviours in blazars are still a topic of debate. A bluer-when-brighter (BWB)
trend is commonly observed in blazars (e.g., Raiteri et al.\ 2001; Villata et al.\ 2002; Papadakis et al.\ 2003; Clements et al.\ 2001;
Papadakis, Villata, \& Raiteri 2007; Rani et al.\ 2010; Agarwal \& Gupta 2015), in particular those
of the HSP class, in which the optical continuum is generally believed to be entirely dominated by the
non-thermal jet synchrotron emission. In LSP blazars, in which the accretion disc can provide a substantial
contribution to the optical continuum, a redder-when-brighter trend sometimes indicates the increasing
thermal contribution at the blue end of the spectrum, with decreasing non-thermal jet emission (e.g., Miller 1981; Villata et al.\ 2006;
Raiteri et al.\ 2007; Gaur, Gupta, \& Wiita 2012). 
For each night, we have calculated a linear fit of the colour indices, CI, against
V magnitude: CI =$m$V + $c$. The corresponding fit values of the slope, $m$, and the constant,
$c$, are reported in Table 5. The linear Pearson correlation coefficient, $r$, and the corresponding null 
hypothesis probability, $p$, are also reported there.  A positive slope indicates a positive correlation 
between the colour indices and apparent magnitude of the blazar which implies the general trend of BWB 
(or redder-when-fainter) whereas a negative correlation is observed for a negative slope, indicating a 
redder-when-brighter behaviour (RWB).

For 3C 454.3, we found significant positive correlations ($p \le 0.05$) between the V magnitude and most of the
colour indices on almost all nights except on 2013 Oct 24 when the object exhibited a negative correlation. 
Thus, the BWB trend was prominent during our observations. That the spectrum steepens as the brightness 
decreases can be clearly seen in the plots of 2013 Sept 25 (V-R), Sept 28 (V-R), Oct 25 (V-R), Oct 27 
(V-R), Oct 28 (V-R and B-I), Oct 29 (B-I and V-R) and Nov 04 (V-R). However, a RWB trend was found on 24 Oct (B-V and B-I) when 
the spectrum becomes steeper as the source brightens, as evident from the plot. For other colours in the 
respective nights no correlation above the 95 per cent confidence level was present. Thus the FSRQ showed 
multiple instances of positive correlations and one single instance of a negative correlation, along 
with several colour-magnitude sets showing no correlations at all. In general, the BWB trend was prominent
during our observations.

\begin{figure}
\hspace{-0.2in}
\epsfig{figure= 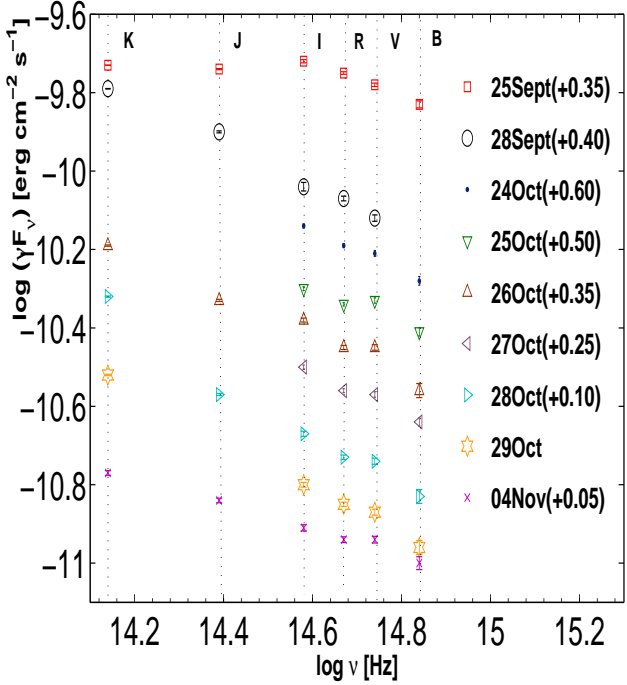,height=3.8in,width=3.6in,angle=0}
\caption{SED results for 3C 454.3 in near-IR-optical frequency range.}
\end{figure}
According to Sasada et al.\ (2010), the colour--magnitude relationship varies among the outburst state, the
active state and the faint state. 3C 454.3 followed a BWB trend in the outburst state, whereas it exhibited 
a RWB trend in the faint state and again a BWB trend was associated with an active state. From Section 4, 
during the time span of our observations 3C 454.3 was $\sim 3.4$~mag fainter than the brightest known magnitude of 
$R \sim 12$, but brighter by $\sim 1.6$ mag than the faintest level of $R = 17$ mag. Hence it seems to be 
classified best as being in an active state (out of the three states above) and the colour behaviour 
obtained by us is consistent with those reported in Sasada et al.\ (2010) and Gu \& Ai (2011), according to which 
the RWB trend is rarely seen for FSRQs.

Though usually dominated by jets, the optical emission of blazars is generally a combination of jet and 
accretion disc photons. The optical emission from 3C 454.3 is possibly contaminated by thermal emission 
from the accretion disc, providing a slowly and weakly variable ``blue'' emission component, which may
make a substantial contribution to the blue -- UV continuum, as FSRQs are usually low-frequency synchrotron peaked 
sources (e.g., Abdo et al.\ 2010). The BWB trend, prevailing when the jet synchrotron emission completely
outshines the accretion-disk emission, can be explained through an episode of more efficient acceleration
of relativistic particles in the jet. In a relatively simple model, this can be represented an injection 
of fresh electrons with an energy distribtion harder than that of previous, partially cooled electrons (e.g., Kirk, Rieger, \& Mastichiadis
1998; Mastichiadis \& Kirk 2002). More detailed simulations of such a scenario, based on the solution of a
Fokker-Planck equation of the electron distribution including first- and second-order Fermi acceleration,
have been done (e.g., Chen et al.\ 2011; Diltz \& B\"ottcher 2014). The latter work has demonstrated
that, if an increased level of optical synchrotron emission is caused by a temporary decrease of the
stochastic particle acceleration time scale, optical and (synchrotron self-Compton dominated) X-ray
emissions are expected to be anti-correlated, with the X-ray variations lagging behind the optical
variations by several hours. { \bf Instead, Raiteri et al.\ (2011) found a good correlation with a delay
of about 1 day.} This is a prediction that could be tested with future, co-ordinated
optical and X-ray monitoring observations of 3C 454.3. 

We found significant positive correlation for S5 0716+714
e.g. on 2012 March 20 (V-R and R-I), March 22 (V-R), (R-I) \& (B-I),March 23 (V-R) and also on Dec 14 (V-R). No significant negative correlations are found, thus indicating
that the BL Lac tends to be bluer (flatter) when brighter on timescales of days, which may be due to the contribution of two
components to the optical emission: one is the variable with a flatter slope ($\alpha_{1}$) (f$_{\nu}$ $\propto$ $\nu^{-\alpha}$)
and the second one being the stable one with $\alpha_{const}$ $>$ $\alpha_{1}$. The BWB trend for this source was also given by Villata
et al.\ (2000) during the 1999 WEBT campaign in a 72h optical LC. In the case of BL Lacs, any colour changes owing to accretion disc
are highly unlikely, as the disc radiation is overwhelmed by that from the Doppler boosted jet emission. Therefore,
shock based jet models are most likely to explain the observed colour variations 
(e.g., Marscher \& Gear 1985; Marcher et al.\ 2008), since radiation at higher frequencies is expected to emerge first followed by the flux at lower frequencies,
thus producing a bluer colour during the early phase of a flare while a redder one appears during the later observation of the same flare,  as it decays.

\section{\bf Spectral Energy Distributions}

SED properties of a source are an excellent way to investigate theoretical models. Since simultaneous multi-frequency
observations are difficult to attain it is much simpler to focus on a narrow region of EM spectra like NIR/optical, which
could provide vast amount of information on flux and spectral variability that in turn can provide evidence about the emitting regions of
the relativistic electrons, constrain the gamma ray emission models (Ghisellini el al.\ 1997) and the contributions of emissions from jets
(synchrotron), BLR, accretion disc, host galaxy and surrounding regions.

Using our B,V,R and I observations along with the J \& K band data points for 3C 454.3 available on the SMARTS webpage, 
for each particular night, we generated 9 NIR -- optical SEDs spanning  2013 Sept 25 - Nov 04.
For that purpose, we have subtracted Galactic extinction, $A_{\lambda}$, obtained from the NASA/IPAC Extragalactic 
Database\footnote{\tt http://ned.ipac.caltech.edu} from the optical/NIR data points, to obtain extinction-corrected 
fluxes, which were then converted to $\nu F_{\nu}$ fluxes. The resulting SEDs are displayed in Fig. 9.
From the figure it appears that a bump peaking in the V-B frequency range is present in the lower-brightness
optical SEDs. It most likely corresponds to the little blue bump observed in quasars in the rest wavelength 
range $\sim$200 -- 400 nm for which the strong emission lines from the BLR seem to be 
responsible (e.g., Raiteri et al.\ 2007). As expected, this feature disappears in the two brightest SEDs.
The brightest SED for 3C 454.3 was measured on 2013 Sept 25. Minor variations were seen in the SEDs for 
25 -- 28 Oct after which it decayed day by day, reaching a minimum on 2013 Nov 4.

\begin{figure}
\hspace{-0.2in}
\epsfig{figure= 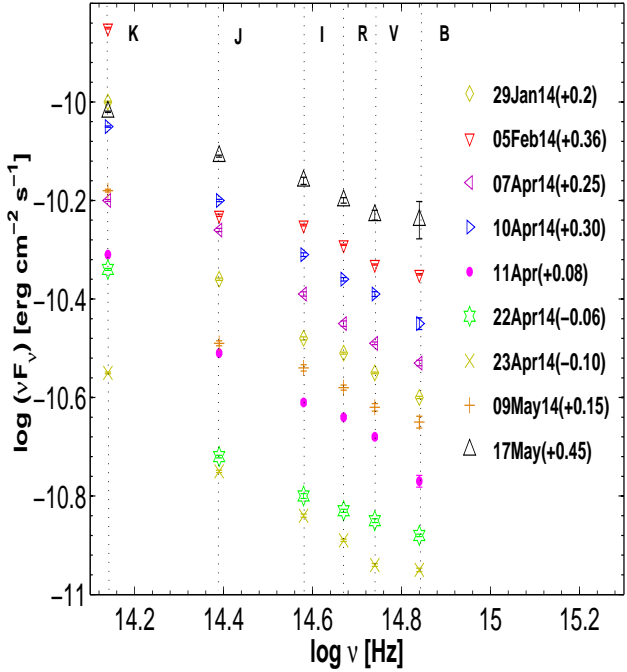,height=3.8in,width=3.6in,angle=0}
\caption{SED results for 3C 279 in near-IR-optical frequency range.}
\label{sedfigure2}
\end{figure}

Fig. \ref{sedfigure2} shows the SED of 3C 279 during 2014 Jan to May covering NIR to optical. We selected 9 days when there were quasi-simultaneous
observations in B, V, R, I, J and K to perform a limited spectral analysis of the source. The NIR and optical
data have been corrected for Galactic extinction following Cardelli, Clayton, \& Mathis (1989) using total to selective extinction ratio
$A_v/B_{B-V}$ = 3.1 (Rieke \& Lebofsky 1985). We find significant changes in NIR region. We apparently are seeing a signature of one of the broad bumps
peaking in the NIR region, which should be the synchrotron component as it peaks in IR/optical bands in case of red blazars (located between
10$^{12.5}$ -- 10$^{14.5}$ Hz). We get the brightest and faintest SEDs for 2014 Feb 05 and Apr 23, respectively.

Fig. 11 shows optical-NIR SED for the source S5 0716+714, built with our B, V, R, I data sets for
7 different epochs with different brightness levels. Magnitudes in all bands were corrected for Galactic extinction.
The synchrotron peak frequency has  values  between 10$^{13}$Hz and 10$^{17}$Hz.
The SEDs seem to indicate a synchrotron peak in the near-IR.
Due to the lack of multi-wavelength data here, a detailed description is difficult.

\section{\bf discussion}

The three blazars discussed in this paper are highly active, displaying outbursts across the whole electromagnetic spectrum. 
We explored the variability properties of 3C 454.3, 3C 279 and S5 0716+714 using seven telescopes in Bulgaria, India and Greece during 
the 2011 through 2014 observing seasons. We also searched for colour variations on IDV and short-term timescales.

Models to explain diverse time scale variability are broadly classified as intrinsic and extrinsic. Intrinsic 
origins of AGN variability are those associated with variations in the accretion flow and the relativistic jet.
Extrinsic mechanisms include interstellar scintillations which, due to its frequency dependence, are only 
relevant in low frequency radio observations and can therefore not be the case of optical INOV. Another 
external variability cause is gravitational microlensing, which is achromatic and is unlikely to be the sole cause of IDV
(Wagner \& Witzel 1995). Possible IDV mechanisms in radio-loud AGN are generally believed to be 
related to conditions in the jet, while in radio quiet quasars variability is likely due to the intrinsic 
variability of the accretion disc or related to a weak blazar component (e.g., Stalin et al.\ 2005). The 
absence of IDV may be due to a relatively stable relativistic jet with no irregularities in the jet flow.

The optical and multiwavelength flux variations in blazars, particularly in their high state, are often 
interpreted in terms of relativistic shocks in the Doppler boosted relativistic jets pointing at or nearly 
along our LOS (e.g., Marscher \& Gear 1985). In an unsteady jet, various instabilities,
including turbulence behind shocks, changes in the direction from the LOS or other geometrical effects 
occurring within the jets may contribute to the variability of the non-thermal emission from blazars (e.g., Camenzind et al.\ 1992; 
Gopal-Krishna \& Wiita 1992).
However, the observed variability in the low state, when the non-thermal jet emission 
is less dominant over thermal emissions from the accretion disk, can sometimes be explained by accretion-disk-based 
models (Wagner \& Witzel 1995; Urry \& Padovani 1995; including hotspots or instabilities in or above the accretion discs itself (e.g., Chakrabarti \& Wiita 1993).
As the blazars are AGNs seen nearly face on, any irregularities related to the accretion
disc will be visible directly in the low state. 

Blazar emission in the post-outburst stage may be plausibly described in the framework of a shock-in-jet model
(e.g., Marscher \& Gear 1985; Spada et al.\ 2001; Graff et al.\ 2008; Joshi \& B{\"o}ttcher 2011).
Most of the blazar emission, especially in the active phase, is
believed to be due to their jets lying close to the LOS and hence relativistically Doppler boosted. Shocks propagating down these
relativistic jets are likely to play important roles in explaining the observed IDV with emissions due to shock regions emitting over
multiple wavebands with variability timescale decreasing as frequency increases. Disc induced fluctuations when advected into the jet could be responsible for the changes in the physical
parameters of the jet like its velocity, density or magnetic field (e.g.\ Wiita 2006).  Even slight variations caused in that way are Doppler boosted by
a factor of $\delta^{2}$ to $\delta^{3}$ (e.g.\ Blandford \& Rees 1978) whereas the observed variability timescale
is compressed by $\delta^{-1}$ (e.g.\ Gopal-Krishna et al.\ 2003).

\begin{figure}
\hspace{-0.2in}
\epsfig{figure= 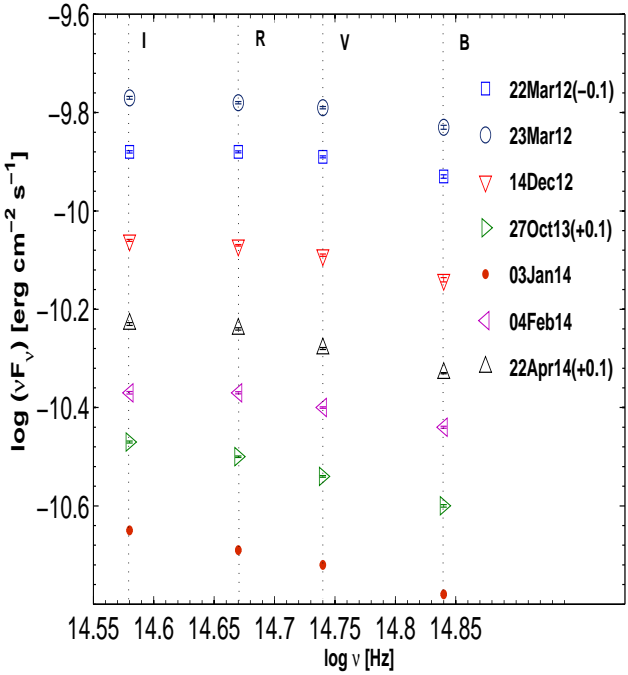,height=3.8in,width=3.6in,angle=0}
\caption{SED results for S5 0716+714 in optical frequencies.}
\end{figure}

The presence of the quasi-power-law continuum is an indication of the synchrotron nature while flux variability was associated with a
hardening of the synchrotron spectrum, implying a hardening of the underlying non-thermal electron distribution. 
The lack of any time delays among the optical bands is well explained by the small difference of the respective 
observing frequencies, implying small differences in the dynamical time scales of the radiating electrons, which
are likely shorter than the expected light-travel time across the jet, with typical radial dimensions of the order
$R \sim 10^{15}$ -- $10^{16}$~cm. Thus, any microphysical time delays would be washed out by light-travel time
effects, even if the observational sampling rate would enable the detection of such delays. The hardening of
the underlying electron distribution, leading to the observed BWB trend, may be well explained by a correlation
between increasing synchrotron flux and more efficient electron acceleration, leading to an increased population
of the highest-energy electrons. A plausible physical mechanism to achieve this would be an increased level of
hydro-magnetic turbulence in the active region of the blazar jet, which would, to first order, lead to a shortening
of the characteristic particle acceleration time scale. Such a scenario has recently been modeled by Diltz \& B\"ottcher (2014)
who showed that individual flaring events caused by such  parameter variations would lead to the
observed BWB trend, along with an anti-correlation between optical and X-ray fluxes, with X-ray variations
lagging the optical variability by up to several hours. This could be tested with future, co-ordinated optical
and X-ray monitoring observations of $\gamma$-ray blazars, in particular 3C 454.3.
Of course optical
variability of BL Lacs could be due to more than one mechanism, i.e., a ``mild chromatic'' longer term component and a strongly chromatic
shorter term component which are due to Doppler factor variations, and intrinsic variations caused by particle acceleration in the Doppler
boosted relativistic jets (e.g., Mastichiadis \& Kirk 2002), respectively.

\section {\bf Conclusions} 

We have performed multiband optical photometry for the blazars 3C 454.3,  3C 279, and  S5 0716+714 between 2011 and 2014 to study flux and spectral
variability on intraday, short and long time-scales.

1. During our $\sim$ 37 nights of monitoring for IDV we have found genuine flux variations, using C- and F-tests, on intraday time-scales 
for $\sim$ 21 nights in B, V, R or I with amplitudes of variability up to 31 per cent for  3C 454.3, 9.2 per cent for 3C 279 and
13.5 per cent for S5 0716+714. 
 Each of the blazars seemed to be in an active stage and highly variable
during the time span of our studies. The two FSRQs, 3C 454.3 and 3C 279, both were recently in flaring states during the middle of the year 2014.

2. The STV was relatively weak for each blazar during our whole monitoring period.

3. Strong intra-night colour variations were found in some colour indices for some nights, but no consistent
colour variations were found on a yearly basis.

4. No significant time lags between the optical wavebands were detected, in agreement with expectations due
to the small frequency separations of the optical wavebands.

5. A BWB trend was observed to be dominant in our objects during our observations. Such flattening of
the optical spectrum with increasing flux may be interpreted as being dominated by jet synchrotron emission,
where increasing flux is related to a hardening of the underlying non-thermal electron spectrum, possibly
indicating an enhanced particle-acceleration efficiency. 

6. The NIR -- optical spectral properties of each blazar were studied by generating SEDs for each night with 
sufficient data points in B, V, R, I, J and K filters.

\section*{Acknowledgments}
We thank the referee for very useful comments which helped us to improve the manuscript. 
This paper has made use of up-to-date SMARTS optical/NIR light curves that are available at 
www.astro.yale.edu/smarts/glast/home.php. SMARTS observations of LAT monitored blazars are supported by 
Yale University and Fermi GI grant NNX 12AP15G, 
and the SMARTS 1.3m observing queue received support from NSF grant AST-0707627. 
The Steward Observatory AGN studies are supported by Fermi Guest Investigator grants NNX08AW56G, NNX09AU10G, 
and NNX12AO93G. 
ACG  is partially supported by the Chinese Academy of Sciences Visiting Fellowship for Researchers from 
Developing Countries (grant no. 2014FFJA0004). HG  is sponsored by the Chinese Academy of Sciences Visiting Fellowship 
for Researchers from Developing Countries (grant No. 2014FFJB0005), and supported by the NSFC Research Fund for International 
Young Scientists (grant no. 11450110398). MFG acknowledges the support from the National Science Foundation of China (grant no. 
11473054) and the Science and Technology Commission of Shanghai Municipality (grant no. 14ZR1447100).
This research was partially 
supported by Scientific Research Fund of the Bulgarian Ministry of Education and Sciences under grant 
DO 02-137 (BIn 13/09) and by Indo-Bulgaria bilateral scientific
exchange project INT/Bulgaria/B-5/08 funded by DST, India. The Skinakas Observatory is a collaborative project of 
the University of Crete, 
the Foundation for Research and Technology -- Hellas, and the Max$-$Planck$-$Institut f{\"u}r 
Extraterrestrische Physik. The work of MB is supported by the South African Research Chair Initiative 
by the National Research Foundation and the Department of Science and Technology of South Africa through 
SARChI Chair grant No. 64789. 

{}

\vspace{0.1in}
{\bf SUPPLEMENTARY MATERIAL} \\

The following supplementary material is available for this article online.
Additional Supporting Information regarding them may be found in the online version of this article. \\

{\bf Table 1, 2 and 3.} Complete details 
of the observations are listed in Tables 1, 2 and 3 of supplementary material available for this article.

{\bf Table 2.} Results of IDV observations of the blazars. The full version of this
table is available in the online version of this manuscript. \\

{\bf Figure 7.} IDV LCs of blazars, their respective SFs and DCFs (sample). The full version of this Figure is available in the online
version of this manuscript in the Supplementary Materials section. \\

{\bf Figure 8.} Colour magnitude plots on intra-day timescales (sample). The full version of this Figure is available in the
Supplementary Materials section. \\

\clearpage

\end{document}